\documentclass{elsarticle}
\usepackage[caption = false]{subfig}
\usepackage[multiple]{footmisc}
\usepackage{eurosym}
\usepackage[final]{changes}
 \usepackage{hyperref}   

\usepackage{amsmath,amssymb,amsthm,mathrsfs,amsfonts,dsfont} 

\usepackage{amsfonts}
\usepackage{array}
\usepackage{amsthm}
\usepackage{epstopdf}
\usepackage{amsmath}
\usepackage{setspace}
\usepackage{algorithm,algorithmic}
\usepackage{natbib}
\usepackage{color}
\usepackage{rotating}
\usepackage{tikz}
\usepackage{titlesec}
\usepackage[toc,page]{appendix}
\usepackage[font=small,skip=0pt]{caption}

\newcommand*\colvec[3][]{
    \begin{pmatrix}\ifx\relax#1\relax\else#1\\\fi#2\\#3\end{pmatrix}
}

\newcommand*\colvecfive[5][]{
    \begin{pmatrix}\ifx\relax#1\relax\else#1\\\fi#2\\#3 \\#4\\#5\end{pmatrix}
}

\numberwithin{table}{section}
\numberwithin{figure}{section}

\titleformat{\paragraph}
{\normalfont\normalsize\bfseries}{\theparagraph}{1em}{}
\titlespacing*{\paragraph}
{0pt}{3.25ex plus 1ex minus .2ex}{1.5ex plus .2ex}

\graphicspath{{figures/}}  

\journal{}
\begin{document}

\begin{frontmatter}
\title{The multilayer architecture\\ of the global input-output network and its properties}

\author[Bic1]{Paolo Bartesaghi}
\author[Catt]{Gian Paolo Clemente}
\author[Bic1]{Rosanna Grassi\corref{cor1}}
\author[Kiel]{Duc Thi Luu}
		
\cortext[cor1]{\emph{Corresponding author. email: rosanna.grassi@unimib.it}} 
\address[Bic1]{University of Milano - Bicocca, Via Bicocca degli Arcimboldi 8, 20126 Milano, Italy\\ email: paolo.bartesaghi@unimib.it; rosanna.grassi@unimib.it}
\address[Catt]{Universit\`{a} Cattolica del Sacro Cuore di Milano, Largo Gemelli 1, 20123 Milano, Italy\\ email: gianpaolo.clemente@unicatt.it}
\address[Kiel]{Department of Economics,  University of Kiel,  Germany\\ email: d.t.luu@economics.uni-kiel.de}
		
\begin{abstract}
We analyze the multilayer architecture of the global input-output network using sectoral trade data (WIOD, 2016 release). With a focus on the mesoscale structure and related properties, our multilayer analysis takes into consideration the splitting into industry-based layers in order to catch more peculiar relationships between countries that cannot be detected from the analysis of the single-layer aggregated network.

We can identify several large international communities in which some countries trade more intensively in some specific layers. 
However, interestingly, our results show that these clusters can restructure  and evolve over time.  In general, not only their internal composition changes, but the centrality rankings of the members inside are also reordered, \added{industries from some countries diminishing their role and others from other countries growing importance.}
These changes  in the large international clusters may reflect the outcomes  and the dynamics of cooperation, \added {partner selection} and competition among industries and among countries in the global input-output network. 
\end{abstract}

\end{frontmatter}

\section{Introduction}

\label{Intro}

The network architecture plays a central role in explaining the propagation of shocks and the robustness of \added {different financial and economic systems}.  
\added{The 2007-2009 global recession and the subsequent European debt crisis have shown that the distress of some financial institutions in a particular country could be easily transmitted across borders into other countries, because of international inter-dependencies and the very high degree of financial integration \citep[see, e.g.,][]{Cetorelli_Goldberg_2011, Shin_2012, Stefanie_et_al_2013,  Bostandzic_et_al_2018, Hale_et_al_2019, Park_Shin_2020}.}
From the sphere of real economic activities,  
in an economic system of heterogeneous \added{and} interdependent agents (e.g., industries or firms), 
\added{the understanding of the structure of their interactions }
plays a crucial role in exploring how shocks of a specific agent can be propagated to the others \added{possibly leading} to a large aggregate fluctuation or a systemic failure in the economy   \citep [see, e.g.,][]{Acemoglu_et_al_2012, Carvalho_2014, Acemoglu_et_al_2016, Acemoglu_et_al_2017, Atalay_2017, Eppinger_et_al_2020}.  
\added{It has been also suggested that such understanding at an international scale is useful in examining the cross-country transmission and the propagation of local shocks \citep [see, e.g.,][among others] {Carvalho_et_al_2016, Acemoglu_et_al_2016,  Luu_et_al_2018b, Luu_et_al_2018a,  Boehm_et_al_2019, Eppinger_et_al_2020, Giammetti_et_al_2020}.}
In summary, through input-output linkages in a production network, shocks to a particular node may have two potential 
effects on other nodes:  (i)  supply-side shocks generate  propagation to the downstream  customers,  and (ii) demand-side shocks  generate effects on upstream   suppliers \citep[e.g. see][]{Acemoglu_et_al_2016}. Furthermore, direct and indirect propagations capture both the first-order impact on the ``nearest \added{neighbors}'' of the affected industry as well as the higher-order impacts on those who are the  customers of  the customers or the  suppliers of  the suppliers, and so on.
 As shown in \cite {Acemoglu_et_al_2012}, 
not only lower-order structural properties among sectors, \deleted{such as the node degrees or strengths,}
but also more complex properties of the economic network can play a defining role in explaining cascade effects in the whole system. More specifically, even if  two different networks have the same first-order characteristics (e.g. 
same degree or strength sequence) different higher-order interactions among industries may lead to different severity levels of aggregate downturns \citep[][]{Acemoglu_et_al_2012}.\footnote{Different higher-order of connectivity coefficients can be defined to capture more complex patterns of cascades in an input-output network. For example, at the third-order level of inter-connection, higher aggregate fluctuations could be observed if in the network the sectors having high strengths also share many common suppliers \cite [see][for further details] {Acemoglu_et_al_2012}.}

More in general, once the network structure is taken into consideration, the understanding of  topological properties, from a microscopic to a macroscopic perspective, is crucial 
\added{to assess }how a local shock or disruption to a specific node or cluster of nodes can be propagated and amplified to the rest of the system.  \added{This is why the understanding of the higher-order topological properties \added{(such as those at the mesoscale level like the community structure studied in the present work)} may have meaningful implications in managing complex economic systems.} 
In contrast to homogeneous random \added{and regular} graphs, it is often observed that real networks properties are more heterogeneous \citep [e.g. see][]{ Newman_2003}. 
For example, some nodes may have intensive interactions with many other nodes while some others only concentrate on few partners (heterogeneity in the degree and strength sequences). Furthermore, at a broader scale, nodes and edges may fall into different \added{sets} such that the internal interactions within these groups are stronger than those between different \added{ones} (heterogeneity in clustering \added{behaviors}).
 The latter feature is often referred to as the community structure, which has been intensively studied in the literature for a wide range of networked systems  \citep[e.g. see][and the literature therein for the detection of this structure in various networks from different fields]{FORTUNATO_2010}. 
 
 \added{From a perspective of policy implications, } the identification of such a mesoscale structure can be successfully utilized to assess the functioning, the stability and the evolution of networks. Let us consider an economic system of interacting agents as an example.  
 First, if agents can be classified into different blocks, one could examine whether a local shock \added{or a ``communication"} will spend most of the time or will even be trapped in a particular block or will be spread widely into the whole system. \added{However, in the presence of large blocks (e.g.  referring to the present work, large international trading communities), the diffusion of a shock may have a pervasive effect on many other internal nodes.} 

\added{Moreover,} further analysis of the importance and position of nodes within each of these groups may also provide meaningful implications, e.g. (i) central nodes that have  many links to the other partners may play a crucial role in the functioning of that group, and (ii) boundary nodes that are mutually connected with those from different blocks may play an important role in blocks' communication and act as the ``gatekeepers'' or ``middle men'' that spread shocks \added{and carry on them from a block to another one}. 

On top of this, such an identification can be used to explore the network origin of the business cycle co-movements among sectors or countries,  since agents who belong to a tightly connected cluster \added{(or community)}  may tend to synchronize more their economic activities  \citep [e.g. see][among others] {Long_Plosser_1983, Burstein_et_al_2008, diGiovanni_Levchenko_2010, Johnson_2014,  diGiovanni_et_al_2018}.  
For example, in a micro-founded model of business cycles, \cite{Long_Plosser_1983} show that input-output interdependencies could be an important factor responsible for the co-movements among the outputs of different industries in the economy. At the international scale, \cite{Burstein_et_al_2008} suggest that countries more engaged in sharing international productions exhibit a higher level of co-movement of business cycles. \added{In another study,  \cite{diGiovanni_Levchenko_2010} find that nations with strongly trading relations among themselves also often show a higher level of business cycle synchronization. In a similar vein, from a micro perspective, \cite{diGiovanni_et_al_2018} report that French firms,  that have stronger trade linkages with a particular foreign country, also tend to have with it a higher correlation. 
}  
\added{Among different observed economic networks, perhaps the global trade system has received a more remarkable attention, due to its economic importance and data availability. It has been widely suggested that the world trade system exhibits a heterogeneous structure at different orders  of interconnection.
For example, using the methods of network analysis and analyzing  the network of  merchandise trade imbalances between countries over the period from 1948 to 2000, \cite{Serrano_et_al_2007} find the presence of both local heterogeneity (when only few trade linkages possess a large percentage  of the country’s total flow) and global heterogeneity (when only a small proportion of all the trade linkages in the whole network carries most of its total flow). \cite{Fagiolo_et_al_2008}, carrying out a weighted network analysis over the period 1981-2000, show that many weak trade linkages coexist with some much stronger ones, and  countries with intense trade relationships tend to be more clustered together. \cite{Benedictis_Tajoli_2009} analyze  the World Trade Network (WTN) over the years from 1950 to 2000 and confirm the presence of trading blocks with a strong heterogeneity in the way countries select their trade partners. 
Focusing on the mesoscale structure of the WTN, \cite{Bartesaghi_et_al_2020}  show that the global trade system can be decomposed into different communities in which  the internal interactions among members  in each group are more intensive than the external ones. In addition, within each group, there exist some ``main, central" countries ``surrounded" by the other ``satellite" members. The persistence of communities in the global trade system has been also investigated  during the COVID-19 pandemic, focusing on centrality measures that are meaningful in capturing possible structural modifications \citep[][]{Antonietti2022}.}
\added{Furthermore, although the global trade system has a certain degree of persistence over a definite time horizon, if we investigate deeper the temporal dynamics of different structural properties of the network, we can observe an evolution over time \cite[e.g., see] []{Serrano_et_al_2007, Benedictis_Tajoli_2009}.  The driving forces of such temporal  dynamics can be traced back to several concepts and factors suggested by trade theories. For example, due to the extensive margin effects, the system can expand if new trade linkages are established.
In contrast, with the intensive margin effects, the network dynamics can be driven by  changes in the strength of the trade relationships over time}\footnote{In this case, the binary links already exist but the weights can become more (or less) intensive over time.} \citep[e.g., c.f.][and the literature therein]{Felbermayr_Kohler_2006, Helpman_et_al_2008}.
\added{There could be other factors governing the evolution of the global trade system, such as changes in the country fitness \citep[e.g.,][]{Anderson_1979, Garlaschelli_Loffredo_2005}, new partnerships of bilateral and multi-bilateral trade agreements or currency unions \citep[e.g.,][]{Subramanian_Wei_2007, Helpman_et_al_2008, Benedictis_Tajoli_2009}. \cite{Mundt2021} proposes an empirical model of network formation that takes into account the endogeneity of structural network characteristics. Applied to the Input-Output data, the model shows significant fluctuations of network ties over time, inducing changes also in the aggregate network structural properties.}
While much of empirical analysis of  economic networks has so far either focused on single types of relations separately or combined  all  of them together into an aggregated one, less attention has been paid to a much richer  structure   of these networks, i.e. the so-called multilayer architecture \citep [e.g.][]{Biancon_multilayer_book_2018, Battiston_et_al_2018}. In fact, in many real networks, various types of interactions may coexist: \textit {linkages among different nodes of the same type, linkages among different nodes from different types, and those between nodes with their copies (replicas)}.
\added{For example, in the financial system, banks often interact among themselves in different layers via different channels and various markets \citep[e.g. see][]{Carlos_et_al_2014, Poledna_et_al_2015,  Bargigli_et_al_2015, Luu_Lux_2019}. In a similar vein, the global trade system exhibits a multilayer \added{structure}, which should be considered as one of the critical factors for the emergence of multiple channels of cross-border propagation of economic crises \citep[e.g.][]{Lee_Goh_2016}. In the increasing  integration of  global economies,  each country may simultaneously trade with other countries on many different commodities-based layers   \citep[e.g. see][]{Barigozzi_et_al_2010, Gemmetto_Garlaschelli_2014, Gemmetto_et_al_2016}. \cite{Barigozzi_et_al_2010} study the properties of the multilayer trade network where each of 97 layers represents  trade relations  among 162 countries in a particular commodity over the 1992-2003 period. Interestingly, their findings suggest that  the distributions of link-weights (trade intensities),  the averages of connectivity, the clustering coefficients and centrality levels vary across commodity-based layers and are very different from those of the aggregated trade network. The heterogeneity and dissimilarity among layers may reflect the increasing specialization process in the global trade network when countries tend to focus on certain products. \cite{ Gemmetto_Garlaschelli_2014} and  \cite{Gemmetto_et_al_2016} apply a statistical network approach to examine the role of the intrinsic heterogeneity in the local constraints like the degree sequences and/or strength sequences in explaining the overlaps and correlations among layers in the undirected and the directed versions of the international trade network. The network is comprised of more than 200 nodes (countries) with nearly 100 layers representing different traded commodities. Comparing across layers (traded commodities), the weighted analysis shows that while many  pairs of layers display only a negligible degree of similarities between themselves,  few layers exhibit a high level of overlaps, implying that some countries tend to export or import more with a similar set of trading partners in some particular commodities.
These studies also suggest an important role of the distribution of hubs  (i.e., countries with a high level network centrality) in explaining the relations between layers.}  


\added{Empirical analyses, with a focus on \textit{the global production system that comprised supply-chain or input-output trade linkages among industries from different countries}, have been received less attention due to fact that more comprehensive, global data only becomes available only recently. As pointed out in \cite {Baldwin_et_al_2015}, international production relations have been flourishing over the last few decades and they still continue to evolve. The authors also suggest that more work is needed to uncover the complexity and hidden patterns embedded inside the global production system, as the importance and the degree of integration or dependency varies across countries, sectors, types of intermediate and final goods in such system. To our best knowledge, one of the first attempts to analyze the topological properties of the world input-output network of the individual industries from different countries was the study of \cite{Cerina_et_al_2015}. Using the World Input Output Database (WIOD, 2013 release), \cite{Cerina_et_al_2015} study different properties from the local to the global level of the world input-output network where nodes are the individual sectors while weighted links represent the monetary trading flows between them. Their results suggest that, overall, the binary links are very dense but the interaction intensities are highly asymmetric.  At the local level, the authors show that the network analysis, based on different centrality and community coreness measures, can provide deeper insights into identification of the key sectors.   In a recent work, focusing on the production network among industries from 20 country members in the European Monetary Union (EMU), \cite{Luu_et_al_2018b} show that the external (cross-border) input-output linkages exhibit a hierarchical structure in which a small number of  industries from few country members trade more intensely among themselves and form a cohesive core. In contrast, less active ``satellite" industries  mostly trade with the key sectors in the core. From a network perspective,  these key sectors, indeed, play as hubs bridging industries from different country members of the EMU. As shown in \cite{Luu_et_al_2018a}, such an asymmetric network architecture has crucial implications for the understanding of the pathway on how a local shock to a particular sector in a country member of the EMU can be transmitted to those in other countries.}
\added{Using the methods of the multilayer  analysis, \cite{Russo_et_al_2022} investigate the evolution of the international trading network of different components and parts of the automotive industry over the period 1993-2018. According to their study, several denser and more internally cohesive sub-networks can be identified like a cluster comprised by trade relations of Germany with some Central Eastern European countries, another cluster formed by the US and its partners in the preferential trade agreements such as Canada and Mexico, and another one with a prominent role of Japan and China.
\cite{Russo_et_al_2022} also show that the shape as well as the composition of clusters and the relative importance of countries also change over the years, with a remarkable rise of China after its accession to WTO.}

\added{The better understanding of the production networks can, of course, be useful for conducting effective industrial policies. For instance, successful experience and lessons from Japan, South Korea, Taiwan, and other ``Newly Industrialized Country" (NIC) have pointed out that  industrial policies\added{,} that select some ``strategic"  sectors to promote\added{,} can play a prominent role in facilitating the industrialization process  \citep[e.g., see][among others]{Antonio_2002, WB_Industrial_Policy_Survey_2006, Liu_2019, Antras_Gortari_2020}. However,  under the expansion of the global production network in which we often observe such a hierarchical structure dominated by few countries and their industries, the potential role of industrial policies (especially for less developed countries or for infant industries) might be more limited than what was traditionally supposed \citep[][]{WB_Industrial_Policy_Survey_2006}.} 

The main contribution of our study is to provide an analysis of the global trade network from a different perspective, i.e.  the multilayer architecture of the input-output interdependencies among industries and countries. \added{In our multilayer network countries are nodes and sectors represent layers and we focus} on the mesoscale structure and related network properties spanning from nodes to nodes across  layers.  We consider a  comprehensive structure of the network in which different types  of connections (i.e. intralayer as well as interlayer ones)  do exist all together. 
In addition, after classifying different clusters of nodes and associated layers, we further investigate the  internal   structural properties of each cluster and then compare with those of the other clusters.  \added{We also examine the role of individual countries and industries in forming and functioning different large international clusters. Through these analyses,}  we aim to  uncover  hidden structures and complexities inside the detected communities \added{and shed light on the diversification and specialization patterns of countries when they trade their inputs and outputs in and between different industries-based layers}. 
To these ends, we extend various network metrics and community detection methods used for single layer networks to those applicable for  complex  networks with a much richer multilayer architecture. 
In the next steps, we apply them to analyse the input-output relations between different sectors in different countries,  using  the world input-output database (WIOD, 2016 release; see \cite{Timmer2015, Timmer_et_al_2016}).

Our results show that  the multilayer analysis is able to catch more peculiar trade relationships  that cannot be detected at  the \added{mono-layer} aggregate trade network.
Among others, we find that, in the weighted version,  while many layers have dissimilar internal structures, few others  are somewhat more overlapped or correlated.  On top of that, the interlayer interactions are also highly heterogeneous.
Digging deeper at a broader mesoscale structure,  we can identify several large international communities in which some countries trade more intensively in certain specific industry-based layers. 
However, it is worth to emphasize that such a mesoscale structure does evolve over time, which may reflect the competition  as well as cooperation dynamics among  industries and among nations. 
For example,  in 2000, the first largest international community was comprised by sectors from the US, Japan together with several sectors from China, while the  second largest cluster mainly consists of industries from the former Eastern Bloc's countries.  In contrast, as for the data in 2014,  we identify the first largest international community where  relevant sectors of main Asian players (China, India, Japan, South Korea) together with those from Australia are clustered together. On the other hand, sectors from countries involved in the North American Free Trade Agreement (Canada, Mexico and USA) belong to the second largest one.

The remainder of this paper is organized as follows.  In Section \ref{Data and Methods}, we describe the multilayer representation of the global trade system as well as the \added{proposed methodology used} to analyze such a network.
Section \ref {Results} summarizes \added{ preliminary analysis of data and the main results with related economic interpretation}. \added{A focus on temporal evolution is also provided}. We conclude in Section \ref {Conclusions}. Further information on the dataset and additional results are left to the Appendices.

\section{Network representation and methodology} \label{Data and Methods}

\subsection{Multilayer representation and fundamental metrics  of the world input-output network}    \label{Network_Representations}

\added{We shall now introduce the general mathematical representation of a multilayer trade network. Formally, to deal with a multilayer network, one of the main approaches}\footnote{\added{An alternative approach consists in using tensors. For the sake of brevity, we refer here only to supra-adjacency representation.  For further details of the tensor approach, we refer the reader to \cite{DeDomenico_et_al_2013} and \cite{BCG2022}.}}\added{ is to represent the trade interdependencies among $L$ sectors from $N$ different countries as a block matrix, with $L\times L$  blocks, each one of order $N$, called the weighted supra-adjacency matrix $W^{supra}$ with size $N_{supra}\times N_{supra}$, where $N_{supra}=NL$  \citep[][]{Biancon_multilayer_book_2018}:}

  \begin {equation}
 W^{supra}=\left[\begin{array}{cccc}
    W^{[1, 1]}&  W^{[1, 2]}  &  \dots & W^{[1, L]}\\ 
   W^{[2,1]}&  W^{[2, 2]}  &   \dots & W^{[2, L]}\\ 
    \vdots &  \vdots &  \ddots & \vdots\\ 
      W^{[L, 1]}&  W^{[L, 2]}  & \dots & W^{[L, L]}\\ 
\end{array}\right].
\label{weighted_supra}
\end {equation}



Each  $W^{[\alpha, \alpha]}$  
is a $N$-square block matrix representing  the intralayer interactions among $N$ nodes in  layer $\alpha$, while  when $\alpha \neq \beta$, each $W^{[\alpha, \beta]}$ 
with size $N\times N$  is a weighted (adjacency) block  matrix capturing the interlayer interactions between nodes  in layer $\alpha$ and nodes in layer $\beta$.  

Its binary version, \added{indicating the existence of links}, is characterized by a supra-adjacency  matrix $A^{supra}$ with the same size, \added{where the block matrix $W^{[\alpha, \beta]}$ is replaced by the binary matrix $A^{[\alpha, \beta]}$.}

The supra weighted matrix $W^{supra}$ can be further decomposed into two distinct parts
 $W^{supra}=W^{intra}+ W^{inter}$, in which the first part  $W^{intra}$  consists of only intralayer linkages   and  the second part  $ W^{inter}$ consists of only external interlayer linkages:
 \begin {equation}
W^{intra}= \left[\begin{array}{cccc}
    W ^{[1, 1]}&  O    & \dots & O \\ 
     O  &        W ^{[2, 2]}  &  \dots& O \\ 
    \vdots &  \vdots  & \ddots & \vdots\\ 
   O &  O    & \dots &       W ^{[L, L]}
\end{array}\right],
\end {equation}
 and
 \begin {equation}
W^{inter}= \left[\begin{array}{cccc}
    O &  W^{[1, 2]}  & \dots & W^{[1, L]} \\ 
     W^{[2, 1]} &   O & \dots & W^{[2, L]}\\ 
    \vdots &  \vdots &  \ddots & \vdots\\ 
   W^{[L, 1]} &  W^{[L, 2]}  &  \dots  & O\\ 
\end{array}\right],
\end {equation}
where $O$ is the $N$-square matrix 
whose elements are zero.

Similarly, in the binary version, we can also define  $A^{intra}$ and $A^{inter}$ as the intralayer and interlayer adjacency matrices, respectively. 

\added{We can extend the standard vertex centrality measures in the multilayer context. Since each country or sector in the global trade network can be both the buyer and the seller at the same time, it is necessary to distinguish between the \textit{incoming} and \textit{outgoing}  linkages with their partners.
Given the matrix $W^{supra}$ showing the interaction intensities, and the matrix $A^{supra}$ representing the existent interactions, one can compute the different types of degrees and strengths of nodes (countries) across layers (sectors). 
We refer the reader to the  \ref{Appendix A} for the formal definitions of in- and out strengths (in and out-degree, respectively).}

\subsection {Interrelations between layers} \label {layer_interrelation}

We recall that  $A^{[\alpha, \beta]}$ and $W^{[\alpha, \beta]}$ capture the interdependencies between nodes in layer $\alpha$  with nodes in layer $\beta$ in the binary and weighted versions, respectively.  Hence, to examine how strong the interaction between two layers  $\alpha$ and $\beta$ is, we measure the average connectivity and intensity based on the elements of these two matrices.
Let us define the average link and weight  between two layers $\alpha$ and $\beta$ as
\begin{equation}
\langle a^{[\alpha,  \beta]} \rangle= \frac {\sum_{i, j} A^{[\alpha, \beta]}_{ij}} {N^2}, \label{average_connectivity}
\end{equation}
and
\begin{equation}
\langle w^{[\alpha, \beta]} \rangle =\frac{\sum_{i, j} W^{[\alpha, \beta]}_{ij}} {N^2} .
\end{equation}
The average strength $\langle w^{[\alpha, \beta]} \rangle$ can be further normalized by $w^*$, \added{that is} the largest element of the weighted \added{supra-adjacency} matrix $W^{supra}$:
 \begin{equation}
\langle w^{[\alpha, \beta]} \rangle_{norm} =\frac{\sum_{i, j} W^{[\alpha, \beta]}_{ij}} {N^2 w^*},  
\label{average_intensity}
\end{equation}
which will lead to $0\leq \langle w^{[\alpha, \beta]} \rangle_{norm} \leq 1$. 

In the following, we will briefly explain the methods used to measure the overall overlaps and correlations between layers.
Following   \cite{Gemmetto_Garlaschelli_2014, Gemmetto_et_al_2016, Luu_Lux_2019},  we define the overall (normalized) degree of overlaps between every pair of layers $\alpha$ and $\beta$ in the binary version as
\begin{equation}
O^{[\alpha,  \beta]}_{bin}= \frac {2\sum_{i, j} \min (A^{[\alpha, \alpha]}_{ij}, A^{[\beta, \beta]}_{ij} )} {\sum_{i, j}  (A^{[\alpha, \alpha]}_{ij}+ A^{[\beta, \beta]}_{ij} )}. 
\label{overlap_bin}
\end{equation}
Analogously, for the weighted version,  the overall (normalized)  level of overlaps between two layers $\alpha$ and $\beta$ is given by  
\begin{equation}
O^{[\alpha,  \beta]}_{w}= \frac {2\sum_{i, j} \min (W^{[\alpha, \alpha]}_{ij}, W^{[\beta, \beta]}_{ij} )} {\sum_{i, j}  (W^{[\alpha, \alpha]}_{ij} + W^{[\beta, \beta]}_{ij} )}. 
\label{overlap_weight}
\end{equation}
It can be easily shown that $O^{[\alpha,  \beta]}_{bin}$  ranges in $[0, 1]$, with $O^{[\alpha,  \beta]}_{bin}=0$  if and only if two layers $\alpha$ and $\beta$ have no overlaps at all, while 
$O^{[\alpha,  \beta]}_{bin}=1$ if the two adjacency matrices are identical. Similar interpretations apply to $O^{[\alpha,  \beta]}_{w}$ used in the weighted version.

Alternatively, the overall degree of similarity between two layers $\alpha$ and $\beta$ \added{could be measured by} the Pearson correlation coefficient (element by element) between $A^{[\alpha, \alpha]}_{ij}$ and $A^{[\beta, \beta]}_{ij}$ or between $W^{[\alpha, \alpha]}_{ij}$ and $W^{[\beta, \beta]}_{ij}$ over all possible pairs of node indices $(i,j)$:

\begin{equation}
R^{[\alpha,  \beta]}_{bin}= \frac {\langle A^{[\alpha, \alpha]}_{ij} A^{[\beta, \beta]}_{ij}  \rangle- \langle A^{[\alpha, \alpha]}_{ij} \rangle \langle A^{[\beta, \beta]}_{ij} \rangle} {\sigma [A^{[\alpha, \alpha]}_{ij}] \sigma[A^{[\beta, \beta]}_{ij}]},
\label{corre_bin}
\end{equation}
\begin{equation}
R^{[\alpha,  \beta]}_{w}= \frac {\langle W^{[\alpha, \alpha]}_{ij} W^{[\beta, \beta]}_{ij}  \rangle- \langle W^{[\alpha, \alpha]}_{ij} \rangle \langle W^{[\beta, \beta]}_{ij} \rangle} {\sigma [W^{[\alpha, \alpha]}_{ij}] \sigma[W^{[\beta, \beta]}_{ij}]},
\label{corre_weight}
\end{equation}
where in general,  $\langle X  \rangle$ and  $\sigma [X]$ are the notations for the mean and standard deviation of $X$.\\

\added {It should be emphasized that} while the average connectivity and the average intensity \added {defined in Eqs. (\ref{average_connectivity}) and (\ref{average_intensity})} are based on the \textit {interlayer} linkages, the level of overlaps and the degree of similarity between two layers $\alpha$ and $\beta$ only depend on the \textit {intralayer} linkages. \added {In the context of the global input-output analyzed in our current work, the average connectivity and the average intensity for each pair of layers (sectors) $\alpha$ and $\beta$ indicate whether, over all different $N$ countries, these two sectors strongly interact or not. The level of overlap and of degree similarity, in contrast, returns the extent to which trade relations among $N$ countries in a particular industry $\alpha$ are also similar to trade relations among $N$ countries in another industry $\beta$.}

\subsection{Multilayer communicability and its applications to community detection}    \label{Mutilayer_Communicability}


The communicability between every pair of nodes quantifies the number of \textit {all possible connection paths} between them \citep{ESTRADA_et_al_2012}. It should be emphasized that the  basic network metrics such as degrees or strengths can only capture the  second order structural interdependencies (from all other nodes to a node or from a node to all other nodes).    Even the clustering coefficients, which are often used to analyse the clustering \added{behaviors} in complex networks (e.g. see \cite {Luu_et_al_2017} and the literature therein), can explain the third order of the structural correlations among three nodes alone. In contrast, the communicability is able to catch richer information about the direct as well as indirect pathways associated with different orders of interconnectedness between nodes.


\added{The communicability among nodes has already been used to detect the community structure in the context of monoplex networks \citep [e.g. see][]{Estrada_Hatano_2008, Bartesaghi_et_al_2020}.}
As a natural extension, the multilayer communicability  indicates the number of  paths through both possible intralayer  and interlayer links  that connect a given node in a particular layer to the other nodes in the  multilayer architecture.
In the context of global input-output network, communicability indicates the number of different upstream and downstream propagation channels between sectors and between countries, via both \added{immediate} and \added{not immediate} pathways. In fact, it can be somewhat related to the concept of the average propagation length often used to measure the economic distances between industries in the  literature related to input-output analysis \citep[e.g.][]{Dietzenbacher_et_al_2005, Miller_Blair_2009}.

In the binary version, given a supra-adjacency matrix $A^{supra}$, the communicability matrix of the multilayer network is
\begin {equation}
G_{bin}=I + \frac {(A^{supra})^1}{1!}+ \frac {(A^{supra})^2} {2!}+...+ \frac {(A^{supra})^k}{k!}+...=\exp(A^{supra}).
\label{communicability_single_layer}
\end {equation}
Note that  $G_{bin}$ can be further expressed in  the form of a supra matrix as

 \begin {equation}
G_{bin}=\exp(A^{supra})= \left[\begin{array}{cccc}
    G^{[1, 1]}&  G^{[1, 2]}  & \dots & G^{[1, L]}\\ 
   G^{[2, 1]}&  G^{[2, 2]}  &  \dots & G^{[2, L]}\\ 
 \vdots &  \vdots &  \ddots & \vdots \\ 
      G^{[L, 1]}&  G^{[L, 2]}  & \dots & G^{[L ,L]}\\ 
\end{array}\right],
\label{communicability_supra}
\end {equation}

where in general $G^{[\alpha, \beta]}$ (size $N\times N$) is the matrix containing the communicability between pairs of nodes belonging to layer $\alpha$ and layer $\beta$ \citep{Biancon_multilayer_book_2018}. As a special case,  when $\alpha=\beta$,   $G^{[\alpha, \alpha]}$ represents communicability among nodes within layer $\alpha$. However,  it is worth to mention that $G^{[\alpha, \alpha]}$ may differ from  $\exp(A^{[\alpha, \alpha]})$ if there exist couplings among replicas  and/or interlayer connections.

It is worth noting the similarity between the communicability matrix and the Leontief inverse matrix. They are both matrix functions obtained as a sum of a power series expansion. The Leontief inverse matrix is a resolvent-type matrix and unravels the technological interdependence within the productive systems in a given economy \citep[e.g.][]{Silva2017}. Then, the Leontief inverse and the communicability matrix are related measures of aggregate fluctuation capacity and they differ in the way long walks in the network are penalized\footnote{Notice that, as for the Leontief matrix, the input-output interdependencies play a key role in the communicability matrix. Therefore, a similar interpretation based on micro-founded models as in \cite{Long_Plosser_1983} can be further investigated}. While in the Leontief inverse equal weights are assigned to walks of different length, in the communicability matrix a walk of length $k$ is penalized\footnote{Notice that a similar assignment of decreasing weight on walks can be given representing the influence structure induced by a Neumann series.} in the sum by a factor $1/k!$. This choice accounts, in a more realistic way, for the fact that a shock originated in a node will have a reduced impact on the latter nodes in the production chain.

Exploiting the communicability matrix we can define suitable centrality measures for the nodes in the multilayer. In particular, in the directed version, for a node $i$ on layer $\alpha$, it is necessary to distinguish the incoming paths from all nodes in a layer $\beta$ to $i$ and the outgoing paths from $i$ to  all nodes in a layer $\beta$. Hence, we define the \textit {receive (communicability)  centrality} $\text{rc}_{i}^{[\alpha \leftarrow  \beta]}$  and the  \textit {broadcast (communicability) centrality} $\text{bc}_{i}^{[\alpha \rightarrow  \beta]}$ as follows:

\begin{equation}
\text{rc}_{i}^{[\alpha \leftarrow  \beta]}=\sum_{j =1}^{N}  G_{ ji}^{[\beta, \alpha]}, \quad
\text{bc}_{i}^{[\alpha \rightarrow \beta]}=\sum_{j =1}^{N}   G_{ij}^{[\alpha, \beta]}.
\label{layer_layer_rc}
\end{equation}

The total communicability centralities of  node $i$ in layer $\alpha$ from/to all nodes in all $L$ layers are then given by
\begin{equation}
\text{rc}_{i}^{[\alpha, \ total]}=\sum_{\beta=1}^{L} \text{rc}_{i}^{[\alpha, \beta]}, \quad
\text{bc}_{i}^{[\alpha, \ total]}=\sum_{\beta=1}^{L} \text{bc}_{i}^{[\alpha, \beta]}.
\label{total_rc}
\end{equation}


It is straightforward to extend the communicability matrix and the \added{definitions of centrality measures} to their weighed counterparts. 
In the next step, we  can define the communicability matrix $G_w$ for the weighted version  as
\begin {equation}
G_w=I +  \frac { \bar {W}^{supra}}{1!}+ \frac { (\bar {W}^{supra})^2}{2!}+...+ \frac {(\bar {W}^{supra})^k}{k!}+...=\exp( \bar {W}^{supra}).
\label{communicability_multilayer_layer_weight}
\end {equation}

where $ \bar {W}^{supra}$ represents the weighted supra-matrix after a suitable normalization. The normalization helps to avoid the excessive influence of links with higher weights in the network \added{as pointed out in \cite{Crofts_Higham_2009} and \cite{Estrada_book_2011}. The element $\bar{W}^{[\alpha, \beta]}_{ij}$ of the normalised matrix $\bar{W}^{supra}$ represents the incidence of the trade between sector $\alpha$ in economy $i$ and sector $\beta$ in economy $j$ with respect to the average volumes \added{they trade}.} 

Note that the communicability matrix, considering  all possible walks between two nodes, accounts also for the incidence of the trade between sector $\alpha$ in $i$ and sector $\beta$ in $j$ through intermediate sectors/nodes.  As for the inverse of the Leontief matrix, $G_w$ takes into consideration both the effects of indirect and direct trades, but communicability assigns a decreasing weight to indirect connections when their length (i.e., the length of the walk between two nodes) increases.
 
 Based on the weighted communicability matrix  $G_w$, one can also easily derive the weighted \textit {receive (communicability)  centralities}  and  \textit {broadcast (communicability) centralities}, both for layer-layer communicability  centralities and for the total ones, similar to the binary counterparts defined in Eqs. (\ref{layer_layer_rc}) and  (\ref{total_rc}).

In the following, we will explain how to apply communicability centralities to identify communities in a multilayer network. We start by introducing on the network a suitable distance based on the idea of communicability. \added{To ensure that the introduced distance is well defined, we need to work with a symmetrized matrix. To this end, we first symmetrize the matrix $W^{supra}$ by constructing $W^{symm}=\frac{1}{2}({W}^{supra}+{({W}^{supra}})^T)$. To avoid the excessive influence of links with higher weights in the network, we then normalize the new matrix $W^{symm}$ by  $\bar{W}^{symm}=S^{-\frac{1}{2}}W^{symm}S^{-\frac{1}{2}}$, where $S$ is the (supra) diagonal matrix of the strengths of each node in each level.}
As expressed by the definition, the communicability between two nodes is a weighted sum of the number of all walks connecting the pair (see \cite{ESTRADA_et_al_2012}). Indeed, the fact that two nodes can be connected by means of all possible walks, and not only by shortest paths, is implicit in the idea of communicability.

In general, the communicability-based distance $\xi_{ij}^{[\alpha, \beta]}$ between node $i$ in layer $\alpha$ and node $j$ in layer $\beta$ is defined as
\begin{equation}
	\label{communicabilitydistanceonmyltilayer}
	\xi_{ij}^{[\alpha,\beta]}=G_{ii}^{[\alpha,\alpha]}-2G_{ij}^{[\alpha,\beta]}+G_{jj}^{[\beta,\beta]},
\end{equation}
where $G$ is computed as in Eq. (\ref{communicability_single_layer}) in the binary version and as in Eq. (\ref{communicability_multilayer_layer_weight}) in the weighted one \added{by replacing $\bar{W}^{supra}$ with $\bar{W}^{symm}$}. 
Let us also emphasize that  we are assigning the same meaning to distances between nodes in the same layer, distances between versions of the same node in different layers, and distances between different nodes in different layers. 

Following \cite{Chang_et_al_2016} and \cite{Bartesaghi_et_al_2020} we can compute the cohesion function $\gamma_{ij}^{[\alpha,\beta]}$, between any couple of nodes, as

\begin{equation}
	\gamma_{ij}^{[\alpha,\beta]}=\bar{\xi}_{ii}^{[\alpha,\alpha]}+\bar{\xi}_{jj}^{[\beta,\beta]}-{\xi_{ij}^{[\alpha,\beta]}}-\bar{\xi},
\label{cohesiontensor}
\end{equation}
where $\bar{\xi}_{ii}^{[\alpha,\alpha]}$ is  the average communicability distance of node $i$ belonging to layer $\alpha$ from all the other nodes in all layers and $\bar{\xi}$ is the average communicability distance over the whole network.

The cohesion function $	\gamma_{ij}^{[\alpha,\beta]}$ can be interpreted as a cohesion measure between the two nodes. Specifically, when positive (respectively, negative), it represents the gain (respectively, the cost) of grouping nodes $i$ in layer $\alpha$ and $j$ in layer $\beta$ in the same community of a given partition. 
We assume then to maximize the global cohesion function ${\cal Q}$:

\begin{equation}
\label{QmetricMultilayer}
{\cal Q}=\sum_{[i,\alpha],[j,\beta]}\gamma_{ij}^{[\alpha,\beta]}x_{[i,\alpha],[j,\beta]},
\end{equation}

where $x_{[i,\alpha],[j,\beta]}$ is the Kronecker delta function, which is equal to $1$ if two nodes $[i, \alpha]$ and $[j,\beta]$ are in the same cluster and $0$ otherwise.
Hence, we obtain the best possible partition via maximization of the function ${\cal Q}$ defined in (\ref{QmetricMultilayer}).

More specifically, the network partition that allows to identify clusters and the consequent optimal partition are established according to the steps of the following algorithm.

\begin{enumerate}
	\item let ${\mathscr G}$ be the original multilayer (in general, directed and weighted) network that has $L$ layers  and $N$ nodes per layer, and let ${W}^{supra}$ be the corresponding supra weighted matrix;
	\item to apply the communicability-based method, we first  build the undirected weighted network associated with the symmetric supra weighted matrix defined as $W^{symm}=\frac{1}{2}({W}^{supra}+{({W}^{supra}})^T)$;
	\item we then build the undirected weighted network associated with the normalised weighted supra-adjacency matrix $\bar{W}^{symm}=S^{-\frac{1}{2}}W^{symm}S^{-\frac{1}{2}}$; 
	\item we compute the distances according to formula (\ref{communicabilitydistanceonmyltilayer}) and define the threshold interval $[\xi_{\rm min}, \xi_{\rm max}]$, where $\xi_{\rm min}$ and $\xi_{\rm max}$ represent the minimum and the maximum communicability distances between couples of nodes, respectively and set  $\xi_h=\xi_{\rm min}$ with the initial index $h=0$;
	\item we define a $NL \times NL$ matrix ${M}_{h}^{supra}$ whose entries are 1 if $ \xi_{ij}^{[\alpha,\beta]}\leq \xi_h$ and $0$ otherwise and build the undirected unweighted network from the binary supra-adjacency matrix ${M}_{h}^{supra}$;
	\item we select the partition $P_{h}$ given by the components of the network associated to ${M}_{h}^{supra}$ and compute the partition quality function $\cal Q$ in equation (\ref{QmetricMultilayer})
	\item we set the number of iterations $r$, compute the step increment $k=\frac{\xi_{\rm max}-\xi_{\rm min}}{r}$, set $\xi_h=\xi_{h-1} + k$ and $h=h+1$ and repeat steps 5-6 until $\xi_h  = \xi_{\rm max}$;
	and select the optimal partition $P^{\star}_{h}$  as the partition $P_{h}$ that provides the optimal $\cal Q$.
	
\end{enumerate}

\added{It is worth stressing} few key points of the presented methodology. We aim at  clustering nodes and layers $[i,\alpha]$ where $i$ are countries and $\alpha$ are sectors on the basis of a specific communicability distance. 
\added{Our procedure is based on the selection of the optimal threshold. The value of the threshold allows indeed to disentangle the role of very tight relationships between couple of nodes. For instance, a very low threshold leads to a great number of isolated nodes, while higher values produce instead larger communities. In the proposed approach, the value of the threshold is not selected as a prior, but it is computed by the procedure assuring the optimal partition obtained from the maximization of the cohesion function ${\cal Q}$ defined in formula (\ref{QmetricMultilayer}).}



\section {Multilayer  community structure in the global input-output network}
\label{Results}

\subsection{Data}  

We \added{analyze} the mesoscale structure and the related topological properties of the global trade network from the perspective of the multilayer architecture, using an updated version of the world input-output database (WIOD, 2016 release).\footnote {The database is described in detail
by \cite{Timmer_et_al_2016}. It is publicly available at: http://www.wiod.org/database/wiots16.} The updated version released in 2016 is the second wave of WIOD data, providing yearly information for trade in input-output among 56 different sectors in 44 countries and the aggregate of the rest-of-the-world \added {(ROW)} over the period from 2000 to 2014 (see Tables \ref{table_WIOD_industries} and \ref{table_WIOD_countries} in the \ref{Appendix B} for a list of countries and sectors).  
\added {In total, we have $56 \times 44 = 2464$ industries worldwide. Traded amounts (weights) among them are expressed in millions of dollars. We provide more detailed descriptive statistics of the weights in Table \ref{table_descriptive_stats_weights}.}
An initial analysis of the topology of the network has been developed. Since it includes standard techniques of network analysis, for the sake of brevity, we refer the reader to the \ref{Appendix C}. \added{The results naturally lead to  the following related question from a topological perspective at a broader scale  of the global trade system  in input-output.}   Are there clusters of countries whose members interact stronger among themselves simultaneously in different layers associated with different industries?  To answer this question,  we conduct an analysis of a multilayer community structure in the global input-output network.

\subsection {Multilayer  community structure in the global input-output network}    \label{Results_2}

In this section, we report main results obtained by applying the procedure based on multilayer communicability defined in Section \ref{Mutilayer_Communicability}.
The proposed approach has been applied directly to the multilayer network, which has been preliminarily transformed in an undirected one \added{by} substituting each pair of bilateral directed links with \added{an} undirected link, with a weight equal to the average weight.
\\
Starting from a supra-adjacency matrix with 2464 nodes, given by 44 countries and 56 sectors and using \added{volumes of trade} in 2014, the methodology provides 117 communities (except some isolated nodes). Notice that clustered groups of nodes are located both intralayers and interlayers. This means that we can find countries \added{that are} members of the same community in a specific sector, as well as countries sharing the same community in more sectors. 
In particular, two large international communities \added{(namely Community 1 and Community 2)} are detected, with 443 and 318 nodes, respectively. The other ten communities have instead a  smaller size, being between 30 and 50 nodes. A graphical representation of all members of the top twelve communities, in terms of number of nodes, can be found in Figure \ref{comm}, \ref{Appendix D}.

Looking at the clusters obtained, some noticeable elements can be identified. First of all, we observe an Australian-Asian community, where all relevant sectors of main Asian players (China, India, Japan, South Korea) are clustered together. On the other hand, countries involved in the North American Free Trade Agreement (Canada, Mexico and USA) belong to the second community. \\
Furthermore, it is interesting to explore the clustering \added{behavior} of European countries that are ``in the middle'' between these two relevant communities. Indeed,  we find that almost all of them are clustered in Community 1 for some sectors and in Community 2 for other sectors (see, for instance, Germany and Great Britain). To further explain this result, we \added{compute} for each country the number of sectors that belong to these two communities. We also compute\footnote{\added{Detailed results are reported in Table \ref{table_cou} in \ref{Appendix D}.}} the Gini heterogeneity index\footnote{In particular, since we deal with nominal variables, we compute the index $H_{j}$ to quantify the heterogeneity of a country $j$ as $H_{j}=1 - \sum_{i=1}^{c}p^{2}_{i,j}$ where $c$ is the total number of clusters detected by the procedure explained in Section \ref{Mutilayer_Communicability} and $p_{i,j}$ is the proportion of sectors of country $j$ in the cluster $i$ such that $\sum_{i=1}^{c} p_{i,j}=1$. This formula is also known in the literature as the Gini-Simpson index.} to give an indication of the dispersion of sectors for each country along the communities. \added {A higher value of this index for a particular country means that, overall, its sectors are more spread out between different communities.}  

As regard to European countries, \added{we notice} how some of them (Bulgaria, Germany, Greece, Norway, Portugal) show a behavior similar to the Australian-Asian countries, being these countries concentrated in the first community for most trade of sectors. However, looking at the sectors, this participation occurs with a different degree of heterogeneity, as indicated by the Gini index.   
For instance, Germany has 37 sectors in Community 1 and only 6 sectors in Community 2. In contrast,
Czech Republic, France, Great Britain, Ireland, Poland, Russia are instead more concentrated in Community 2. The remaining European countries have  only a limited number of sectors belonging to the top two communities, while the other sectors are concentrated in specific groups.\\
Moreover, it is worth pointing out the presence of countries with a lower level of heterogeneity in the community location of their sectors, \added{characterized} by a concentration in the same community for almost all sectors (except isolated ones). 
Indeed, we observe that besides the top largest international clusters,  the next large communities are \added{mainly} formed by different sectors of the same country. These are, for example,   the cases of Spain (Community 3),  Italy (Community 4),   Brazil (Community 5),  Finland (Community 6), Croatia (Community 7), Turkey (Community 8),  Indonesia (Community 9). \added{This behavior can be explained by stronger domestic trades with respect to relations with sectors of other countries.}

 \added{The presence of large international communities together with that of smaller ones leaning to domestic or regional input-output relations have been also found in the other studies such as \cite{Baldwin_et_al_2015}, \cite{Cerina_et_al_2015}, \cite{Luu_et_al_2018a}, \cite{Luu_et_al_2018b}. Indeed,  industries of different countries seem to have different degrees of internationalization. Some of them have been more engaged in cross-border trade linkages and, hence, can contribute to or even lead the functioning of regional or international clusters. It should be emphasized that this is not necessary related to the size of the country (or the size of domestic market), as we do observe that the largest international communities are comprised of both sectors from large economies like the US and China and those from smaller ones like Taiwan, Ireland or Czech Republic and ROW.} \footnote{\added{Note that ROW consists of different nations not listed in the database, in which important members are, for example, those of OPEC nations and ASEAN 6.}}  \added{Among possible explanations, perhaps the relations between sectors from ``headquarter economies" and sectors from ``factory economies" play a prominent role  \citep[e.g., see][] {Baldwin_2006, Baldwin_et_al_2015}. In line with the findings by \cite{Benedictis_Tajoli_2009} for the World Trade network, our conjecture is that the selection of the trading partners does matters for input-output relations. As discussed in \cite{Baldwin_et_al_2015}, ``headquarter economies" with more advanced technology and higher wages (e.g., the US, Japan, Germany) tend to offshore certain stages of production
to low-wage nations (the factory economies like Mexico,
Poland and Czech Republic, Taiwan), which may  lead to the creation of regional or international input-output chains across the world  such as ``Factory Asia", ``Factory North America" and ``Factory Europe". In between these two distinct types of countries, the ``hybrid economies"  can simultaneously play a certain role as a headquarter economy and a certain role   as a factory one.\\
Furthermore, some industries from ``headquarter economies" act as hubs with many linkages with different industries from different nations in the global input-output network. In contrast, those from ``factory economies" tend to rely more on few key players from ``headquarter economies".  This is also, indeed, inline with the results from basic network analysis in the \ref{Appendix C}, where we find in the weighted version of the network that some nodes have a much higher level of the incoming and outgoing strengths.}

Now, if we look at each single sector-based layer, we can further explore how countries in the same layer are classified in communities\footnote{\added{Detailed results can be found in Table \ref{table_sec}, \ref{Appendix D}}}. Overall, we find that the analysis for sectors shows a higher level of heterogeneity (as shown by the Gini index), \added{compared to that for countries}. This implies that, in the same \added {considered} layer (industry), countries are on average split in several different communities. However, except for sectors ``T" and  ``U" where countries act almost always as isolated communities, \added{the highest} number of countries in each sector belongs to one of the \added {two largest international} communities. It is also interesting to note that sectors with a lower heterogeneity (as \lq\lq Mining and quarrying (B)\rq\rq and \lq\lq Manufacture of coke and refined petroleum products (C19)\rq\rq) are \added{characterized} by a concentration of countries in Community 1.

\begin{figure}[!h]
	\includegraphics[height=7in,width=7in]{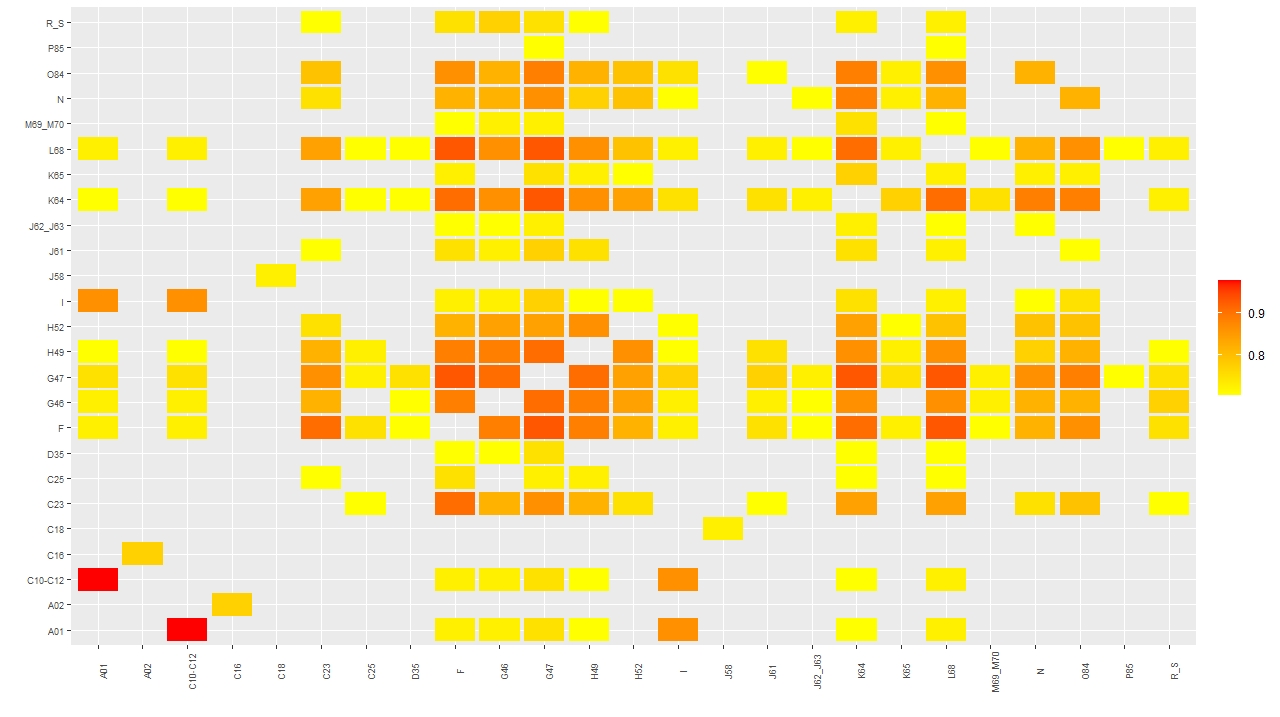}
	\caption{Matrix of Jaccard similarity coefficients between sectors. Only coefficients higher or equal than 0.7 have been reported (it corresponds to 90\% quantile approximatively).}
	\label{corsec}
\end{figure}

To explore the similarity between sectors, we computed the Jaccard similarity index (see Figure \ref{corsec}), displaying only coefficients higher than $0.7$. In this way, we emphasize couple of sectors \added{characterized} by a similar classification of countries in communities. We observe \added{a highest similarity between} the sectors \lq\lq Crop and animal production, hunting and related activities (A01)\rq\rq and \lq\lq Manufacture of food products, beverages and tobacco products (C10-C12)\rq\rq. We have indeed that all countries (except Ireland) have been classified in the same way in these two sectors. This result can be easily explained by the fact that activities involved in these two sectors are closely related \added{(see also the analysis of the interrelations between layer-based sectors in the \ref{Appendix C})}.
Another interesting case is represented by the sectors F and L68 (Construction and Real Estate) that show 41 countries analogously classified (except Malta, Mexico and Ireland). The same \added{behavior} is also observed for the other two couples G47-L68 (Retail Trade - Real Estate) and G47-K64 (Retail Trade - Financial Service).

So far, we have focused on the analysis of  the mesoscale structure in the multilayer version of the global input-output network, with $44$ nodes and $56$ layers. In the next step, we \added{shall} compare previous results with the community structure detected from the mono-layer aggregate version,  where each node is a country and a link now considers the total trade between a couple of countries. For consistency, we apply the communicability-based approach proposed in \cite{Bartesaghi_et_al_2020} to the mono-layer aggregate network.   We obtain seven communities (except five isolated nodes) and report their composition in Table \ref{table_sector}. 
It is noteworthy that two big communities are detected also in this case but the composition of large communities is mainly driven by geographical patterns. This is in line with the results already detected in the literature for aggregate trade network (see \cite{Barigozzi_et_al_2010}, \cite{Bartesaghi_et_al_2020}).
It is interesting to note that the first community includes almost all European Countries. Only some well-known couples are clustered alone (see for instance Spain and Portugal in community 4 or Great Britain and Ireland in community 5). The second community groups instead together North American and Asian players. \\
\added{However, the analysis of the mono-layer aggregate version may lead to an oversimplified conclusion that the input-output network among countries was simply composed by regional blocks. Indeed, we do observe the presence of large international communities in the multilayer version. The aggregate version also ignores the fact that trade inter-dependencies among countries do vary across industry-based slices.  All taken together, the results allow us to emphasize how the multilayer analysis, that takes into consideration distinction among $56$ industry-based layers, is able to catch more peculiar relationships between countries  that cannot  be detected at the aggregate level.} 

\begin{table}[!h]
\renewcommand{\arraystretch}{1.2}
	\scriptsize
	\begin{center}
		\begin{tabular}{ll}
			\hline\hline
			Community & Countries \\ 			
			\hline
              1 & AUT, BEL, BGR, CZE, DEU, DNK, FRA, GRC, HRV, HUN, ITA, LUX, NLD, NOR, POL,  \\
              & ROU, RUS, SVK, SVN, SWE, TUR \\ 
              2 & BRA, CAN, CHN, IND, JPN, MEX, USA, ROW \\ 
              3 &  AUS, IDN \\ 
              4 & ESP, PRT \\ 
              5 & EST, FIN \\ 
              6 & GBR, IRL \\ 
              7 & LTU, LTV \\ 
              Isolated & CHE, CYP, KOR, MLT, TWN \\ \hline
              \hline
	\end{tabular}\end{center}
	\caption  {The composition of the communities computed on the aggregate mono-layer network in 2014.}
	\label{table_sector}
\end{table}

\begin{figure}[h]
	\centering
	\captionsetup[subfloat]{farskip=0pt,captionskip=0pt}
	\subfloat[Top industries in Community 1, 2014]{\includegraphics[height=4.2in,width = 3.2in]{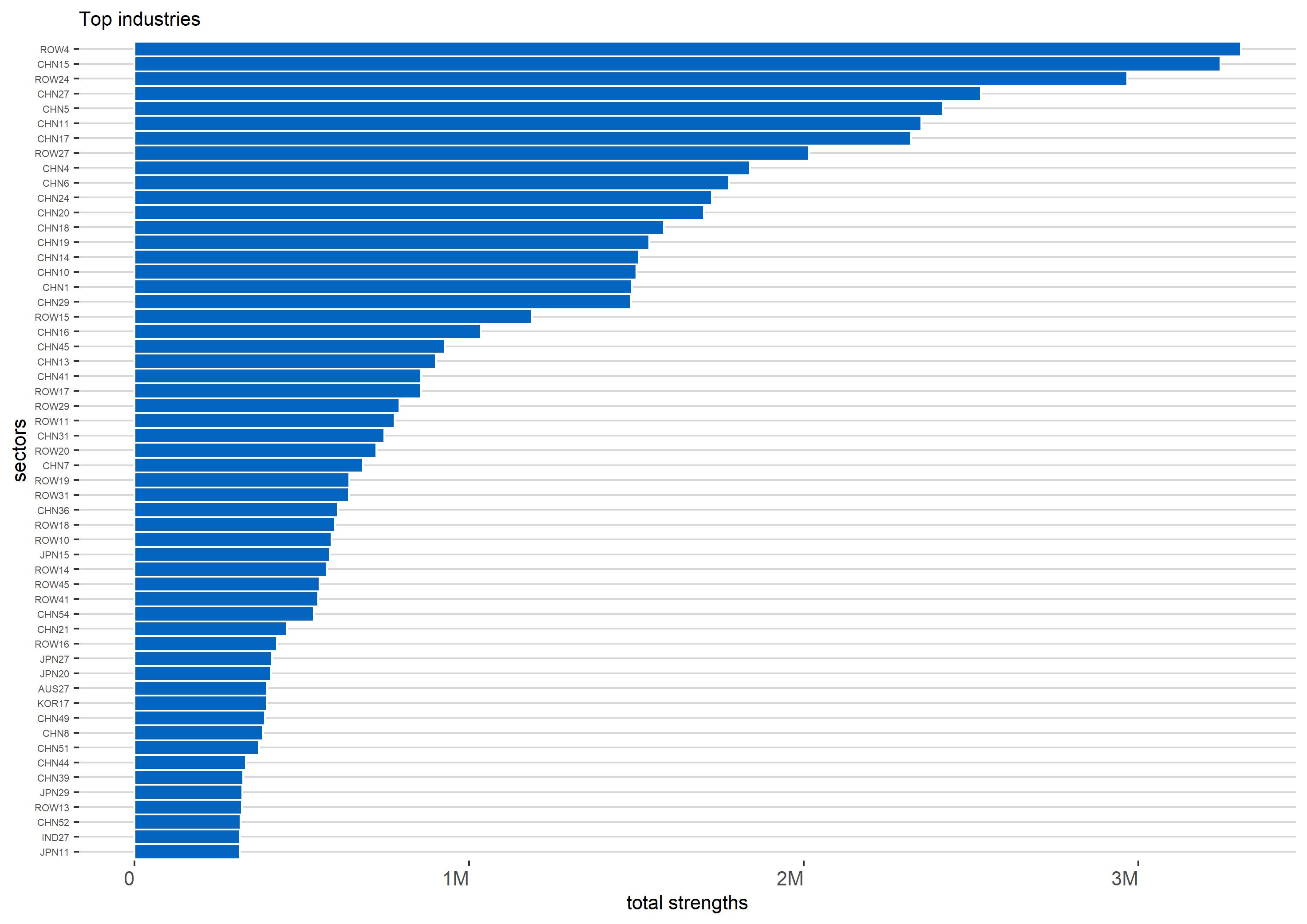}}
	\subfloat[Top industries in Community 2, 2014]{\includegraphics[height=4.2in,width = 3.2in]{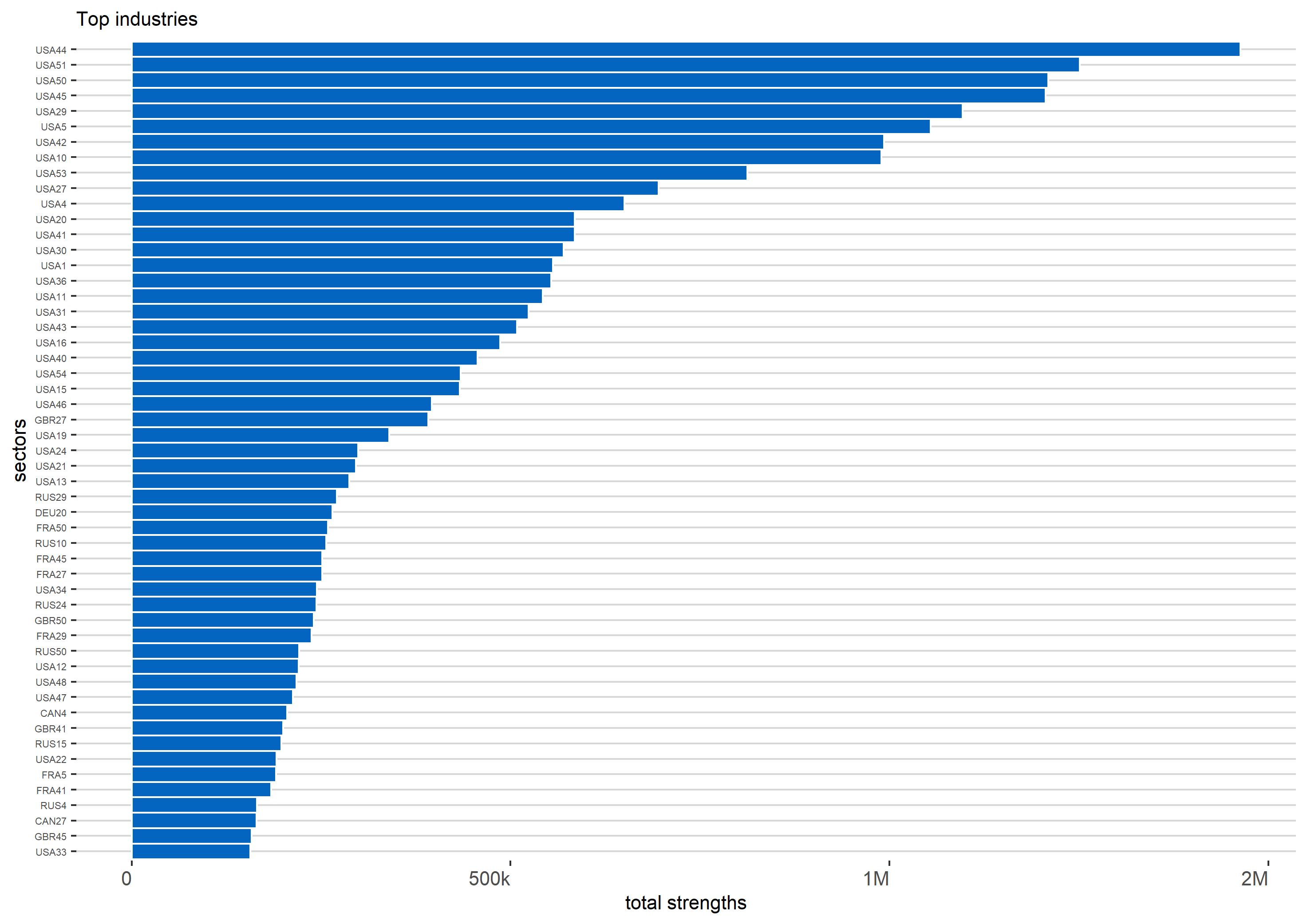}}
	\caption{Ranking of top  industries in two important international communities in 2014, based on the (total) node strengths. Labels on the $y$ axis are given by the country code and by the number related to the sector.}
	\label{top_sectors_two_largest_com_2014}
\end{figure}

\added{We further investigate the internal topological properties of the communities detected in the multilayer version of the network}\footnote{Detailed results can be found in Figure \ref{comm} and Table \ref{table_sec} in \ref{Appendix D}}\added{. Our main purpose here is to identify the ``key actors" in these communities. To this end,} we start selecting countries and sectors that belong to each identified community and we build a sub-network. More specifically, only the links among members of the community are maintained and analyzed, while those externally built with nodes in the other communities are discarded.  In particular,  we focus on the two most representative communities, i.e. the Community 1 and 2, since these two largest clusters consist of different sectors from various countries. 
Figure \ref{top_sectors_two_largest_com_2014} shows the ranking of the top sectors in each community based on the (total) node strengths. 
The results again confirm the dominance of China in the first largest community and that of USA in the second community at the industry levels as many of the influential sectors in the communities 1 and 2 are  actually from these two countries.
Looking at the industry codes of the most influential sectors, we also observe an interesting difference in  economic and \added {industrial} structure between the two communities and hence between the two \added {nations}. While  influential sectors from China are mostly manufacturing-related industries, those coming from the US are more dominated by  real-estate,  finance, technology or service-related sectors.

\subsection{Analysis of temporal evolution of communities}  \label{Results_3}

\begin{figure}[!t]
	\centering
	\includegraphics[height=3 in,width=4in]{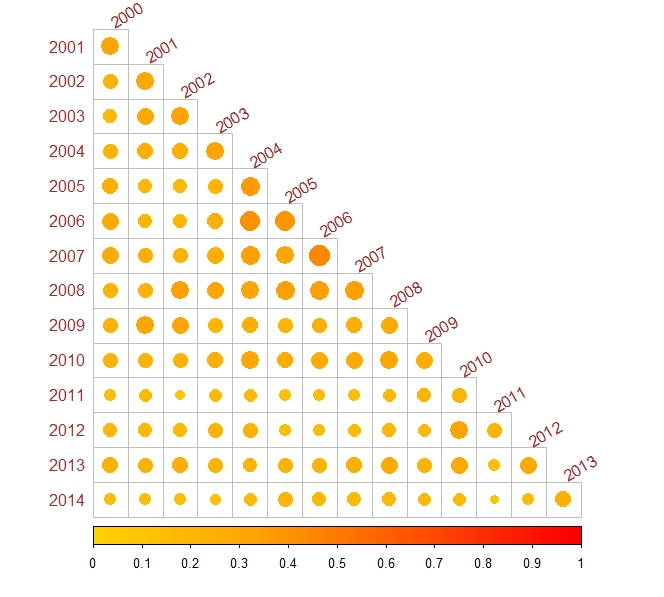}
	\caption{Values of ARI computed between clusters based on couple of years. The color bar and the size of bubbles indicates the value of ARI in the interval $[0,1]$. A higher value of ARI implies more similar community structures in the two corresponding years.}
	\label{ARI}
\end{figure}

\added{Notice that for the sake of illustration, in Section \ref{Results_2} we have focused on mesoscale structure of the considered network in the selected year 2014, which provides the latest available information we can have from the WIOD database.}
\added{We shall now investigate the temporal evolution of the international communities. To this end, we considered the period 2000-2014 and we applied  the method based on the  multilayer communicability distance described in Section \ref{Mutilayer_Communicability} to extract  different clusters for each of these selected years. Remarkably,  we find evidence that the clustering behaviors of trades in input-output among industries and countries do restructure over the period from 2000 to 2014. In general, not only the community composition changes, but the centrality rankings of the members inside each community is  also reordered, with the diminishing of some  industries and  the emergence of new ``key players''. \\
In particular, we start analyzing the similarity between the communities obtained over time. To do this, we computed the Adjusted Rand Index (ARI) between clusters for each couple of years (see \cite{Hubert} and \cite{Rand}). This index falls in the interval $[0,1]$ and it is equal to one only if two partitions are completely identical. Numerous measures for comparing clusterings have been proposed in the literature, however the ARI}\footnote{\added{Given a set of elements and two different partitions, we compute the sum of the number of pairs of elements that are in the same subset in both partitions and the number of pairs of elements that are in the different subsets in both partitions. The ratio between this value and the total number of pairs is the Rand Index and gives a frequency of occurrence of agreements over the total pairs. ARI is the corrected-for-chance version of the Rand index and it gives the overall concordance of two methods taking into account that the agreement between partitions could arise by chance alone.}} \added{remains the most well-known and widely used (\cite{Steinley}). We display in Figure \ref{ARI} values of the index for each pair of years. We often find that while the mesoscale structure exhibits a certain level of persistence  when two subsequent years are compared, it  becomes less and less similar  after longer time periods. Such temporal behavior can be related to substitution and margin effects from international trade theories and the basic network properties that we have analyzed in the \ref{Appendix C}.
Qualitative substitution and extensive margin effects, based on countries and industries that change their trade partners or even obtain new ones, are indeed not very important, since the binary version of the network in the period 2000-2014 is always (and already) very dense. Hence the network can display a certain degree of persistence. Nevertheless, similarly to what has recently been found for the World Trade network \citep[e.g., see] [] {Felbermayr_Kohler_2006, Helpman_et_al_2008, Benedictis_Tajoli_2009}, intensive margin effects play a more important role in explaining the evolution of the global input-output network.
Intuitively, strengthening trade relations may take time and lead to the structural change in the whole network only once the magnitude of traded amount has been sufficiently increased. It should be also emphasized that not all trade relations are simultaneously intensified, as different sectors and countries can play different roles in such a hierarchical system. Furthermore, exogenous shocks and events such as the formation or break of trade agreements, economic zones, monetary unions may also lead to disruptive changes in the community structure of the network.}

\begin{figure} [H]
	\centering
	\captionsetup[subfloat]{farskip=0pt,captionskip=0pt}
	\subfloat[ Top communities in 2000]{\includegraphics[height=3.5 in, width = 3.5 in]{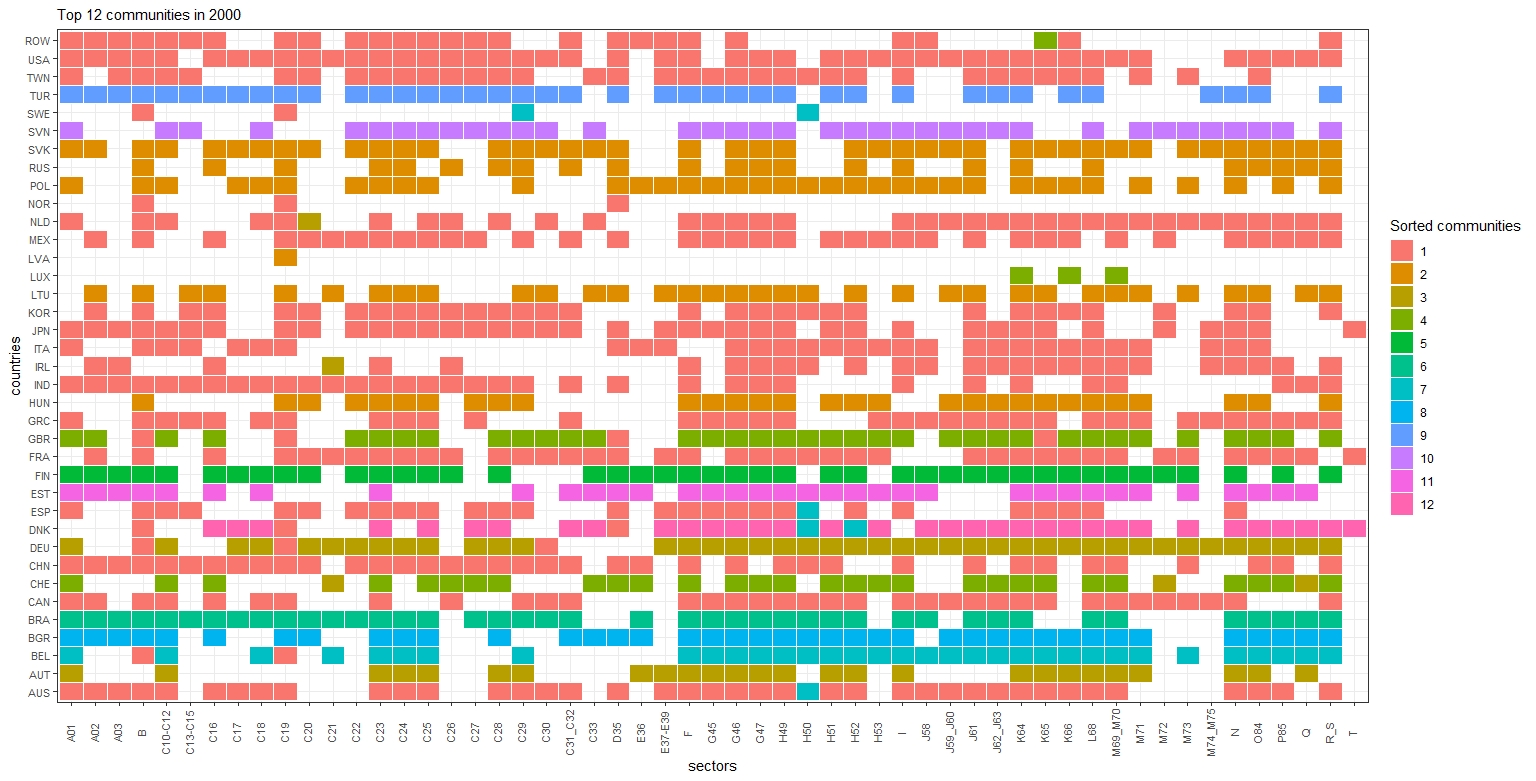}}
\subfloat[ Top communities in 2002]{\includegraphics[height=3.5 in, width = 3.5 in]{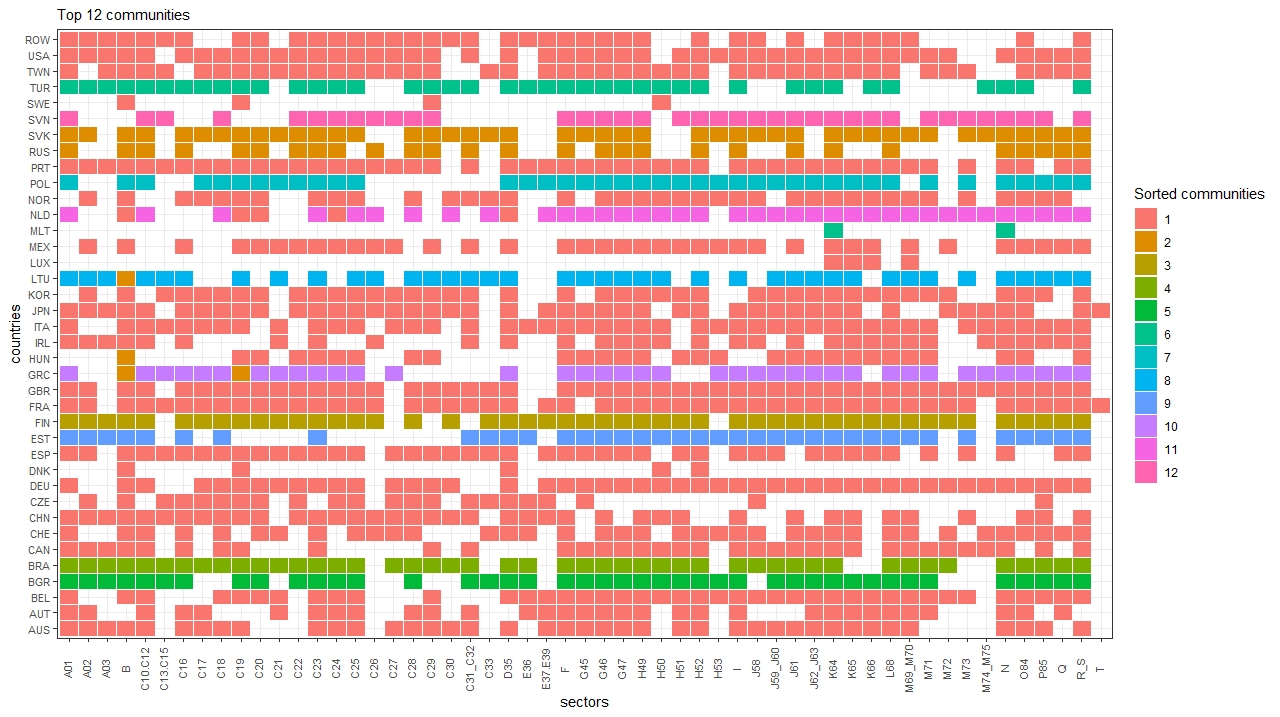}}\\
\subfloat[ Top communities in 2004]{\includegraphics[height=3.5 in, width = 3.5 in]{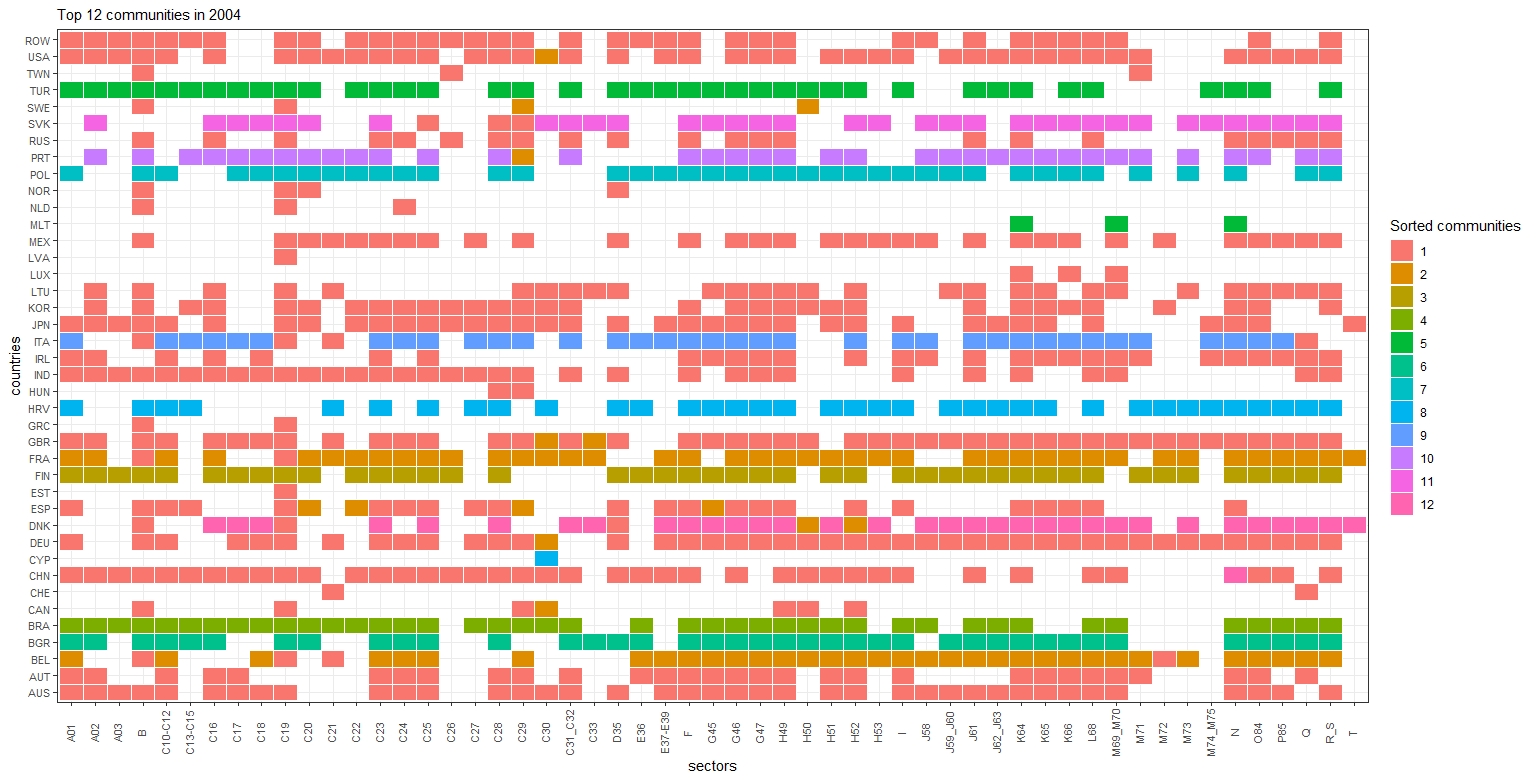}}
\subfloat[ Top communities in 2009]{\includegraphics[height=3.5 in, width = 3.5 in]{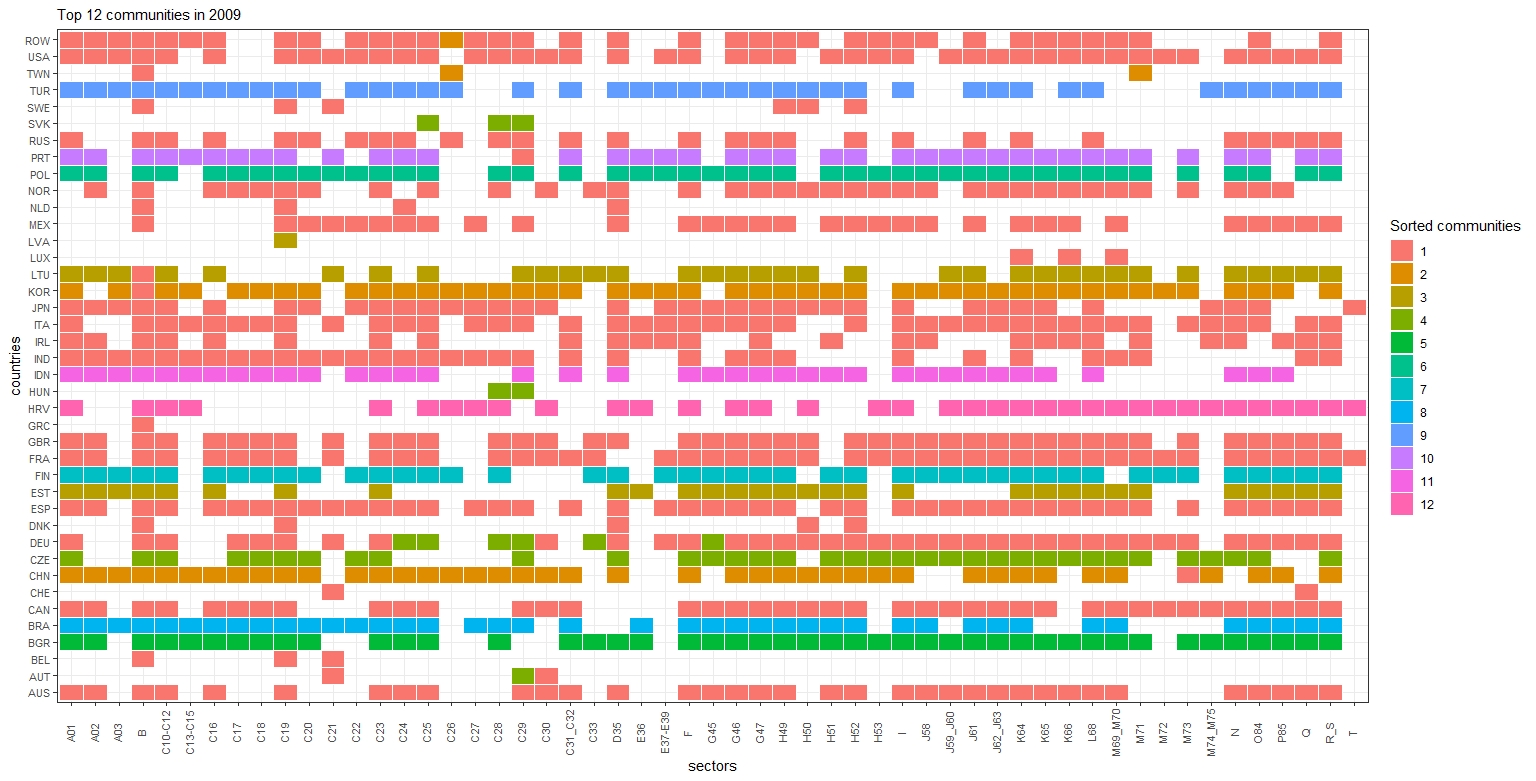}}
	\caption{Detected communities in the global trade network in input-output in selected years 2000, 2002, \added{2004 and 2009}.}
	\label{top_com_2000_2009}
\end{figure}

\added{To provide deeper insights into the changes in the internal structure of the communities, we shall now analyze their composition  over time}\footnote{For the sake of illustration and conciseness, we only report the results related to some selected years before 2014. Results for other years are available upon request to the authors.}.
As shown in Figure \ref{top_com_2000_2009} (panel a)  starting in 2000, the first largest community, with 571 members, is mainly composed by sectors from the US, Japan and interestingly several ones from China. In contrast, the second largest cluster (with 167 members) mainly consists of industries from the former Eastern Bloc's countries, led by those from Russia and Poland\footnote{See also Figure \ref{top_sectors_two_largest_com_2000} in the \ref{Appendix D} for additional results.}

Over the next few consecutive years, first major changes are observed in 2002 (see Figure \ref{top_com_2000_2009} (panel b)). We have indeed an important reduction of the total number of communities and a higher concentration in the largest group with 895 members. In particular, Community 1 is enriched by the inclusion of several sectors from European countries: Austria, Belgium, Germany and Great Britain join the other European countries (Italy, France, Spain, Ireland) in this big community. On the other hand, the presence of Eastern Bloc's country in Community 2 is reducing. The composition of Community 2 is indeed affected by the effects of the disintegration of the Eastern Bloc with the dissolution of the Warsaw Pact and Comecon. We find that the former members of the Warsaw Pact, that joined NATO (as Poland, Hungary and Czech Republic) in 1999, moved from Community 2 (including Russia) to Community 1 (including United States)\footnote{See also Figure \ref{top_sectors_two_largest_com_2002} in the \ref{Appendix D} for additional results.}.
\added{A relevant aspect that could lead to  major changes in the mesoscale structure of the global input-output network  from  2002  is the entrance of China in World Trade Organization (WTO) in December 2001}.\footnote{See, for example, \cite {Subramanian_Wei_2007} for further discussions on the impacts of  WTO on trade.} \added{China's accession to the WTO brought a number of benefits to both China as well as the world. Reduced barriers to trade and larger foreign direct investment inflows boosted  export as well as import  markets of this country.}  \\

\added{Notice that the composition of the first largest community remains almost the same in the next few consecutive years (as can be seen from  Figure \ref{top_com_2000_2009}). However, some Chinese sectors become more and more important.} For example, the role of Chinese manufacturing industries assumes in 2004 a higher relevance in this community (see also Figure \ref{top_sectors_two_largest_com_2004} (a) in the \ref{Appendix D}).

\added{
This result is in line with \cite{Chen} and  \cite{Benedictis_Tajoli_2009} that show  an important increase of labor-intensive manufactured goods' exportation and of importation of raw materials that come from China and the other developing Asian economies (aggregated in ROW) during that period.
In contrast, looking at the internal structure of the second largest community in 2004, we find that sectors from the former Eastern Bloc's countries play a less important role, but those from Belgium and France become more dominant (for details see also Figure \ref{top_sectors_two_largest_com_2004} in the \ref{Appendix D})}.

Overall, we observe that the structure of the  largest international communities still continues to evolve in the subsequent years.  For example, as for the data in 2009, a number of  influential industries from China no longer stay in the first largest community with those from US but  emerge as the ``key players'' in the second largest one. In particular, as displayed in Figure \ref{top_com_2000_2009} (d), a large community dominated by Asian countries is noticeable (in particular formed by sectors of South Korea, China and Taiwan)\footnote{Additional results can be found in Figure \ref{top_sectors_two_largest_com_2009}, \ref{Appendix D}.}. The role of Asian countries \added{still continue} to increase over time,  leading China in 2014 \added{(as already reported and discussed in Section \ref{Results_2})}  to form the largest international community together with  some sectors from the  other Asian countries  such as  South Korea, Taiwan, India, Japan. \added{Altogether,  the  aforementioned  observations may be the signals that the global trade in input-output system has been moving from a former US-Russia  bipolar  system to another bipolar (or multipolar) one, with the rising role of sectors from  China in the recent years.}

\vspace{1cm}

\begin{figure} [H]
	\centering
	\captionsetup[subfloat]{farskip=0pt,captionskip=0pt}
	\subfloat[Heterogeneity index computed for each  country]{\includegraphics[height=3.5 in, width = 3.5 in]{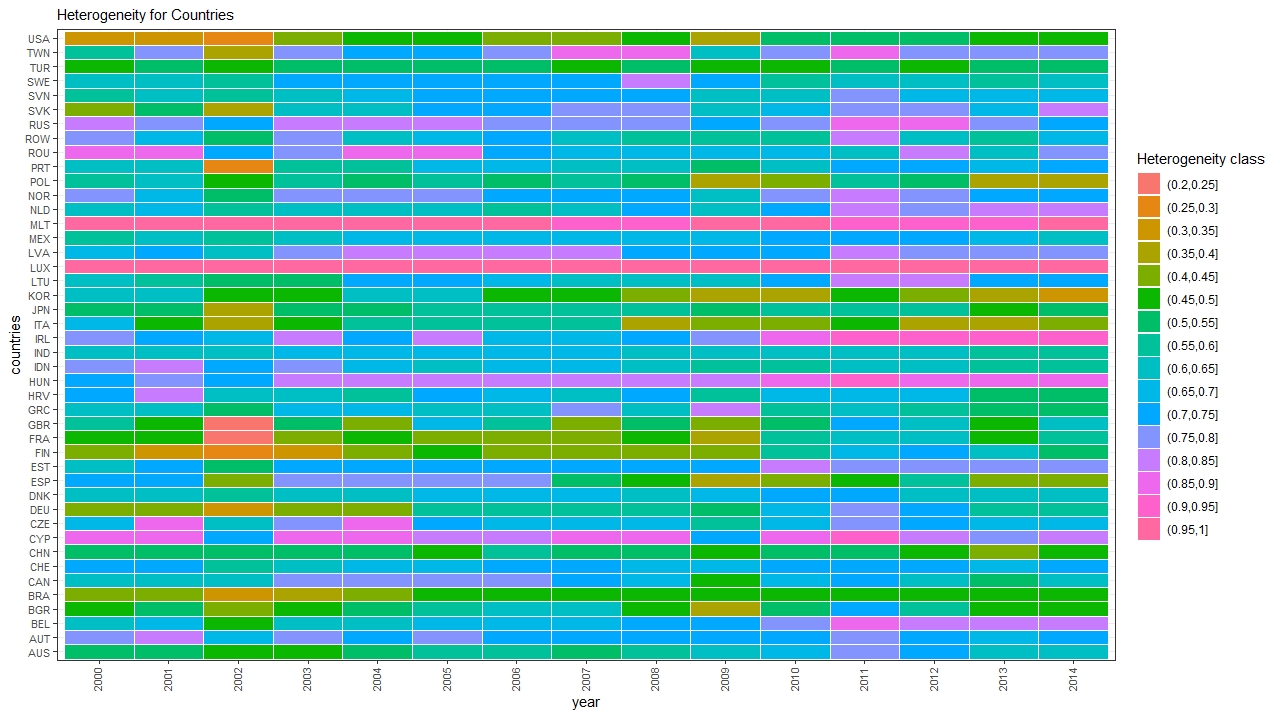}}
	\subfloat[Heterogeneity index computed for each  sector]{\includegraphics[height=3.5 in, width = 3.5 in]{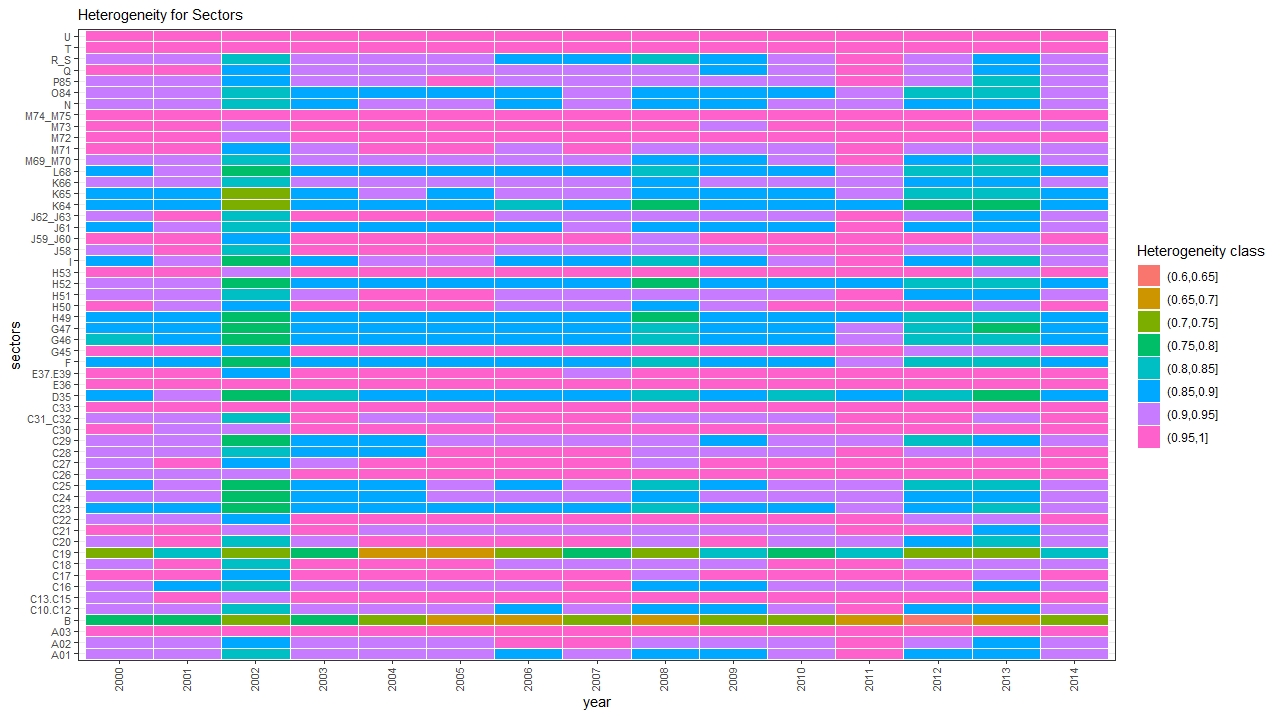}}
	\caption{\added{Heterogeneity index computed for each year at country and sector level, respectively. The color bar indicates the value of Gini index in the interval $[0,1]$. The index provides an indication of the dispersion of sectors (countries) for each country (sector) along the communities. A larger value of this index implies a higher degree of dispersion.}}
	\label{Het_SC}
\end{figure}

\added{To bring our temporal analysis of the mesoscale structure of the global trade network in input-output to a close, we report in Figure \ref{Het_SC} the Gini index, computed for each year, at either country or sector level.}\footnote{\added{Notice that we have computed the Gini index for each country as well as for each sector in 2014 (Tables \ref{table_cou} and \ref{table_sec} in \ref{Appendix D}, last column). Then we compute the index for each country and sector in the other years, to examine its evolution over time from 2000 to 2014. As explained previously, a higher value of this index for a country means that on average its sectors are more spread out between different communities. In contrast, a higher value of this index for a sector (layer) implies that, in the same traded layer, countries are on average split in various communities.}}

\added{We notice, on average, a lower level of heterogeneity for countries, confirming the presence of relevant domestic trades \added{(see Figure \ref{Het_SC}, panel (a)). As reported and discussed in Section  \ref{Results_2},  industries from a number of countries engage less actively in the global  input-output network, and many of them still mainly rely on their domestic counterparts. The Gini indices for these countries, therefore, are relatively lower.
On the other hand,  a higher level of heterogeneity \added{is observed  for several Eastern European and  Mediterranean Sea countries}. We argue that for these countries, a relatively stronger international diversification occurs primarily at the (quantitative) intensive margin effects rather than the extensive ones, since the binary version  of the network capturing the existence of trade links is already very dense and does not change very much.} Indeed, these countries derived great benefits from their proximity to the European Union, large amounts of investment from their Western European \added{neighbors} and the far-ranging domestic reforms on which their accession was conditional (see for details \cite{Ait}). \added{In a similar vein,} some Asian developing countries benefited from a strong export orientation and the increased intra-regional integration, also due to the proximity to the Chinese growth pole. Relevant players (as United States, Germany, Japan and China) show instead a moderate level of heterogeneity due to a greater balance between domestic and international trade linkages.}

\added{Concerning sectors (see Figure \ref{Het_SC} panel (b)),  we notice relatively higher values for their Gini indices, with few exceptions. As already observed for 2014 in Section \ref{Results_2}, the two sectors ``Mining and quarrying (B)" and ``Manufacture of coke and refined petroleum products (C19)" also show systematically a lower heterogeneity and are mainly involved in Community 1 in the previous years. This is not very surprising since from a perspective of the factors of production, over different countries worldwide, these two industries are very closely connected to each other.} 

\added{In addition,  major changes in the mesoscale structure of the global input-output network are again confirmed by a  significant drop in the Gini index in 2002 for all sectors, where a very high concentration of countries and sectors in the  largest community is observed. As discussed before, the restructure of trade relations among European countries, the disintegration of the Eastern Bloc, the rising role of sectors from China after its accession to WTO, etc. may explain such noticeable changes in the network in that year.}

\section {Conclusions}    \label{Conclusions}

This work analyses the global trade network in input-output, using the recently released WIOD database. We show that, by viewing the network through the lens of a multilayer architecture, we are able to extract richer information on interdependencies between countries and industries around the world that cannot be easily detected at the aggregate mono-layer level.

We first analyse the heterogeneity and the diversification in input-output relationships and find that although the list of trading partners is generally broad,   some sectors trade more intensively with a number of other sectors in different countries. We then view countries  as nodes and industries as  layers and explore the similarities and interactions among the layers. Again,  in the weighted version, we observe that the similarity  levels and the interaction strengths are varied across pairs of layers and that few couples of layers tend to be more overlapped and/or more strongly interacting with each other.
Such first insights motivate us to examine the multilayer architecture at a broader scale rather the node-node or layer-layer interrelations alone.

As investigated in our work, at the mesoscale level there exist several large international communities  in which some countries trade their inputs and outputs more intensively in some specific industry-based layers. In such a network structure, the world somewhat seems to have  bipolar or multipolar trade system with different  clusters that are more internally cohesive.

 \added{Furthermore,} interestingly, we also observe that the clustering \added{behaviors} of trades in input-output among industries and countries restructure over the period from 2000 to 2014. In general, not only the internal composition changes, but the centrality rankings of the members inside each community  also reorder, with the diminishing role of  industries from some countries  and  the growing importance of those from some other countries.\\
As in 2000, the first largest community is still mainly comprised by sectors from the US, Japan together with several sectors from China. In contrast, the second largest cluster consists of industries mostly from the former Eastern Bloc's countries (e.g. Russia and Poland).
As in the data for 2014, the most recent year available in the WIOD database,  we identify an Australian-Asian community, where relevant industries of main Asian players (China,
India, Japan, South Korea) and Australia are clustered together. On the other hand, those from countries involved in the North American Free Trade Agreement (Canada, Mexico and USA) belongs to the next largest  community.

\added{The empirical results obtained from our present work provide several meaningful implications.  First,  it is worth mentioning that large international clusters comprised by different industries from different nations detected in our present work may pave the potential pathways for large cascading failures. In particular, a shock to a ``key" sector (hub) of  a ``key" country in one of these clusters could be transmitted and propagated to many other foreign sectors located in the same cluster via cross-border input-output relations, and possibly lead to a large downturn in the global trade.  
As shown during the time of the Covid pandemics, the shortages of inputs provided by some industries can lead to disruptions in a number of subsequent dependent industries in other countries \citep[e.g.,][]{Eppinger_et_al_2020}.}
\added{In addition, the presence of such large international  clusters, in which members of each cluster interact more strongly among themselves, may also imply meaningful network origins (i.e., those related to international input-output relations) of business cycle synchronizations among different countries in the world \citep[e.g.,][]{Burstein_et_al_2008,  Johnson_2014}. Moreover, as pointed out in our present work, the multilayer analysis helps to uncover the complexity of the relations among countries when they trade within and between different layers-based industrial sectors. The proposed approach also allows us to explore the interdependencies among layers rather than just the node-node pairwise relations in the aggregate mono-layer analysis.
As shown in \cite{Lee_Goh_2016}, \cite{Korniyenko_et_al_2018} and in \cite{Tzavellas_2022},  more severe  cascading effects can occur in an architecture of multiple layers. This is mainly because the simultaneous presence of  intralayer as well as interlayer linkages in such an architecture can lead to collective dynamics of interdependent layers and create multiple potential channels of the diffusion of shocks among countries that are ignored by the analysis of the aggregate version or the analysis of single layers alone.}

 Our present work opens several directions for future research. First, the findings on the emergence of several international communities and their structural changes over time need to be investigated further to determine the underlying  economic and other potential mechanisms.
 Second, since in this study we focus on the mesoscale structure and the related properties of the global input-output network,  in our future work, we plan to extend our analysis to study and quantify the other important network measures and properties such as the multilayer clustering coefficients and  the multilayer centrality  \citep[e.g. see][]{DeDomenico_et_al_2013, BCG2022}.
Third, we believe that incorporating additional layers representing other economic relations such as  financial links, trade in final goods and services on top of the input-output  interrelations  among countries would be able to   give a more comprehensive and richer analysis of the multilayer architecture of the world-wide economic network.  \added{Such an analysis certainly provides useful inputs to the study of different potential channels of cascading failures and transmission of shocks in the global economy \citep[e.g., see][]{Lee_Goh_2016, Korniyenko_et_al_2018}}.

\bibliographystyle{model5-names}\biboptions{authoryear}


\clearpage

\appendix

\section{Node centrality in a multilayer architecture}\label{Appendix A}

We report here the definitions of node centrality measures applied in the text. \\
The \textit{intralayer} in-strength  and in-degree for each node $i$ in a layer $\alpha$ are given by 
\begin{equation*}
	s_{i,in}^{[\alpha \leftarrow \alpha]}=\sum_{j \neq i} W^{[\alpha, \alpha]}_{ji}, 
\end{equation*}
and 
\begin{equation*}
	k_{i,in}^{[\alpha \leftarrow \alpha]}=\sum_{j \neq i} A^{[\alpha, \alpha]}_{ji}. 
\end{equation*}
In a similar vein, we can define the \textit{intralayer} out-strength and out-degree  of node $i$ in layer $\alpha$ as
\begin{equation*}
	s_{i,out}^{[\alpha  \rightarrow \alpha]}=\sum_{j \neq i} W^{[\alpha, \alpha]}_{ij}, 
\end{equation*}
\begin{equation*}
	k_{i,out}^{[\alpha  \rightarrow  \alpha]}=\sum_{j \neq i} A^{[\alpha, \alpha]}_{ij}.
\end{equation*} 

In addition, for every pair of different layers $\alpha$ and $\beta$, one can quantify the \textit{interlayer} in-degree and in-strength of interactions from all nodes in a layer $\beta$ to  a specific node $i$ in  layer $\alpha$  as
\begin{equation*}
	s_{i,in}^{[\alpha  \leftarrow  \beta]}=\sum_{j \neq i} W^{[\alpha, \beta]}_{ji}, 
\end{equation*}
\begin{equation*}
	k_{i,in}^{[\alpha  \leftarrow  \beta]}  =\sum_{j \neq i} A^{[\alpha, \beta]}_{ji}.
\end{equation*}
Similarly, the \textit {interlayer} out-strength and out-degree from a particular node $i$ in the layer $\alpha$ to all nodes in a layer $\beta$ are given by
\begin{equation*}
	s_{i,out}^{[\alpha \rightarrow  \beta]}=\sum_{j \neq i} W^{[\alpha, \beta]}_{ij},
\end{equation*}
and
\begin{equation*}
	k_{i,out}^{[\alpha \rightarrow  \beta]}=\sum_{j \neq i} A^{[\alpha, \beta]}_{ij}.
\end{equation*}

In the context of an input-output network, the aforementioned degrees and strengths represent the distributions of the numbers of trading partners and of the magnitudes of trade in input-output across nodes, respectively. A higher incoming (outgoing) degree of  a node indicates that it has a more diversified portfolio of  input sellers (output buyers) in  \added{a particular} layer. Nodes with higher incoming (outgoing) strengths trade their inputs (outputs) more intensively  with the rest of nodes in a certain layer. 

The \textit {total} in-strength and in-degree of a node $i$ in a layer $\alpha$ \textit {from nodes in all $L$ layers} can be straightforwardly extended as 
\begin{equation*}
	s_{i,in}^{[\alpha, \ total]}=\sum_{\beta} s_{i,in}^{[\alpha \leftarrow \beta]},
\end{equation*}
\begin{equation*}
	k_{i,in}^{[\alpha, \ total]}=\sum_{\beta} k_{i,in}^{[\alpha \leftarrow \beta]}.
\end{equation*}

Analogously,  we can define the \textit {total} out-strength and out-degree of a node $i$ in a layer $\alpha$ \textit {across all layers} as 
\begin{equation*}
	s_{i,out}^{[\alpha, \ total]}=\sum_{\beta} s_{i,out}^{[\alpha \rightarrow  \beta]},
\end{equation*}
\begin{equation*}
	k_{i,out}^{[\alpha, \ total]}=\sum_{\beta} k_{i,out}^{[\alpha \rightarrow \beta]}.
\end{equation*}

Notice that from the definitions of  the total in-strength $s_{i,in}^{[\alpha, \ total]}$ and  the intralayer in-strength $s_{i,in}^{[\alpha \leftarrow \alpha]}$, we can easily obtain the \textit {total interlayer} in-strength, which is given by  
\begin{equation*} 
	s_{i,in}^{[\alpha, \ inter]}= s_{i,in}^{[\alpha, \ total]}-s_{i,in}^{[\alpha \leftarrow \alpha]}=\sum_{\beta  \neq \alpha} s_{i,in}^{[\alpha \leftarrow \beta]}.
\end{equation*}
This captures the overall intensity of the dependency of node $i$ in the layer $\alpha$ on the inputs provided by nodes from the other layers.
Similarly, one can derive the \textit {total interlayer} in-degree, the \textit {total interlayer} out-strength and the \textit {total interlayer} out-degree:
\begin{equation*} 
	k_{i,in}^{[\alpha, \ inter]}= k_{i,in}^{[\alpha, \ total]}-k_{i,in}^{[\alpha \leftarrow \alpha]}=\sum_{\beta  \neq \alpha} k_{i,in}^{[\alpha  \leftarrow \beta]},
\end{equation*}

\begin{equation*} 
	s_{i, out}^{[\alpha, \ inter]}= s_{i,out}^{[\alpha, \ total]}-s_{i,out}^{[\alpha \rightarrow \alpha]}=\sum_{\beta  \neq \alpha} s_{i,out}^{[\alpha \rightarrow \beta]},
\end{equation*}

and

\begin{equation*} 
	k_{i, out}^{[\alpha, \ inter]}= k_{i,out}^{[\alpha, \ total]}-k_{i,out}^{[\alpha \rightarrow  \alpha]}=\sum_{\beta  \neq \alpha} k_{i,out}^{[\alpha \rightarrow  \beta]}.
\end{equation*}




\clearpage

\section{Lists and tables}
\label{Appendix B}

In this subsection, we summarize the list of 56  sectors and the list of 44 countries and areas in the WIOD table (2016 release version).

\begin{table} [!hbt]
	\centering
	\scalebox{1.1}{
	\resizebox{\textwidth}{!}{
\begin{tabular}{lll}
\hline\hline
\multicolumn{1}{l}{no.}&\multicolumn{1}{c}{sector name}&\multicolumn{1}{c}{sector code}\tabularnewline
\hline
1&Crop and animal production, hunting and related service activities&A01\tabularnewline
2&Forestry and logging&A02\tabularnewline
3&Fishing and aquaculture&A03\tabularnewline
4&Mining and quarrying&B\tabularnewline
5&Manufacture of food products, beverages and tobacco products&C10-C12\tabularnewline
6&Manufacture of textiles, wearing apparel and leather products&C13-C15\tabularnewline
7&Manufacture of wood and of products of wood and cork, except furniture; manufacture of articles of straw and plaiting materials&C16\tabularnewline
8&Manufacture of paper and paper products&C17\tabularnewline
9&Printing and reproduction of recorded media&C18\tabularnewline
10&Manufacture of coke and refined petroleum products &C19\tabularnewline
11&Manufacture of chemicals and chemical products &C20\tabularnewline
12&Manufacture of basic pharmaceutical products and pharmaceutical preparations&C21\tabularnewline
13&Manufacture of rubber and plastic products&C22\tabularnewline
14&Manufacture of other non-metallic mineral products&C23\tabularnewline
15&Manufacture of basic metals&C24\tabularnewline
16&Manufacture of fabricated metal products, except machinery and equipment&C25\tabularnewline
17&Manufacture of computer, electronic and optical products&C26\tabularnewline
18&Manufacture of electrical equipment&C27\tabularnewline
19&Manufacture of machinery and equipment n.e.c.&C28\tabularnewline
20&Manufacture of motor vehicles, trailers and semi-trailers&C29\tabularnewline
21&Manufacture of other transport equipment&C30\tabularnewline
22&Manufacture of furniture; other manufacturing&C31 C32\tabularnewline
23&Repair and installation of machinery and equipment&C33\tabularnewline
24&Electricity, gas, steam and air conditioning supply&D35\tabularnewline
25&Water collection, treatment and supply&E36\tabularnewline
26&Sewerage; waste collection, treatment and disposal activities; materials recovery; remediation activities and other waste management services &E37-E39\tabularnewline
27&Construction&F\tabularnewline
28&Wholesale and retail trade and repair of motor vehicles and motorcycles&G45\tabularnewline
29&Wholesale trade, except of motor vehicles and motorcycles&G46\tabularnewline
30&Retail trade, except of motor vehicles and motorcycles&G47\tabularnewline
31&Land transport and transport via pipelines&H49\tabularnewline
32&Water transport&H50\tabularnewline
33&Air transport&H51\tabularnewline
34&Warehousing and support activities for transportation&H52\tabularnewline
35&Postal and courier activities&H53\tabularnewline
36&Accommodation and food service activities&I\tabularnewline
37&Publishing activities&J58\tabularnewline
38&Motion picture, video and television programme production, sound recording and music publishing activities; programming and broadcasting activities&J59 J60\tabularnewline
39&Telecommunications&J61\tabularnewline
40&Computer programming, consultancy and related activities; information service activities&J62 J63\tabularnewline
41&Financial service activities, except insurance and pension funding&K64\tabularnewline
42&Insurance, reinsurance and pension funding, except compulsory social security&K65\tabularnewline
43&Activities auxiliary to financial services and insurance activities&K66\tabularnewline
44&Real estate activities&L68\tabularnewline
45&Legal and accounting activities; activities of head offices; management consultancy activities&M69 M70\tabularnewline
46&Architectural and engineering activities; technical testing and analysis&M71\tabularnewline
47&Scientific research and development&M72\tabularnewline
48&Advertising and market research&M73\tabularnewline
49&Other professional, scientific and technical activities; veterinary activities&M74 M75\tabularnewline
50&Administrative and support service activities&N\tabularnewline
51&Public administration and defence; compulsory social security&O84\tabularnewline
52&Education&P85\tabularnewline
53&Human health and social work activities&Q\tabularnewline
54&Other service activities&RS\tabularnewline
55&Activities of households as employers; undifferentiated goods- and services-producing activities of households for own use&T\tabularnewline
56&Activities of extraterritorial organizations and bodies&U\tabularnewline
\hline \hline
\end{tabular}}}
\vspace{1cm}
\caption  {List of 56  sectors \added {in each of 44 countries} in the WIOD table.}
\label{table_WIOD_industries}
\end{table}

\begin{table}[!tbp]
\begin{center}
\scalebox{0.9}{
\begin{tabular}{lll}
\hline\hline
\multicolumn{1}{l}{no}&\multicolumn{1}{c}{country name}&\multicolumn{1}{c}{country code}\tabularnewline
\hline
1&Australia&AUS\tabularnewline
2&Austria&AUT\tabularnewline
3&Belgium&BEL\tabularnewline
4&Bulgaria&BGR\tabularnewline
5&Brazil&BRA\tabularnewline
6&Canada&CAN\tabularnewline
7&Switzerland&CHE\tabularnewline
8&China&CHN\tabularnewline
9&Cyprus&CYP\tabularnewline
10&Czech Republic&CZE\tabularnewline
11&Germany&DEU\tabularnewline
12&Denmark&DNK\tabularnewline
13&Spain&ESP\tabularnewline
14&Estonia&EST\tabularnewline
15&Finland&FIN\tabularnewline
16&France&FRA\tabularnewline
17&United Kingdom&GBR\tabularnewline
18&Greece&GRC\tabularnewline
19&Croatia&HRV\tabularnewline
20&Hungary&HUN\tabularnewline
21&Indonesia&IDN\tabularnewline
22&India&IND\tabularnewline
23&Ireland&IRL\tabularnewline
24&Italy&ITA\tabularnewline
25&Japan&JPN\tabularnewline
26&Korea&KOR\tabularnewline
27&Lithuania&LTU\tabularnewline
28&Luxembourg&LUX\tabularnewline
29&Latvia&LVA\tabularnewline
30&Mexico&MEX\tabularnewline
31&Malta&MLT\tabularnewline
32&Netherlands&NLD\tabularnewline
33&Norway&NOR\tabularnewline
34&Poland&POL\tabularnewline
35&Portugal&PRT\tabularnewline
36&Romania&ROU\tabularnewline
37&Russian Federation&RUS\tabularnewline
38&Slovak Republic&SVK\tabularnewline
39&Slovenia&SVN\tabularnewline
40&Sweden&SWE\tabularnewline
41&Turkey&TUR\tabularnewline
42&Taiwan&TWN\tabularnewline
43&United States&USA\tabularnewline
44&Rest of the World&ROW\tabularnewline
\hline \hline
\end{tabular}} 
\end{center}
\vspace{0.5cm}
\caption  {List of 44 countries \added{(nations)} and areas in the WIOD table.}
\label{table_WIOD_countries}
\end{table}

\begin{table} [!hbt]
	\centering
	\scalebox{0.5}{
	\resizebox{\textwidth}{!}{
\begin{tabular}{rrrrrr}
  \hline \hline
 & mean & median & max & min & skewness \\ 
  \hline
2000 & 4.14 & 0.00036 & 108243.68 & 0.00 & 194.82 \\ 
  2001 & 4.09 & 0.00040 & 112267.14 & 0.00 & 196.01 \\ 
  2002 & 4.19 &  0.00048 & 107060.84 & 0.00 & 196.77 \\ 
  2003 & 4.74 & 0.00070 & 122185.72 & 0.00 & 199.23 \\ 
  2004 & 5.46 &  0.00098 & 137786.65 & 0.00 & 204.64 \\ 
  2005 & 6.14 & 0.00120 & 153480.65 & 0.00 & 218.59 \\ 
  2006 & 6.81 &  0.00144 & 177839.87 & 0.00 & 226.11 \\ 
  2007 & 7.85 & 0.00194 & 224846.10 & 0.00 & 240.24 \\ 
  2008 & 8.90 &  0.00229 & 292710.91 & 0.00 & 281.04 \\ 
  2009 & 8.00 &  0.00203 & 269025.66 & 0.00 & 279.77 \\ 
  2010 & 9.00 &  0.00205 & 295837.80 & 0.00 & 300.23 \\ 
  2011 & 10.34 & 0.00241 & 378726.83 & 0.00 & 337.85 \\ 
  2012 & 10.63 & 0.00235 & 444664.66 & 0.00 & 345.17 \\ 
  2013 & 11.10 & 0.00242 & 509681.09 & 0.00 & 362.35 \\ 
  2014 & 11.45 & 0.00243 & 541342.11 & 0.00 & 367.45 \\ 
   \hline \hline
\end{tabular}}}
\vspace{0.5cm}
\caption  {Descriptive statistics for the weights  (traded amounts in  millions of dollars)  among different  2464 industries from 44 countries over the period 2000-2014. Mathematically, the weights are the elements of the $W^{supra}$ matrix.}
\label{table_descriptive_stats_weights}
\end{table}

\clearpage

\section {Basic network properties of the global input-output network}    \label{Appendix C}

\textit {Trade diversity and concentration   from the total degrees and strengths of nodes:}

\begin{figure} [H]
\centering
\captionsetup[subfloat]{farskip=0pt,captionskip=0pt}
\subfloat[]{\includegraphics[height=2.5in,width = 2.5in]{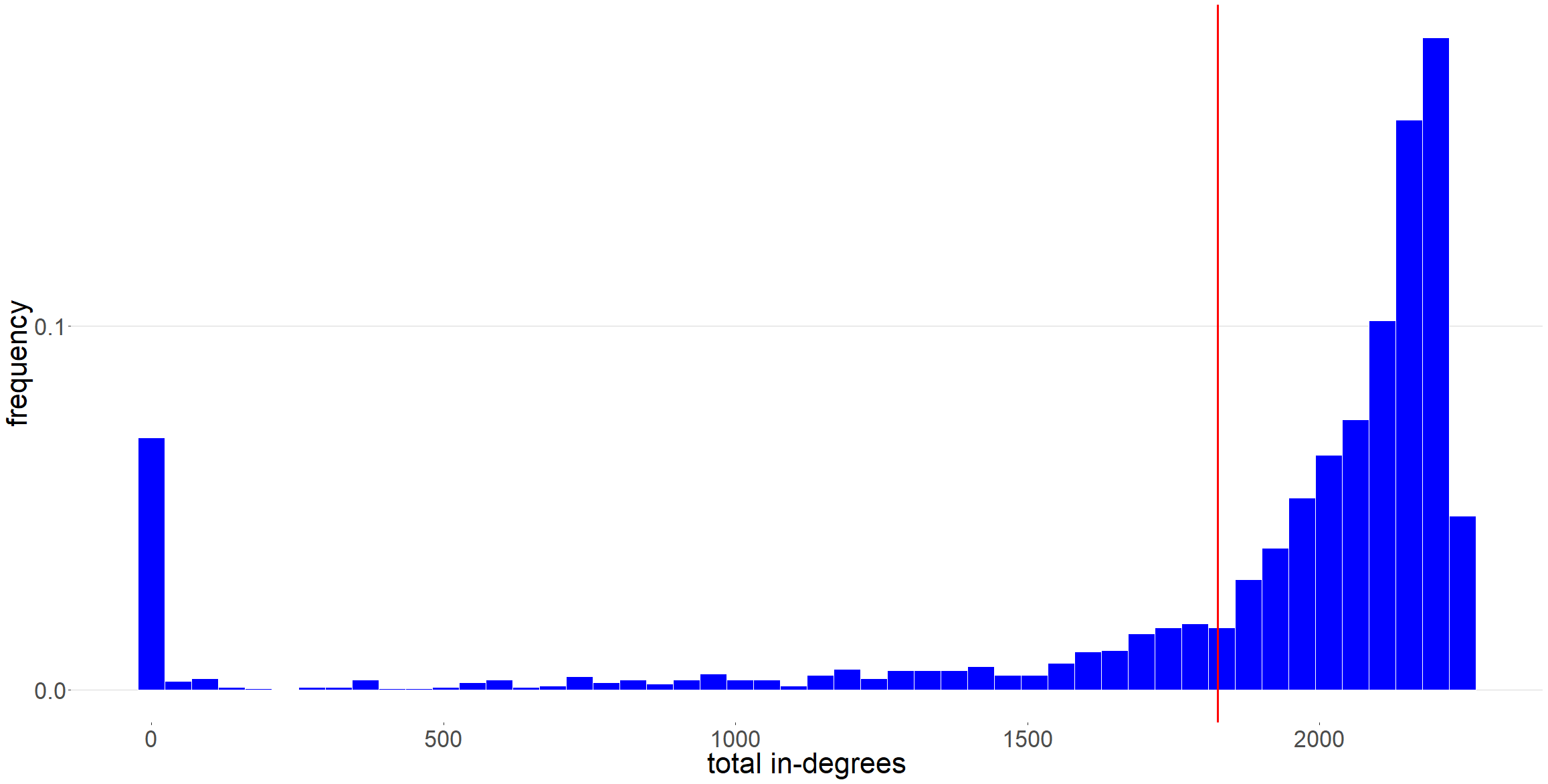}}
\subfloat[]{\includegraphics[height=2.5in,width = 2.5in]{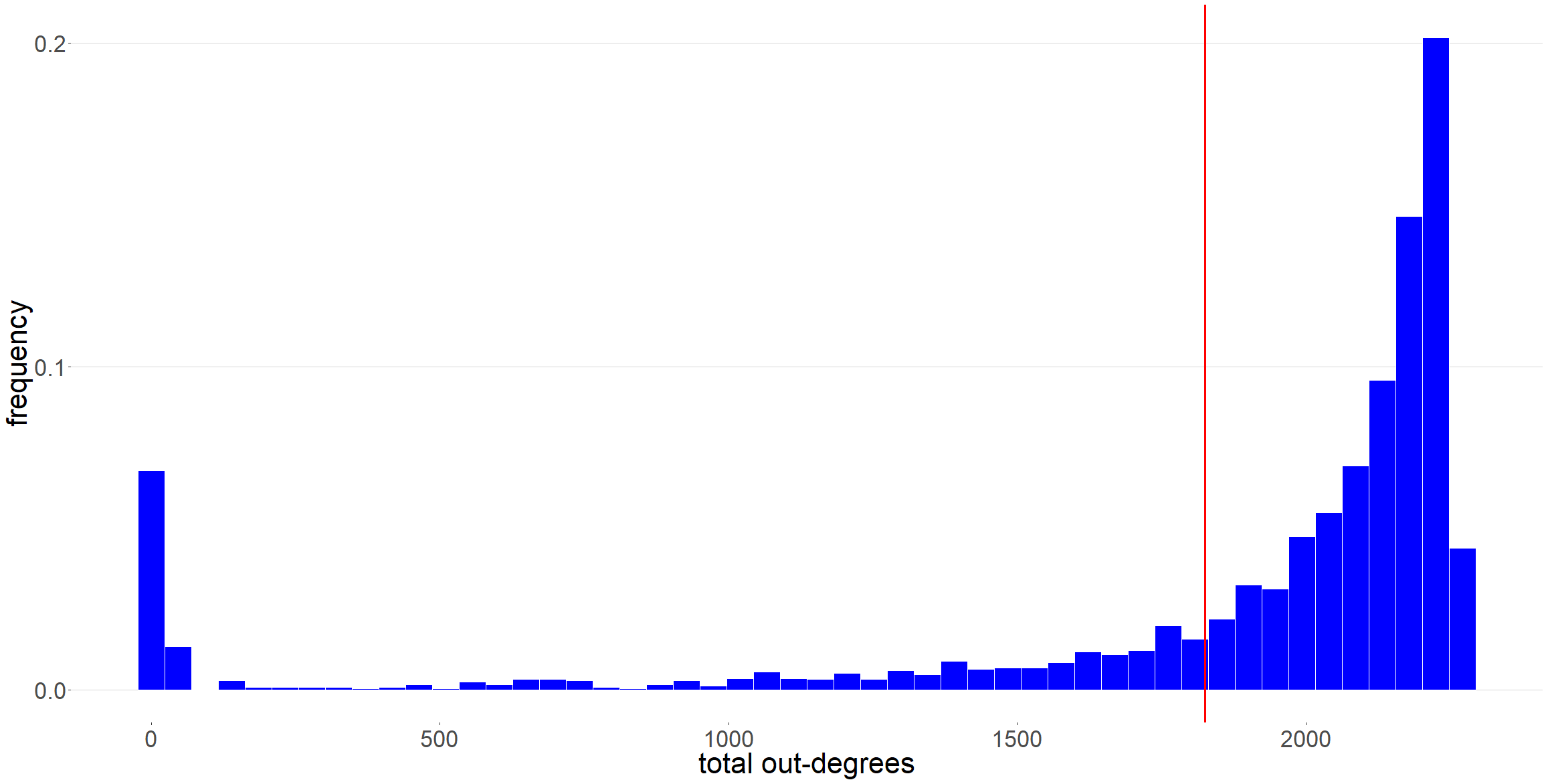}}\\
\subfloat[]{\includegraphics[height=2.5in,width = 2.5in]{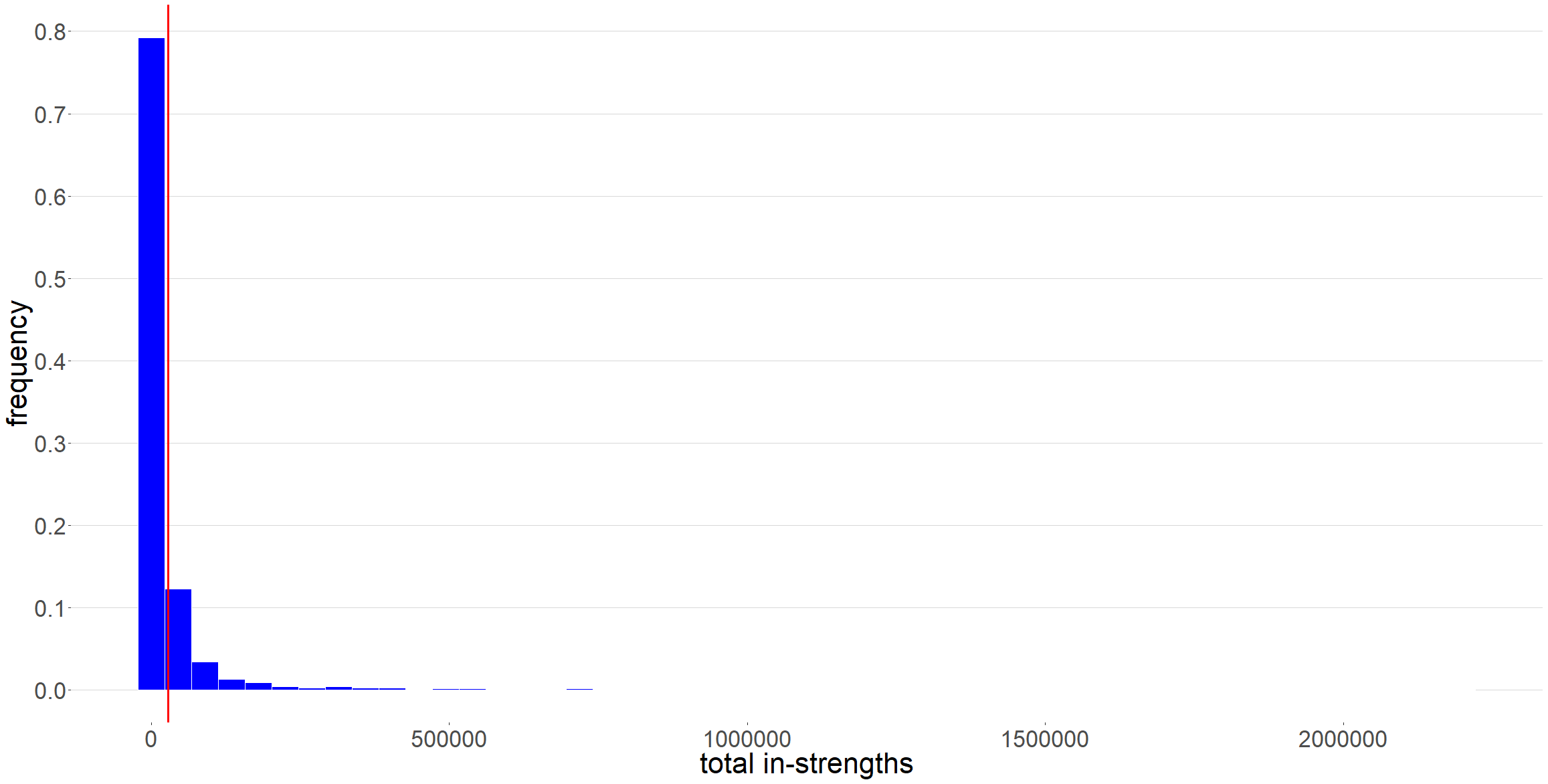}}
\subfloat[]{\includegraphics[height=2.5in,width = 2.5in]{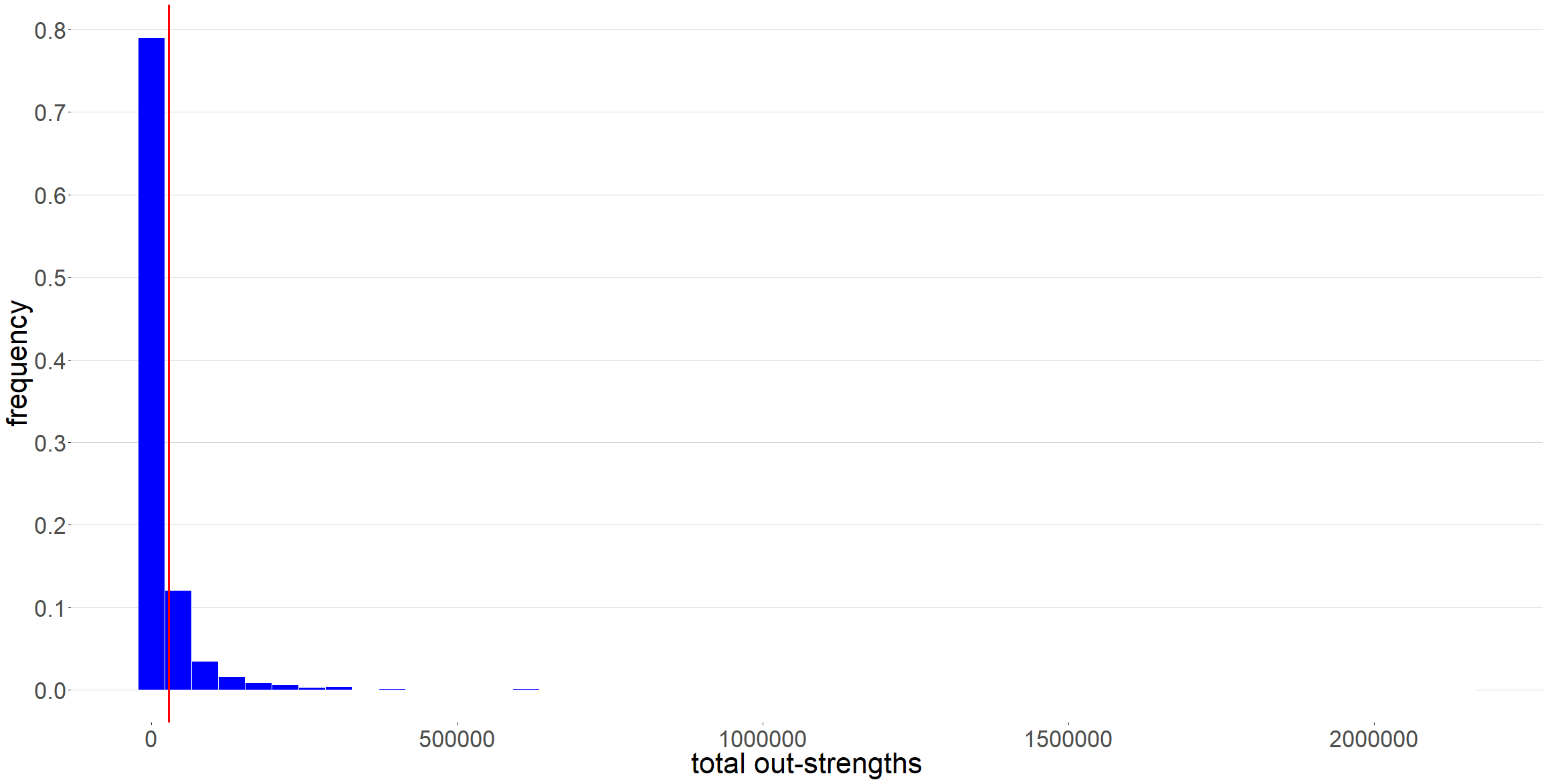}}
  \caption{Distributions of total in- and out-degrees \added{(panels (a) and (b))} and distributions of total in- and out-strengths \added{(panels (c) and (d))}. The red vertical line indicates the mean value.}
  \label{hist_total_in_out_degrees_strengths}
\end{figure}

Figure \ref{hist_total_in_out_degrees_strengths} shows the distributions of the total degrees and strengths of  \added{$2464$ industries from 44 countries} based on the original input-output table \added{in 2014}. \added{Each degree indicates the total number of trading partners while each strength represents the total volume of trades for a particular sector. Note that we also distinguish the in-coming degrees (strengths) and the out-going degrees (strengths) to illustrate  input  and output dependencies, respectively. Further descriptive statistics for the degrees and strengths of sectors in different years are reported in Tables  \ref{table_descriptive_stats_degrees} and \ref{table_descriptive_stats_strengths}.}

It is interesting to observe \added{that, in the binary version, almost} all of nodes have degrees larger than $1500$ and a negative skewness, i.e. a longer tail on the left side of the distribution, is observed \added{(see Figure \ref{hist_total_in_out_degrees_strengths} (a) and (b))}. This implies that, \added{overall}, industries have  a relatively  diversified list of trading partners from different countries. This is consistent with the fact that the binary version of the network is very dense, for both intra-country and inter-country links.

In contrast, when the intensity of trading relationships is taken into account in the weighted version, \added{as can be seen from Figure \ref{hist_total_in_out_degrees_strengths} (c) and (d), both  total in-strengths and out-strengths} exhibit a right-skewed distribution, where only few nodes have strengths much larger than those of the rest.
To further investigate the overall heterogeneity level of trading intensities across partners for each sector, we compute the  Herfindahl–Hirschman \added{(H-H)} indices (e.g., see \cite{KVALSETH2018, Hirschman1964}) associated with  total in- and out-strengths, which are, respectively, defined as 
\begin{equation}
h_{i,in}^{[\alpha]}=\sum_{j, \beta} (w_{ji}^{[\alpha, \beta]}/s_{i,in}^{[\alpha, \ total ]})^2
\quad
{\rm and}
\quad
h_{i,out}^{[\alpha]}=\sum_{j, \beta} (w_{ij}^{[\alpha, \beta]}/s_{i,in}^{[\alpha, \ total ]})^2.
\label{Herfindahl_in_out}
\end{equation}

Generally speaking, the  H–H  indices defined in (\ref{Herfindahl_in_out}) will range from $1/N_{supra}=1/(NL)$ to $1$, and a larger value indicates a higher level of the concentration \added{(or specialization)} of the inputs (purchased from fewer seller-sectors) or outputs (distributed across limited buyer-sectors).

\begin{figure} [H]
\centering
\captionsetup[subfloat]{farskip=0pt,captionskip=0pt}
\subfloat[]{\includegraphics[height=3in,width = 3in]{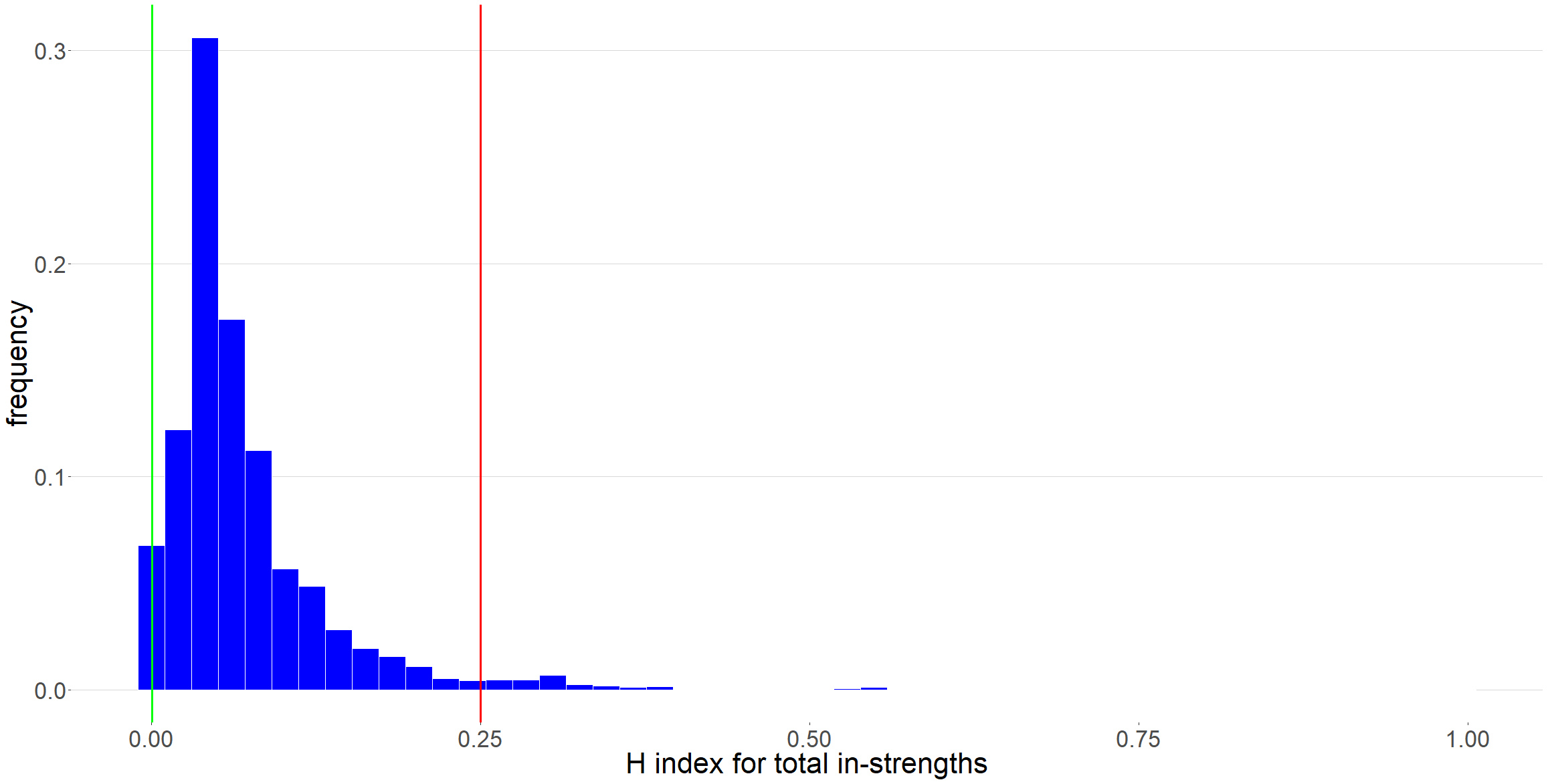}}
\subfloat[]{\includegraphics[height=3in,width = 3in]{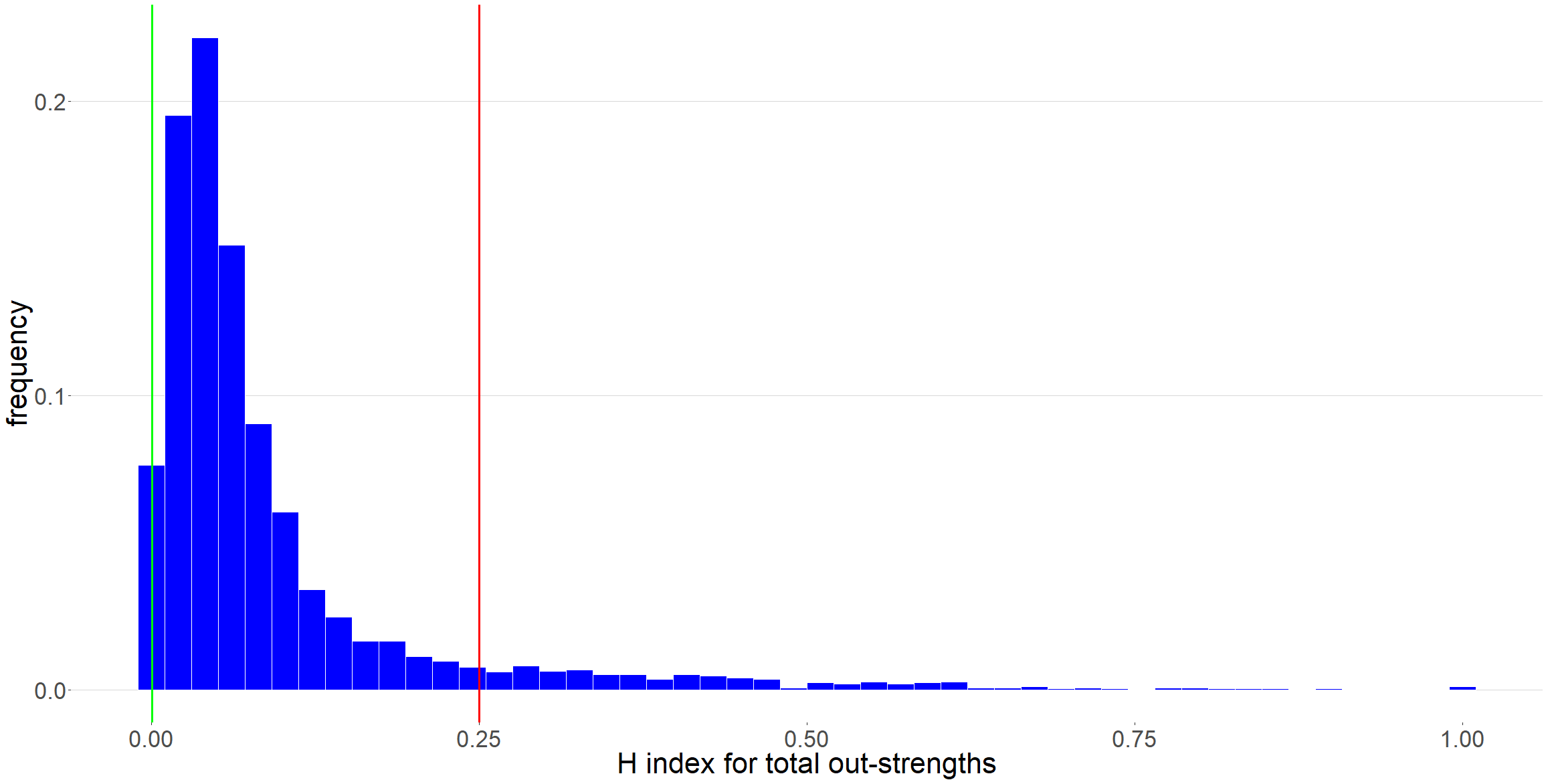}}
  \caption{Distributions of  the \added{H-H} indices  of total in- and out-strengths. The vertical  green line correspond to the minimum $1/N_{supra}$. The vertical red  line represents the critical value $0.25$, which indicates a high level of concentration if $h>0.25$.}
  \label{hist_H_total_in_out_strengths}
\end{figure}

\begin{figure} [H]
	\centering
	\captionsetup[subfloat]{farskip=0pt,captionskip=0pt}
	\subfloat[Based on H-H concentration in out-strengths, 2014]{\includegraphics[height=4.2in,width = 3.2in]{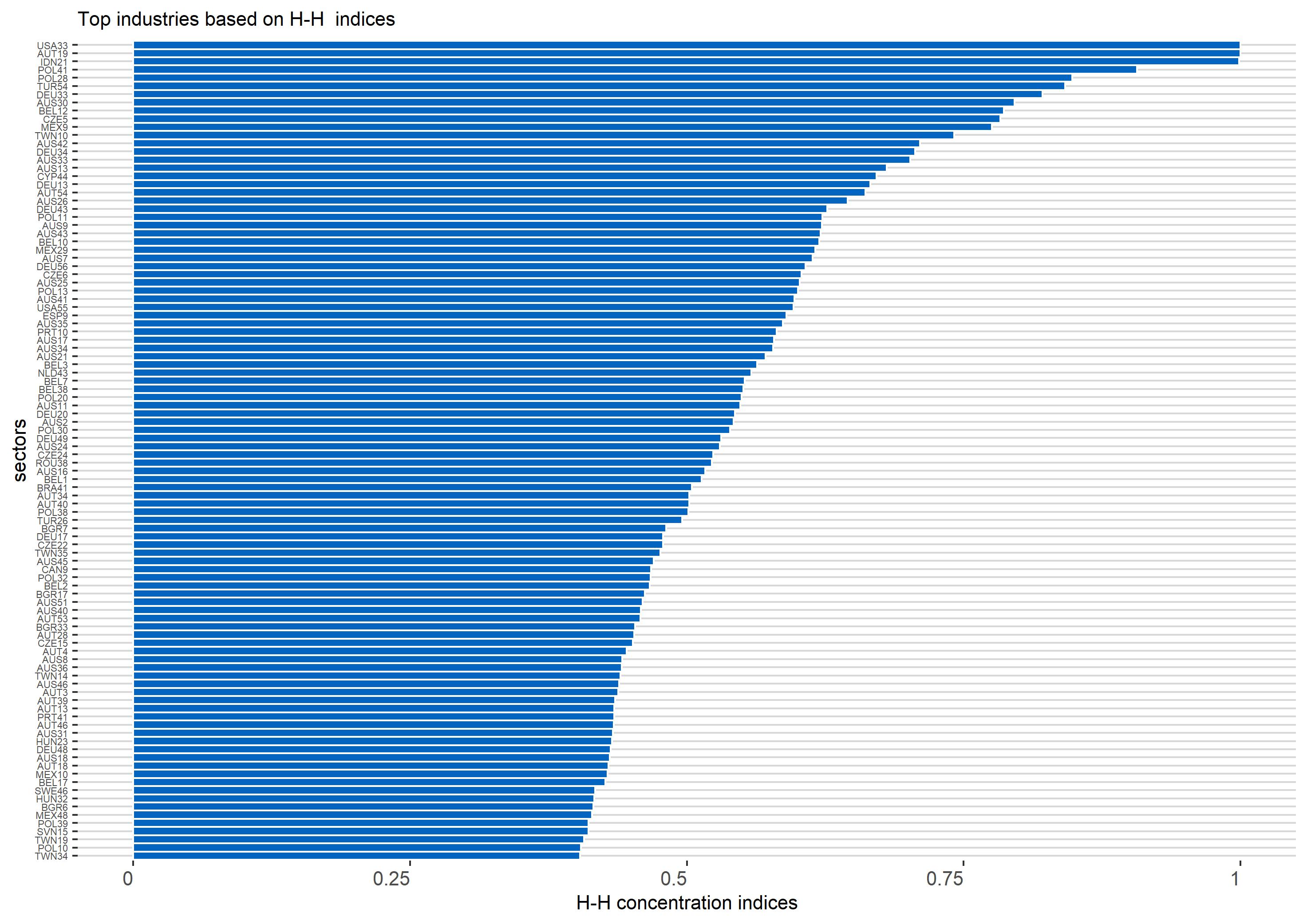}}
	\subfloat[Based on H-H concentration in in-strengths, 2014]{\includegraphics[height=4.2in,width = 3.2in]{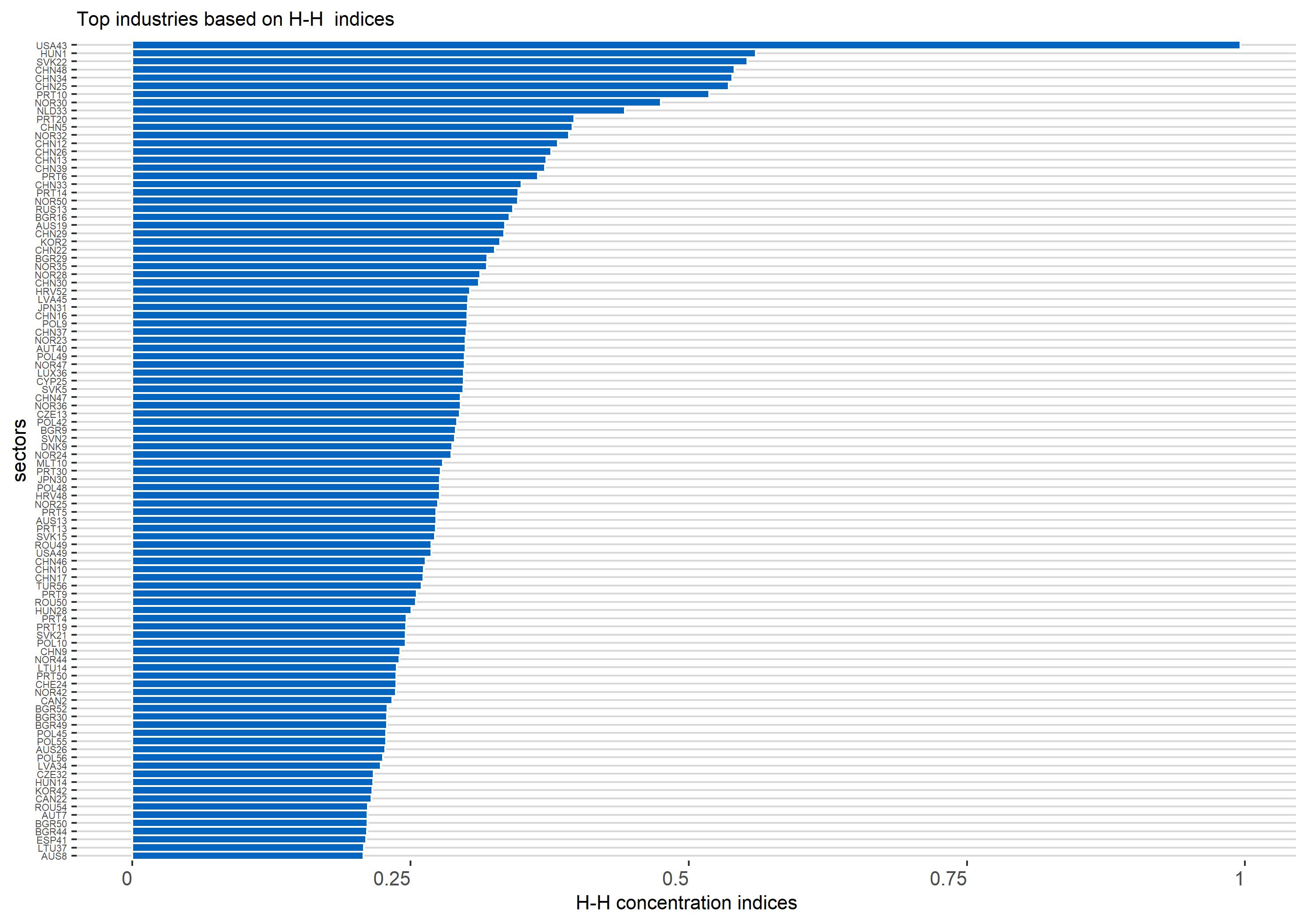}}
		\vspace{0.5cm}
	\caption{Ranking of top  industries based on H-H concentration indices: 
A larger value indicates a higher level of the concentration of the inputs (purchased from fewer seller-sectors) or outputs (distributed across limited buyer-sectors). Labels on the $y$ axis are given by the country code and by the number related to the sector.}
	\label{top_sectors_HH_concentration_indices}
\end{figure}

In our work, we find that while almost all industries in different countries have  a moderate or negligible level of concentration of trading across partners, some other industries do  intensively trade with few input suppliers or output customers  in the global production network (see Figure \ref{hist_H_total_in_out_strengths} \added{and the list of the top 100 sectors that have highest H-H concentration levels in Figure \ref{top_sectors_HH_concentration_indices}). Digging deeper, we find that indeed these less quantitatively diversified sectors come from different countries with various industrial codes, hence no overall definite conclusion can be drawn. However, since they are more concentrated with a limited number of trading partners, they may not be very active in the large international industrial clusters that will be discussed later.}  

\begin{figure} [H]
	\centering
	\captionsetup[subfloat]{farskip=0pt,captionskip=0pt}
	\subfloat[The average of degrees]{\includegraphics[height=3in,width = 3in]{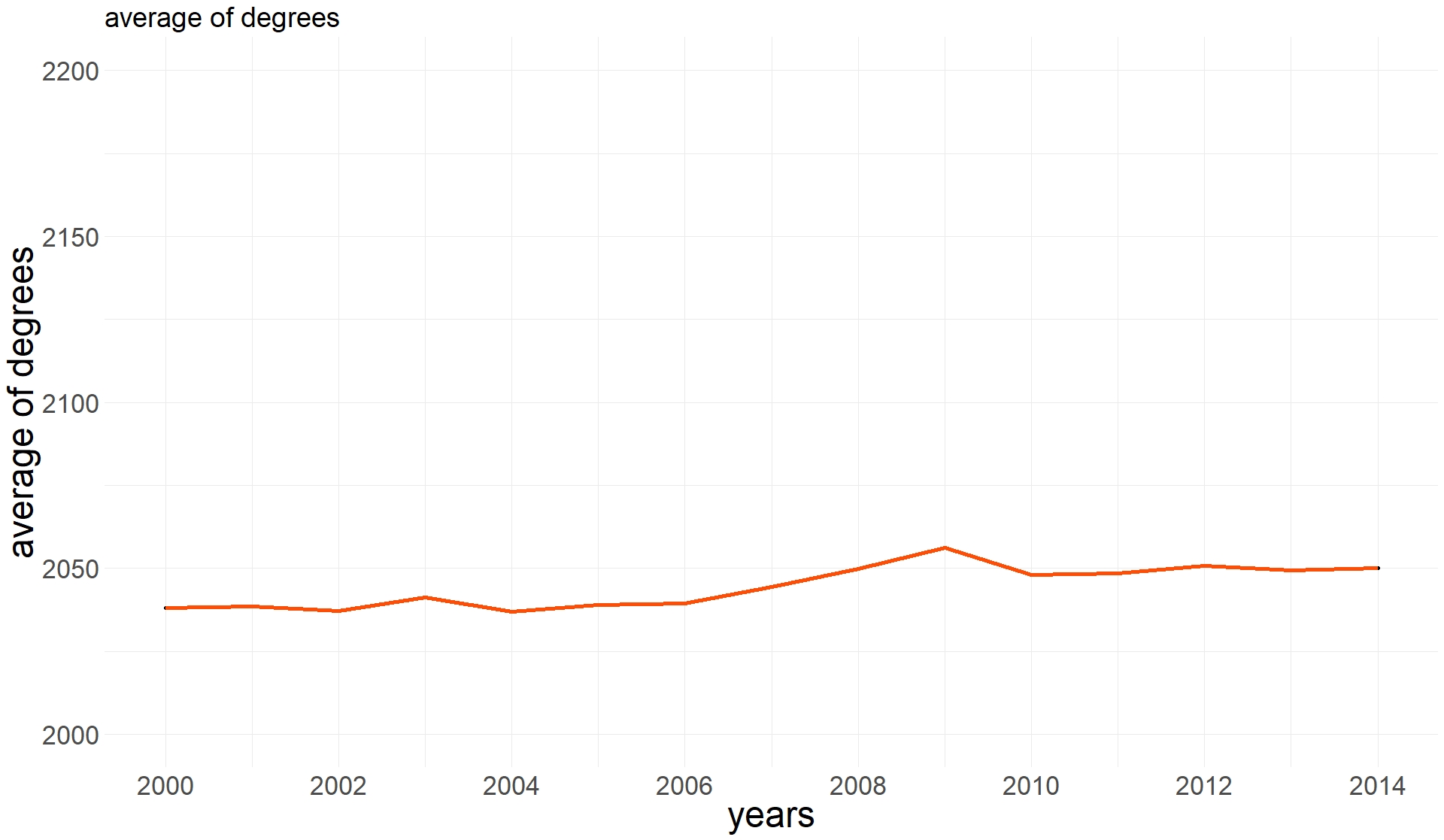}}
	\subfloat[The average of strengths]{\includegraphics[height=3in,width = 3in]{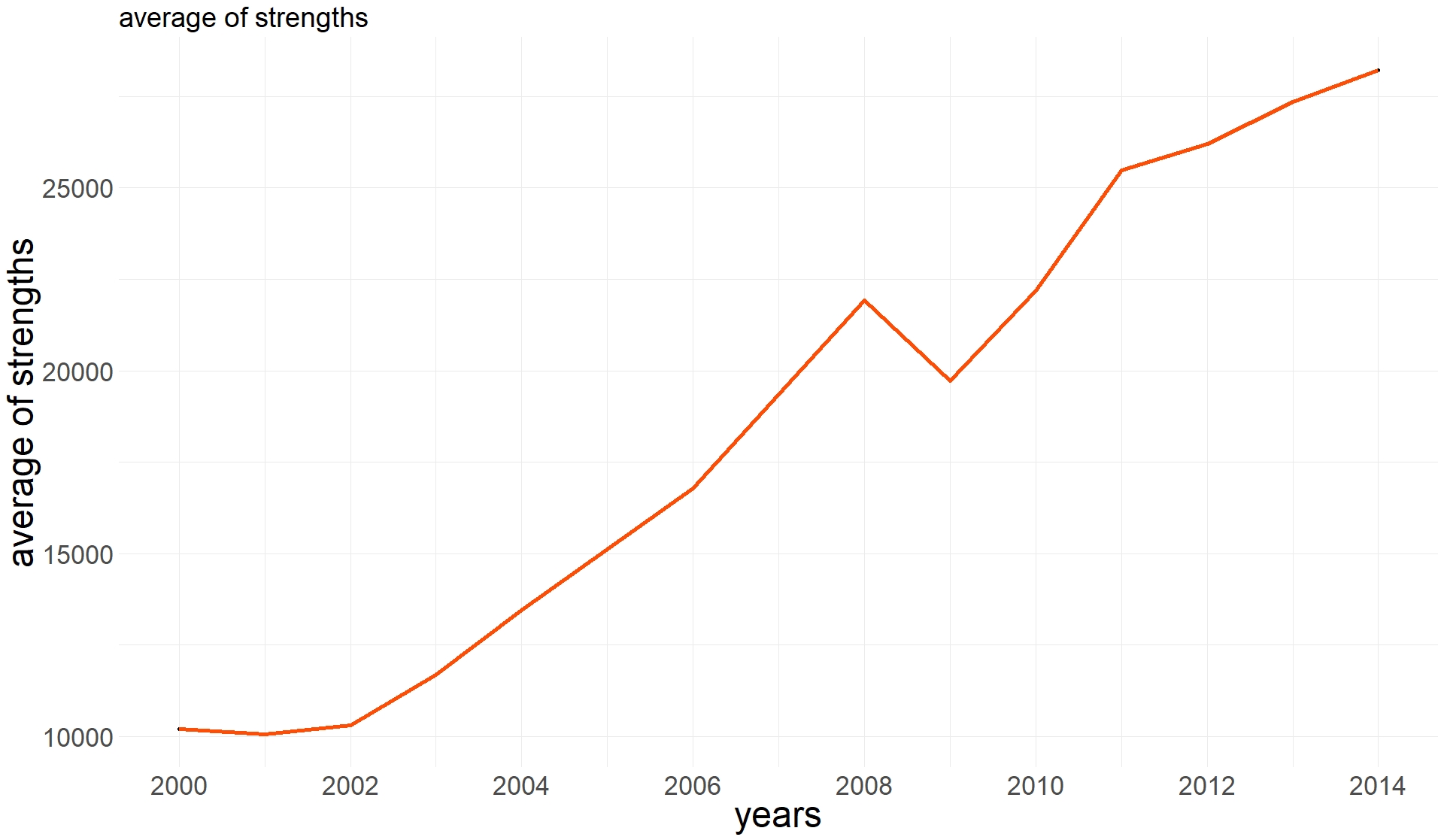}}
	\vspace{0.5cm}
	\caption{The average of degrees (numbers of trading partners) and the average of the strengths (the total traded amounts in millions of dollars) of industries over the period 2000-2014.}
	\label{average_degree_strength}
\end{figure}

\begin{table} [!hbt]
		\begin{center}
	\scalebox{1}{
	\resizebox{\textwidth}{!}{
\begin{tabular}{rllllllllll}
  \hline \hline
  & \multicolumn{5}{c}{out-degree} & \multicolumn{5}{c}{in-dregree} \\
 & mean & median & max & min & skewness & mean &  median & max & min & skewness\\
 \hline
2000 & 2038.14 & 2292.00 & 2294.00 & 0.00 & -2.59 & 2038.14 & 2198.00 & 2251.00 & 0.00 & -3.36 \\ 
  2001 & 2038.57 & 2293.00 & 2295.00 & 0.00 & -2.59 & 2038.57 & 2198.00 & 2251.00 & 0.00 & -3.37 \\ 
  2002 & 2037.24 & 2293.00 & 2295.00 & 0.00 & -2.58 & 2037.24 & 2197.00 & 2250.00 & 0.00 & -3.37 \\ 
  2003 & 2041.16 & 2293.00 & 2295.00 & 0.00 & -2.61 & 2041.16 & 2201.00 & 2254.00 & 0.00 & -3.37 \\ 
  2004 & 2036.86 & 2294.00 & 2297.00 & 0.00 & -2.57 & 2036.86 & 2196.00 & 2248.00 & 0.00 & -3.38 \\ 
  2005 & 2038.90 & 2293.00 & 2295.00 & 0.00 & -2.60 & 2038.90 & 2198.00 & 2250.00 & 0.00 & -3.36 \\ 
  2006 & 2039.37 & 2294.00 & 2296.00 & 0.00 & -2.59 & 2039.37 & 2198.00 & 2247.00 & 0.00 & -3.38 \\ 
  2007 & 2044.27 & 2294.00 & 2296.00 & 0.00 & -2.63 & 2044.27 & 2202.00 & 2255.00 & 0.00 & -3.37 \\ 
  2008 & 2049.89 & 2295.00 & 2297.00 & 0.00 & -2.66 & 2049.89 & 2208.00 & 2256.00 & 0.00 & -3.39 \\ 
  2009 & 2056.17 & 2294.00 & 2296.00 & 0.00 & -2.71 & 2056.17 & 2215.00 & 2258.00 & 0.00 & -3.39 \\ 
  2010 & 2047.98 & 2294.00 & 2297.00 & 0.00 & -2.64 & 2047.98 & 2206.00 & 2257.00 & 0.00 & -3.39 \\ 
  2011 & 2048.43 & 2295.00 & 2297.00 & 0.00 & -2.64 & 2048.43 & 2204.00 & 2254.00 & 0.00 & -3.40 \\ 
  2012 & 2050.75 & 2295.00 & 2297.00 & 0.00 & -2.66 & 2050.75 & 2207.00 & 2257.00 & 0.00 & -3.40 \\ 
  2013 & 2049.36 & 2295.00 & 2297.00 & 0.00 & -2.64 & 2049.36 & 2205.00 & 2256.00 & 0.00 & -3.40 \\ 
  2014 & 2050.11 & 2295.00 & 2297.00 & 0.00 & -2.65 & 2050.11 & 2206.00 & 2259.00 & 0.00 & -3.40 \\ 
   \hline \hline
\end{tabular}}}
	\end{center}
\vspace{0.5cm}
\caption  {Descriptive statistics for the total in-degrees (in-deg.) and total out-degrees (out-deg.) of industries of different countries over the period 2000-2014. These degrees indicate the total number of input providers and output customers of industries, respectively.}
\label{table_descriptive_stats_degrees}
\end{table}

\begin{table} [ht!]
		\begin{center}
	\scalebox{1}{
	\resizebox{\textwidth}{!}{

\begin{tabular}{rrrrrrrrrrr}
   \hline \hline
  & \multicolumn{5}{c}{out-strength} & \multicolumn{5}{c}{in-strength} \\
 & mean & median & max & min & skewness & mean &  median & max & min & skewness \\ 
  \hline
2000 & 10203.69 & 1482.62 & 520662.94 & 0.00 & 7.76 & 10203.69 & 1528.46 & 601310.93 & 0.00 & 8.11 \\ 
  2001 & 10067.04 & 1505.09 & 520018.16 & 0.00 & 7.89 & 10067.04 & 1544.44 & 662165.45 & 0.00 & 8.94 \\ 
  2002 & 10318.20 & 1683.61 & 516178.66 & 0.00 & 8.04 & 10318.20 & 1703.52 & 710021.48 & 0.00 & 9.53 \\ 
  2003 & 11678.59 & 1968.91 & 550096.73 & 0.00 & 7.81 & 11678.59 & 2018.69 & 757248.20 & 0.00 & 9.23 \\ 
  2004 & 13460.95 & 2312.89 & 590936.72 & 0.00 & 7.74 & 13460.95 & 2366.31 & 814150.19 & 0.00 & 8.76 \\ 
  2005 & 15119.28 & 2581.88 & 829788.19 & 0.00 & 8.56 & 15119.28 & 2621.40 & 880764.57 & 0.00 & 8.49 \\ 
  2006 & 16778.55 & 2922.45 & 1036754.48 & 0.00 & 9.03 & 16778.55 & 2958.67 & 948754.53 & 0.00 & 8.25 \\ 
  2007 & 19347.83 & 3492.05 & 1175484.54 & 0.00 & 8.72 & 19347.83 & 3500.86 & 1006047.59 & 0.00 & 7.69 \\ 
  2008 & 21928.81 & 3895.32 & 1693442.94 & 0.00 & 10.71 & 21928.81 & 4006.56 & 1094105.94 & 0.00 & 7.73 \\ 
  2009 & 19717.81 & 3399.06 & 1208091.06 & 0.00 & 8.31 & 19717.81 & 3453.91 & 1119248.06 & 0.00 & 9.00 \\ 
  2010 & 22184.88 & 3497.00 & 1589086.09 & 0.00 & 9.37 & 22184.88 & 3579.50 & 1250019.40 & 0.00 & 8.67 \\ 
  2011 & 25484.10 & 3932.58 & 2105071.02 & 0.00 & 10.77 & 25484.10 & 4021.78 & 1561448.76 & 0.00 & 8.62 \\ 
  2012 & 26201.51 & 3789.64 & 2237154.63 & 0.00 & 10.77 & 26201.51 & 3758.29 & 1772866.98 & 0.00 & 9.09 \\ 
  2013 & 27347.11 & 3930.67 & 2183110.41 & 0.00 & 9.91 & 27347.11 & 3852.11 & 1995572.56 & 0.00 & 9.46 \\ 
  2014 & 28209.77 & 4011.40 & 2145115.96 & 0.00 & 9.44 & 28209.77 & 3847.23 & 2200291.27 & 0.00 & 9.84 \\ 
   \hline \hline
\end{tabular}}}
\end{center}
\vspace{0.5cm}
\caption  {Descriptive statistics for the total in-strengths (in-str.) and total out-strengths (out-str.)  of industries of different countries over the period 2000-2014. Values of strengths are in millions of dollars. They indicate the total amounts of purchased inputs and sold outputs of industries, respectively.}
\label{table_descriptive_stats_strengths}
\end{table}

\begin{figure} [H]
\centering
\captionsetup[subfloat]{farskip=0pt,captionskip=0pt}
\subfloat[]{\includegraphics[height=4in,width = 4in]{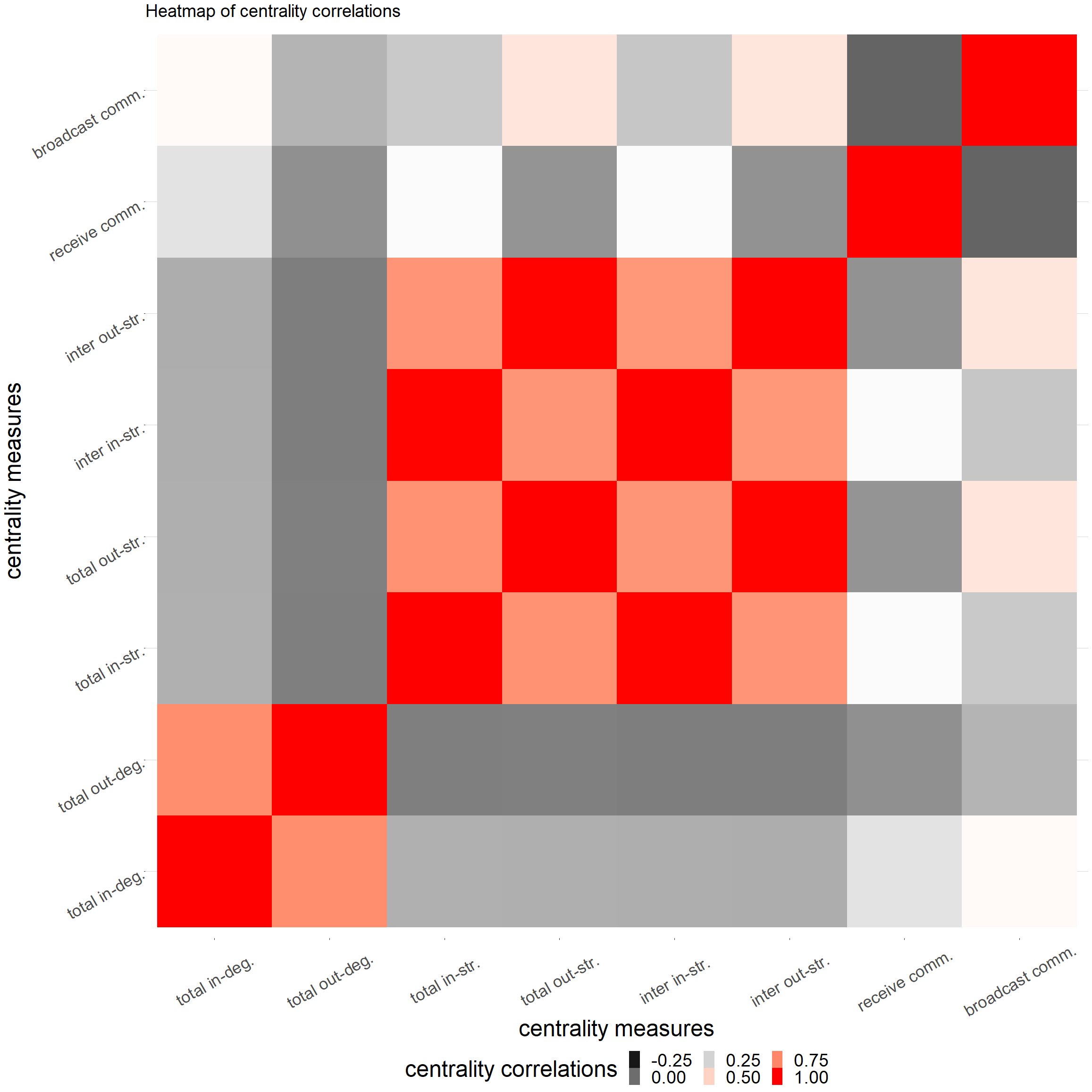}}
  \caption{Correlations among node degrees, strengths and the total (weighted) communicability centralities.  Total in-degrees (total in-deg) and total in-strengths (total in-str): $\{k_{i,in}^{[\alpha, \ total]}\}$, $\{s_{i,in}^{[\alpha, \ total]}\}$.  Total out-degrees (total out-deg) and total out-strengths (total out-str): $\{k_{i,out}^{[\alpha, \ total]}\}$, $\{s_{i,out}^{[\alpha, \ total]}\}$. Interlayer  in-strengths  (inter in-str) and interlayer out-strengths  (inter out-str): $\{s_{i,in}^{[\alpha, \ inter]}\}$, $\{s_{i,out}^{[\alpha, \ inter]}\}$.  The total communicability centralities of  a node  from/to all nodes in all layers (in the weighted version of the network): receive and broadcast communicability. }
  \label{correlations_degrees_strengths_centralities}
\end{figure}

 We report the corresponding Pearson correlation coefficients  in Figure \ref{correlations_degrees_strengths_centralities} to illustrate the overall correlations among different types of node degrees, strengths and the total communicability centralities. First, we can see that the correlation between  in-degrees  and  out-degrees is relatively high.  This is not an unexpected result since many nodes have a high level of in- and out-degrees (see Figure \ref{hist_total_in_out_degrees_strengths} \added{(a) and (b)}) and  \added{the binary network} is very dense.  
 However, the results also show that the correlation levels between degrees and strengths are relatively weaker, which again confirms the distinguished characteristics of the two versions of the network.  \added{As stressed before, from the binary analysis we find that on average each sector trades with a relatively diversified list of partners. However, the results from the weighted counterpart point out that the distribution of traded amounts across partners is not homogeneous and several sectors tend to concentrate higher volumes of trades with selected partners.} \\
Furthermore, given the high degree of heterogeneity in the node strengths, it is interesting to observe that interlayer in-strengths and out-strengths are synchronized together, to a certain extent. This implies that, if a sector in a country exports more its outputs, it also  purchases more inputs from sectors in other countries \added{and vice versa. Hence, the international diversification in the weighted version  seems to be aligned from both the input purchase and output sale perspectives}.
For a more comprehensive assessment, we also include in Figure \ref{correlations_degrees_strengths_centralities}  the \added{total receiving and broadcasting} communicability centralities defined for the weighted version of the network and compute the correlations between them and the various degrees and strengths.  As expected, since  these two additional measures are able to capture all possible travelling paths between nodes, they are less synchronized with the \added{other lower-order} network metrics like the node degrees and strengths.

\textit {Interrelations between layers in a multilayer architecture:}

We shall now reorganize the original input-output table into an equivalent network with  a multilayer  architecture. 
In such \added{structure}, an intralayer link represents a trade link from a country to another one in a particular sector. An interlayer link, in contrast, implies a trade link from a sector in one country to another sector in the same or another country.

\begin{figure} [H]
\centering
\captionsetup[subfloat]{farskip=0pt,captionskip=0pt}
\subfloat[]{\includegraphics[height=2.5in,width = 2.5in]{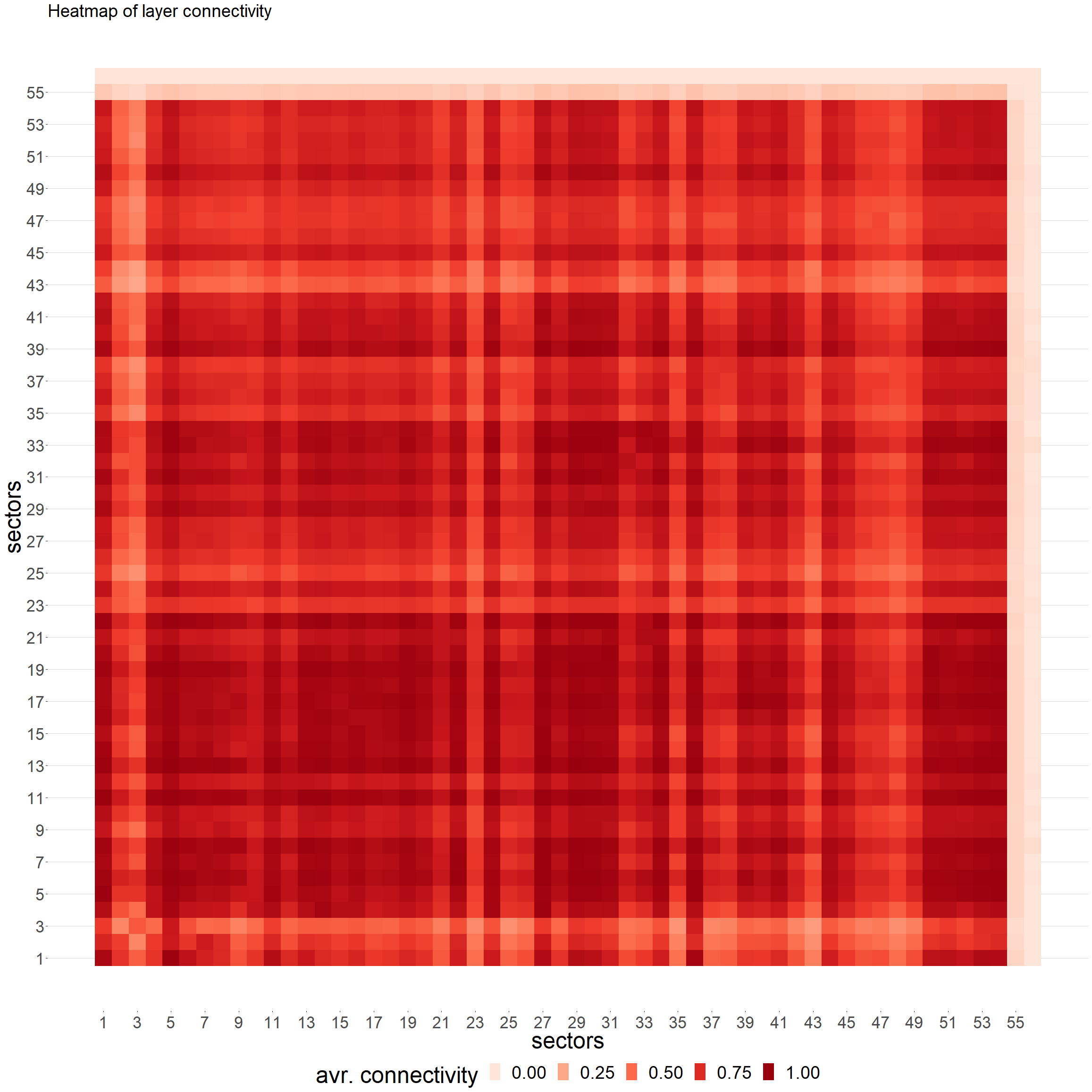}}
\subfloat[]{\includegraphics[height=2.5in,width = 2.5in]{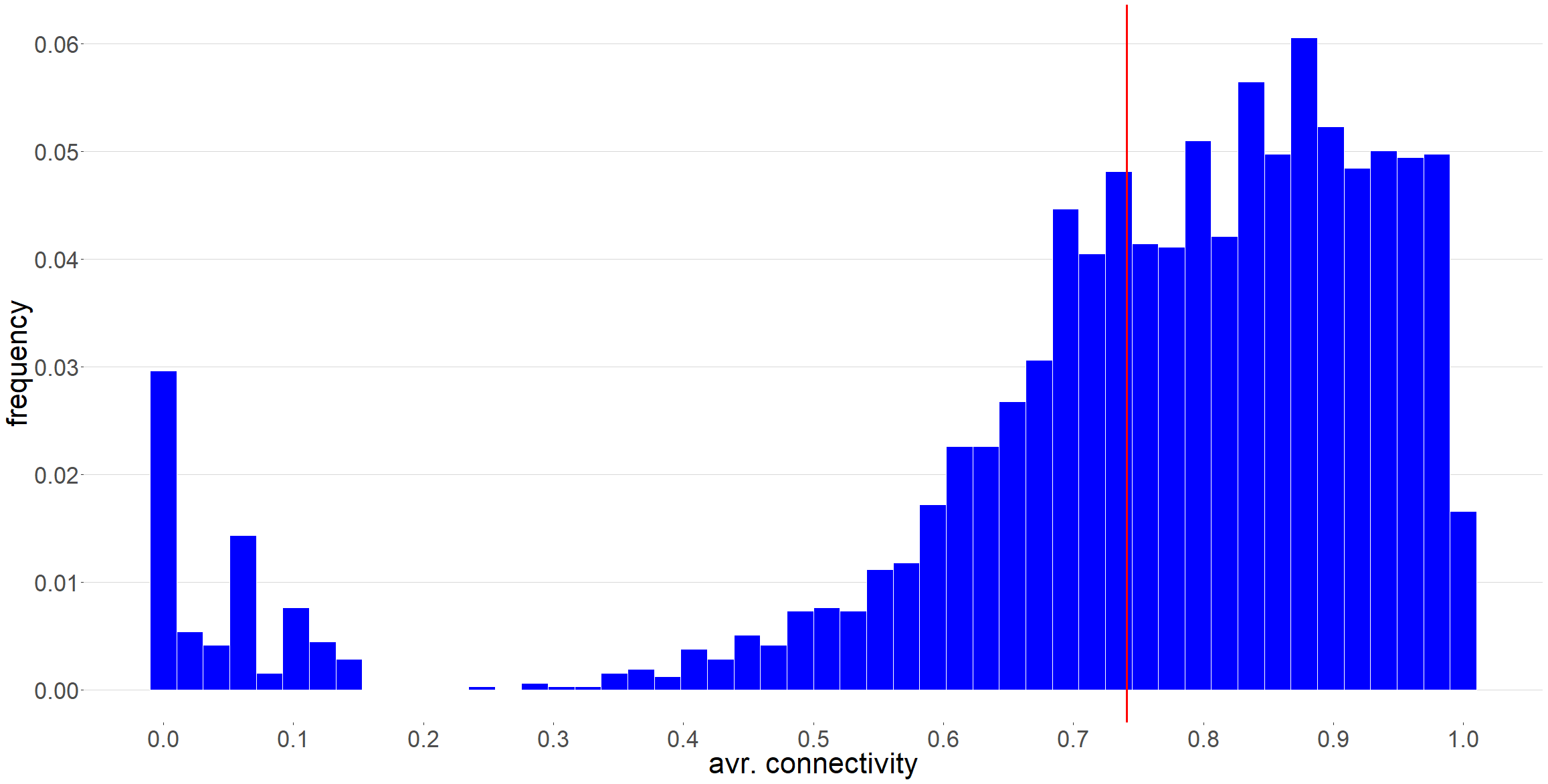}}\\
\subfloat[]{\includegraphics[height=2.5in,width = 2.5in]{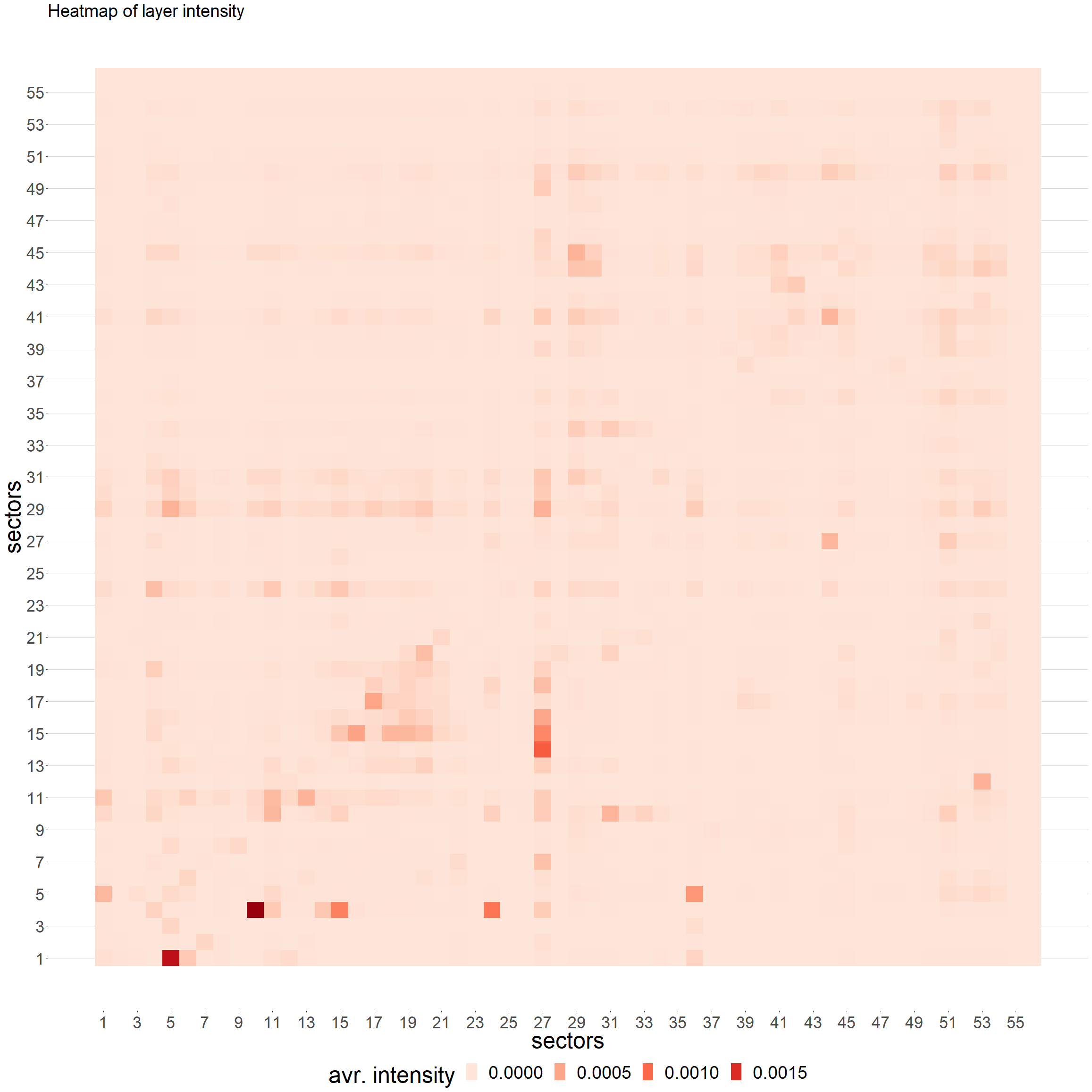}}
\subfloat[]{\includegraphics[height=2.5in,width = 2.5in]{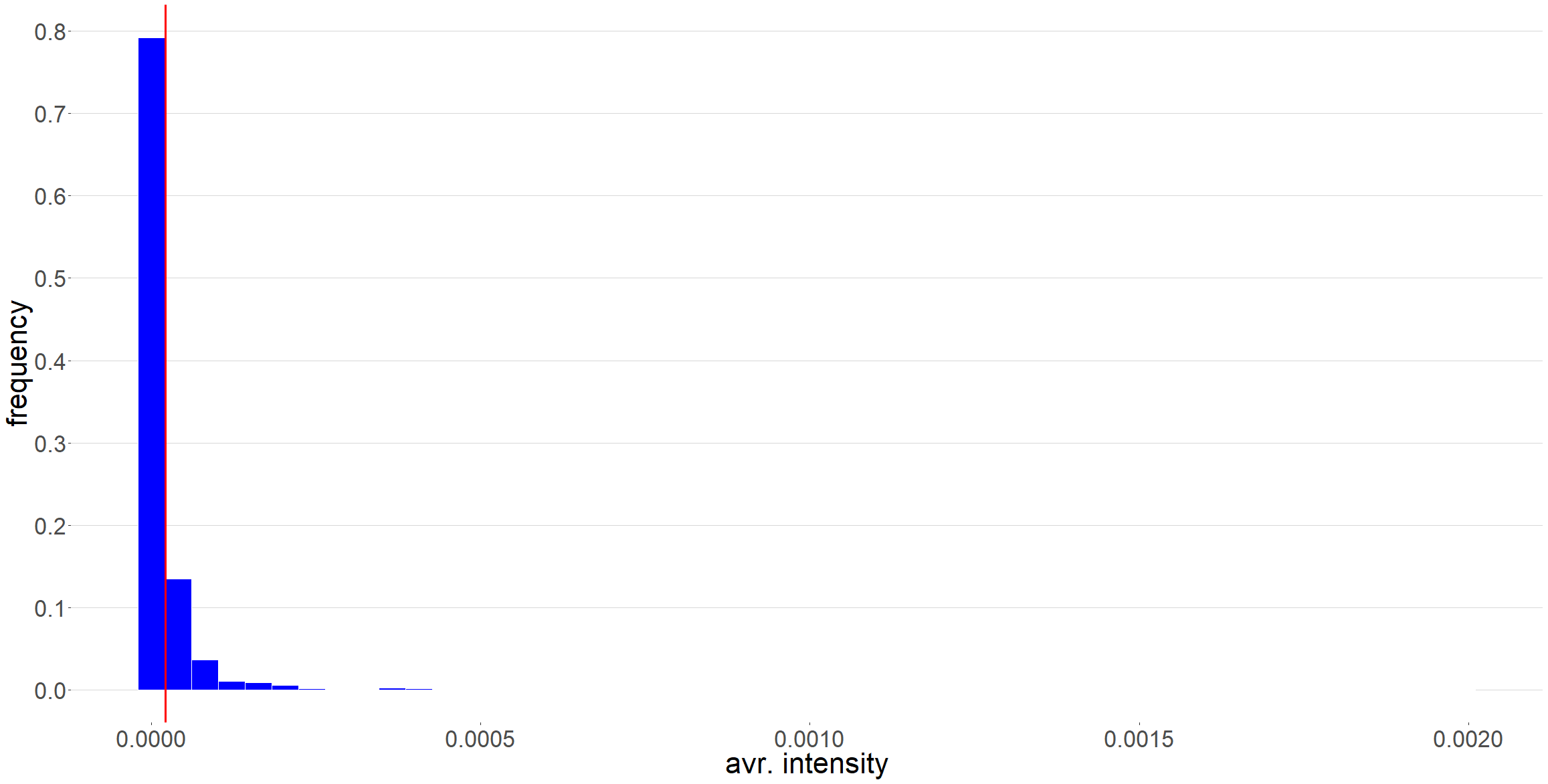}}
  \caption{Average connectivity and average intensity between layers. The measures for these averages are explained in formulas (\ref{average_connectivity}) and  (\ref{average_intensity}). The vertical red  line in each histogram shows the mean value. Note that  here layers represent 56 sectors, while nodes in each layer are 44 countries. \added{The rows and columns of the two color-coded matrices (panel (a) and panel (c)) represent the sector indices. The more detailed} list of sector names and codes is reported in Table \ref{table_WIOD_industries}.}
  \label{average_connectivity_intensity}
\end{figure}

To examine how strongly a layer interacts with another one, we measure the average connectivity and intensity between them. 
 Figure \ref{average_connectivity_intensity} shows again distinct behaviors in the two facets of the multilayer  network. On the one hand,  in the binary version, the density of the connections between many layers is very high \added{(see panels (a) and (b) of Figure \ref{average_connectivity_intensity})}.\footnote{For illustration purpose, in panels (a) and (c) of Figure \ref{average_connectivity_intensity}) we only report the sector indices in the rows and columns of the two color-coded matrices. The list of the corresponding sector names and codes is provided in Table \ref{table_WIOD_industries}} On the other hand, once the link weights (i.e. the traded amounts) are considered, average intensities are highly heterogeneous. In particular, as shown in panels (c) and (d) of Figure \ref{average_connectivity_intensity},  few pairs of layers interact more intensively, suggesting that these industry-based layers trade more between themselves than with the others. \added{We can identify, for example,  “Mining and quarrying  (B)” and “Manufacture of refined petroleum products (C19)”, ``Crop and animal production, hunting and related service activities (A01)"  and ``Manufacture of food products, beverages and tobacco products (C10-C12)", ``Manufacture of other non-metallic mineral products (C23)" and ``Construction (F)", ``Mining and quarrying  (B)” and ``Electricity, gas, steam and air conditioning supply (D35)" as the couples of industries that, on average, interact more intensively. Intuitively, this illustrates the ``natural and universal" input-output dependencies of some industries on  important factors of production provided by their counterparts.}

\textit {Layer overlaps and correlations.}

\begin{figure} [H]
\centering
\captionsetup[subfloat]{farskip=0pt,captionskip=0pt}
\subfloat[]{\includegraphics[height=2.5in,width = 3in]{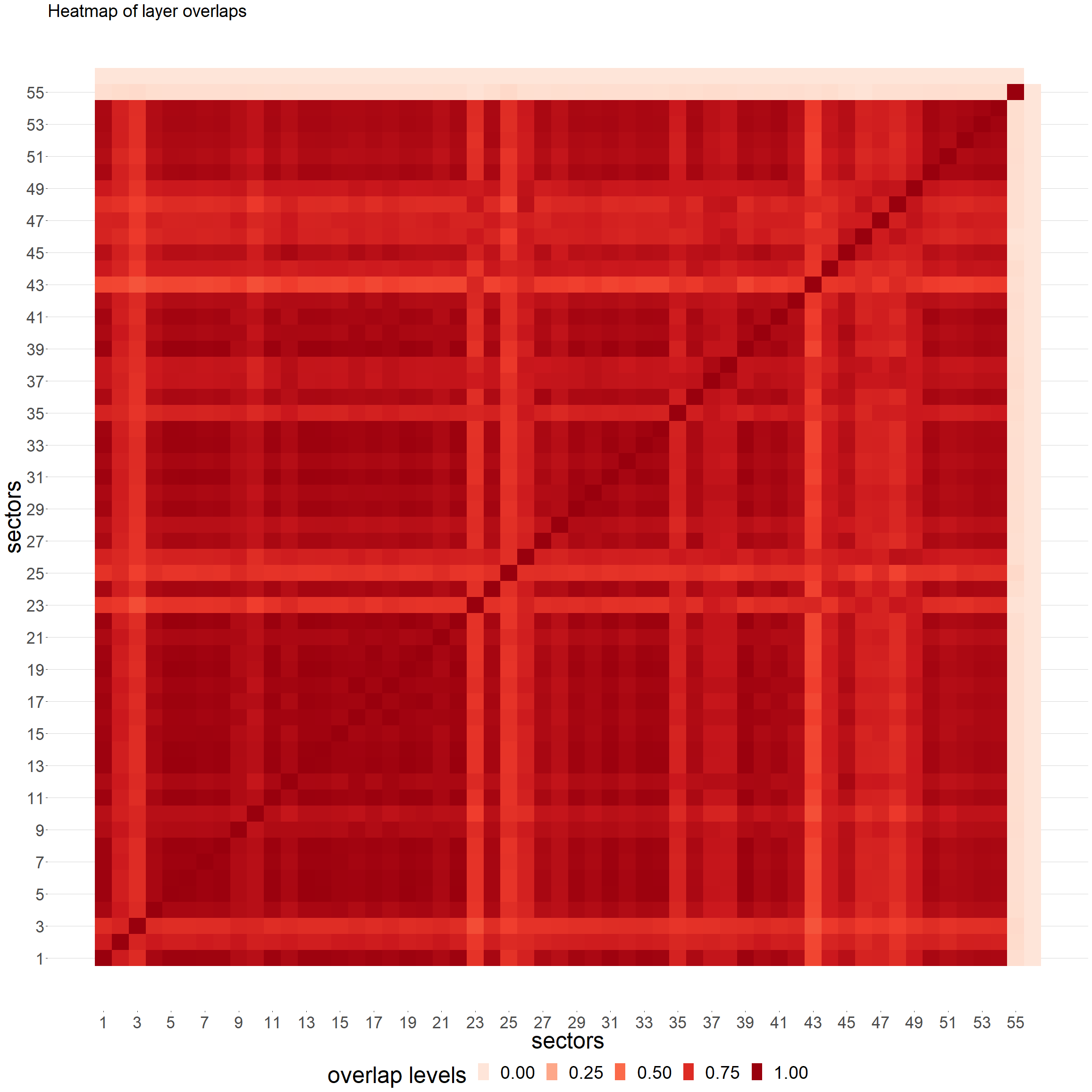}}
\subfloat[]{\includegraphics[height=2.5in,width = 3in]{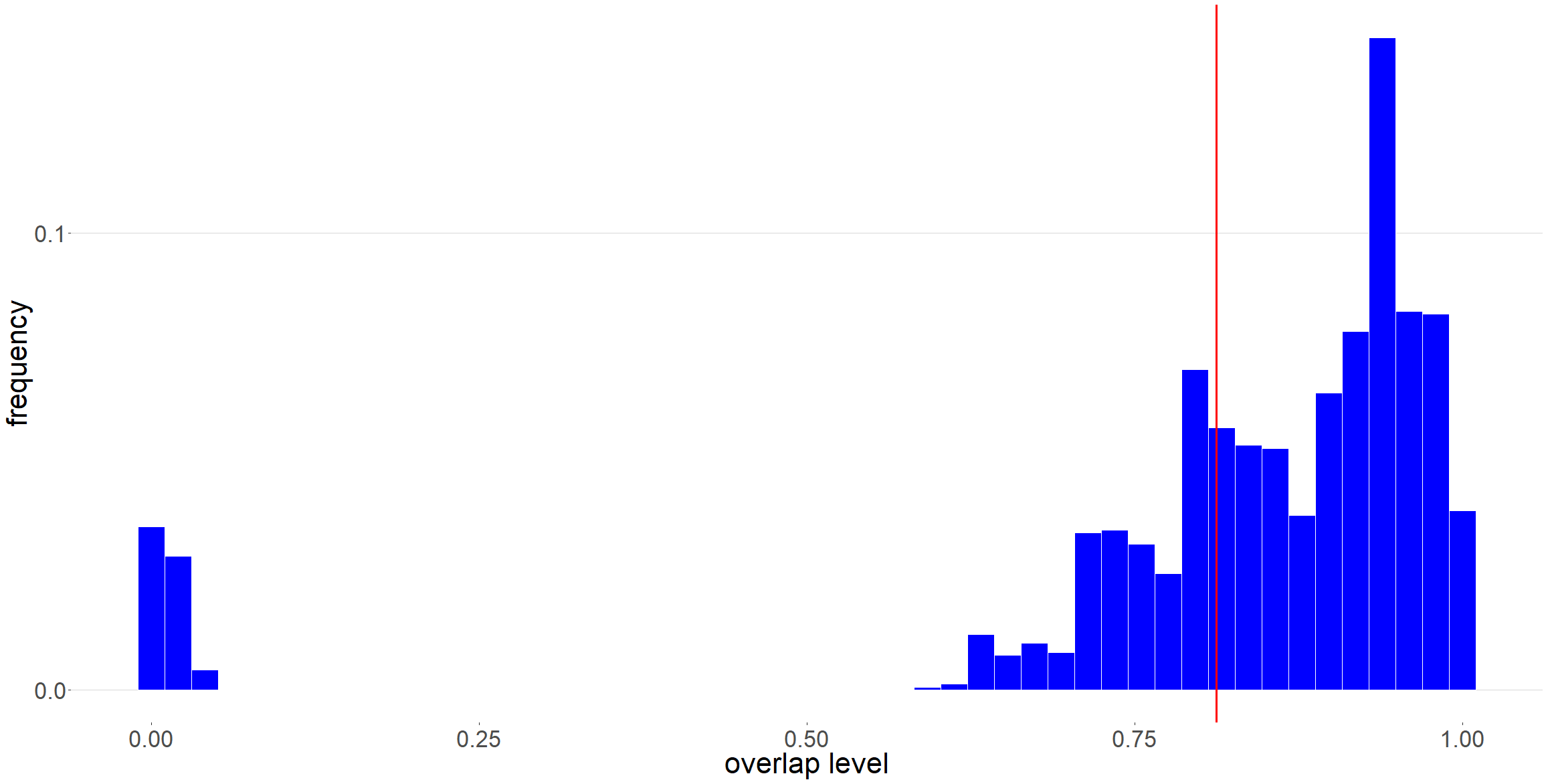}}\\
\subfloat[]{\includegraphics[height=2.5in,width = 3in]{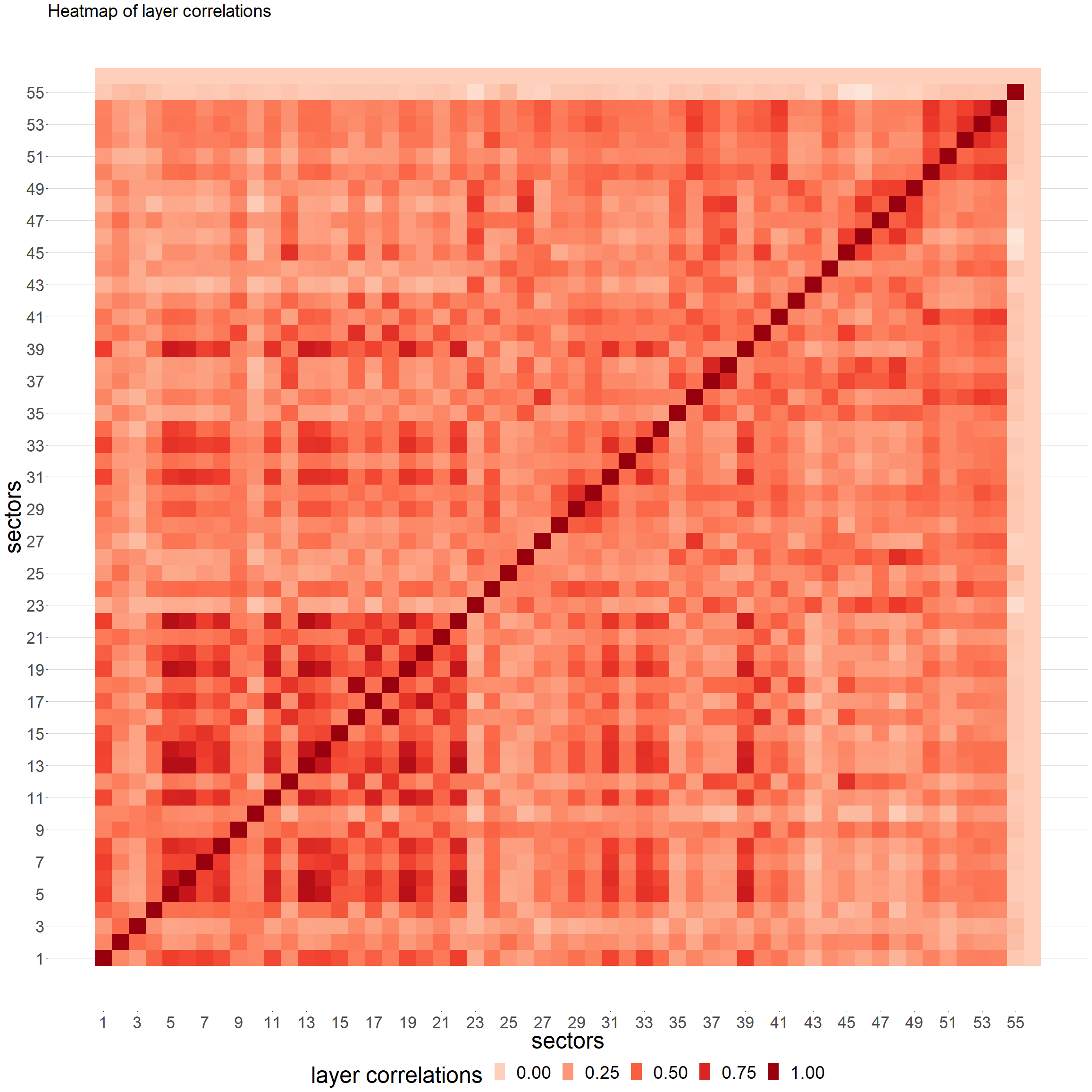}}
\subfloat[]{\includegraphics[height=2.5in,width = 3in]{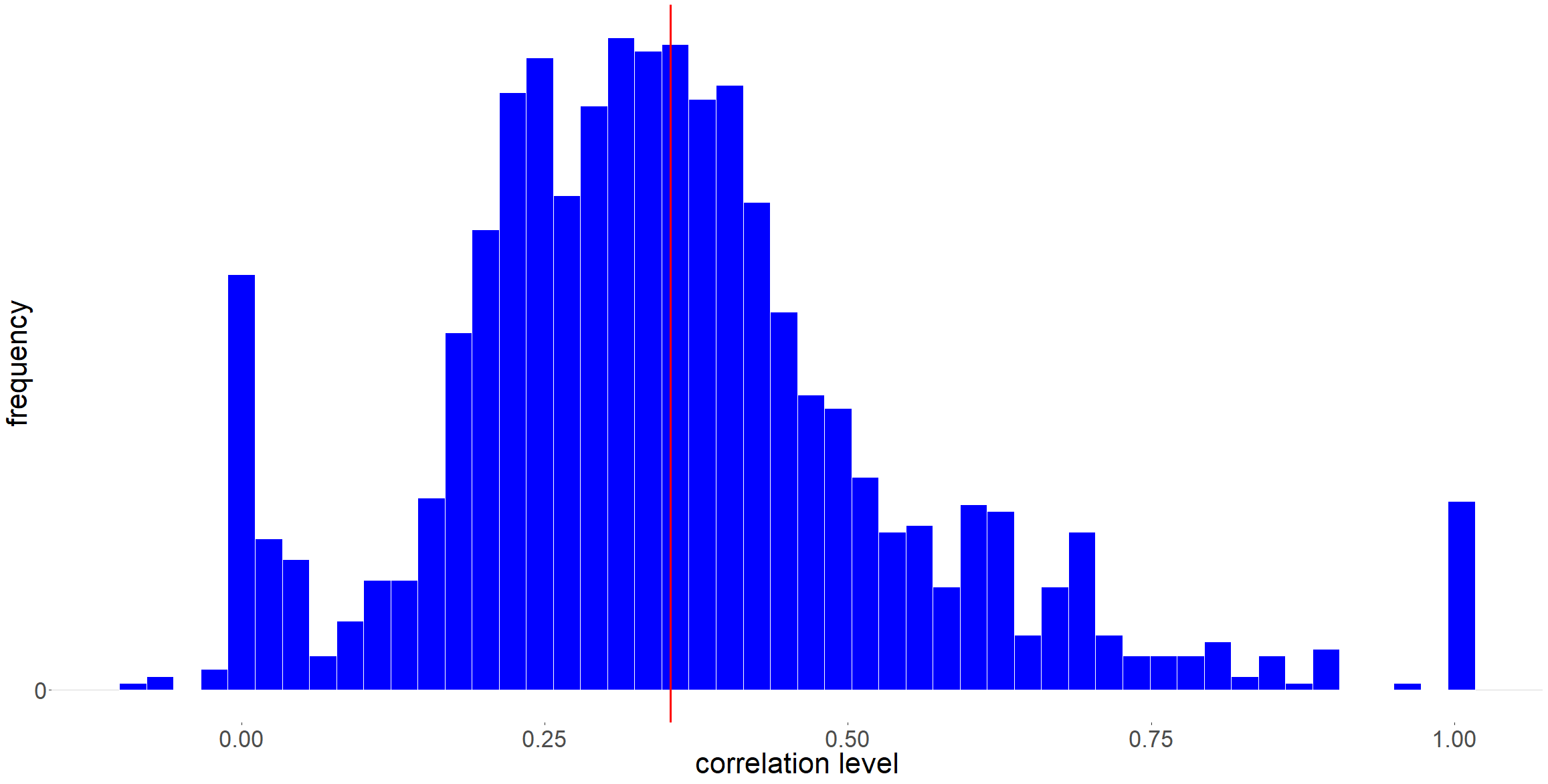}}
  \caption{Overlaps and correlations between layers in the binary version . The methods used to compute the overlap and correlation coefficients are explained in  (\ref{layer_interrelation}). Layers represent 56 sectors, while nodes in each layer are 44 countries.  \added{The rows and columns of the two color-coded matrices (panel (a) and panel (c)) represent the sector indices. The more detailed} list of sector names and codes is reported in Table \ref{table_WIOD_industries}.}
  \label{layers_overlaps_correlations_bin}
\end{figure}

\begin{figure} [H]
\centering
\captionsetup[subfloat]{farskip=0pt,captionskip=0pt}
\subfloat[]{\includegraphics[height=2.5in,width = 3in]{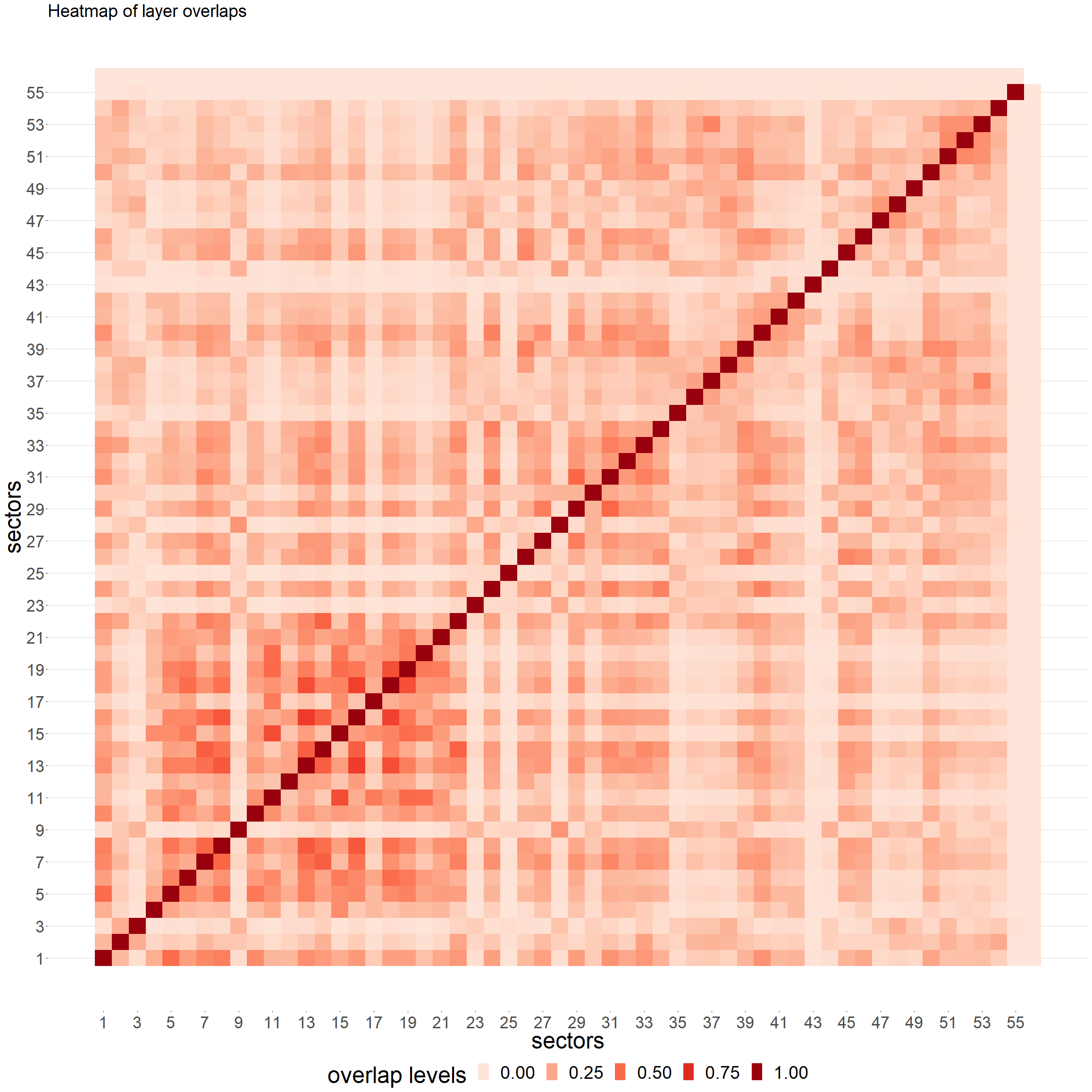}}
\subfloat[]{\includegraphics[height=2.5in,width = 3in]{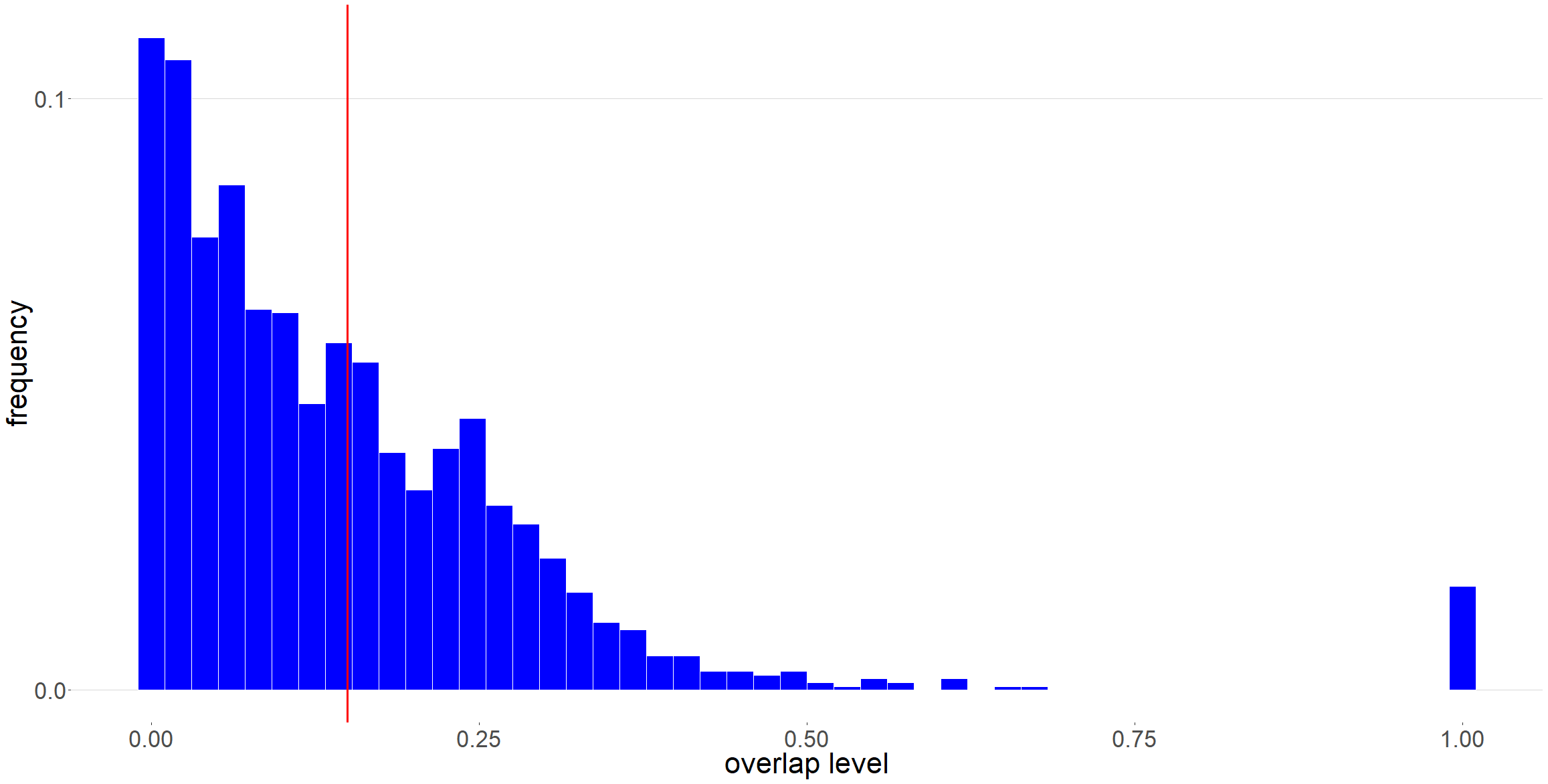}}\\
\subfloat[]{\includegraphics[height=2.5in,width = 3in]{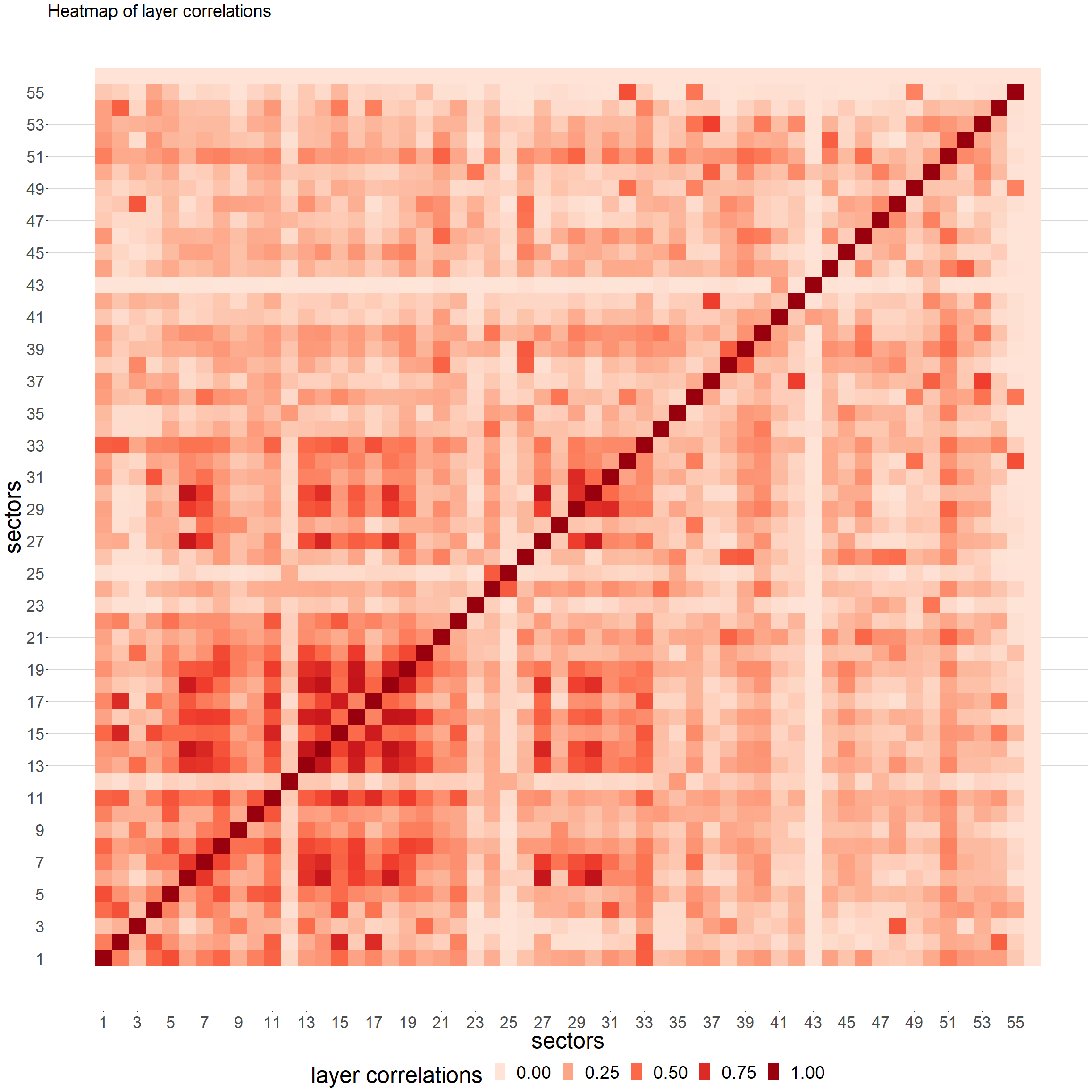}}
\subfloat[]{\includegraphics[height=2.5in,width = 3in]{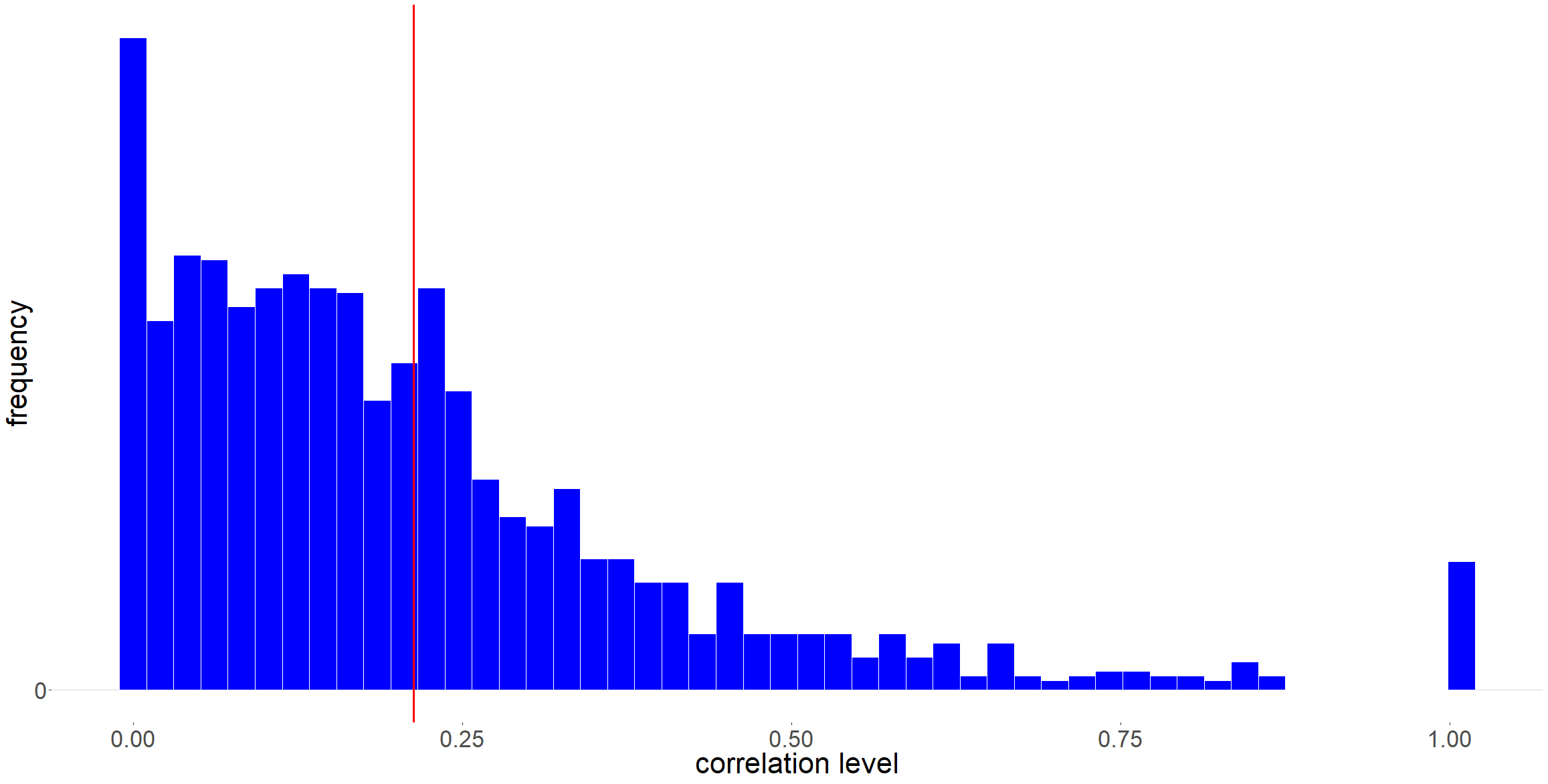}}
  \caption{Overlaps and correlations between layers  in the weighted version. The methods used to compute the overlap and correlation coefficients are explained in (\ref{layer_interrelation}). Layers represent 56 sectors, and nodes in each layer are 44 countries. \added{The rows and columns of the two color-coded matrices (panel (a) and panel (c)) represent the sector indices. The more detailed} list of sector names and codes is reported in Table \ref{table_WIOD_industries}.}
  \label{layers_overlaps_correlations_weight}
\end{figure}

\begin{figure} [H]
\centering
\captionsetup[subfloat]{farskip=0pt,captionskip=0pt}
\subfloat[for binary version]{\includegraphics[height=3.5in,width = 5.5in]{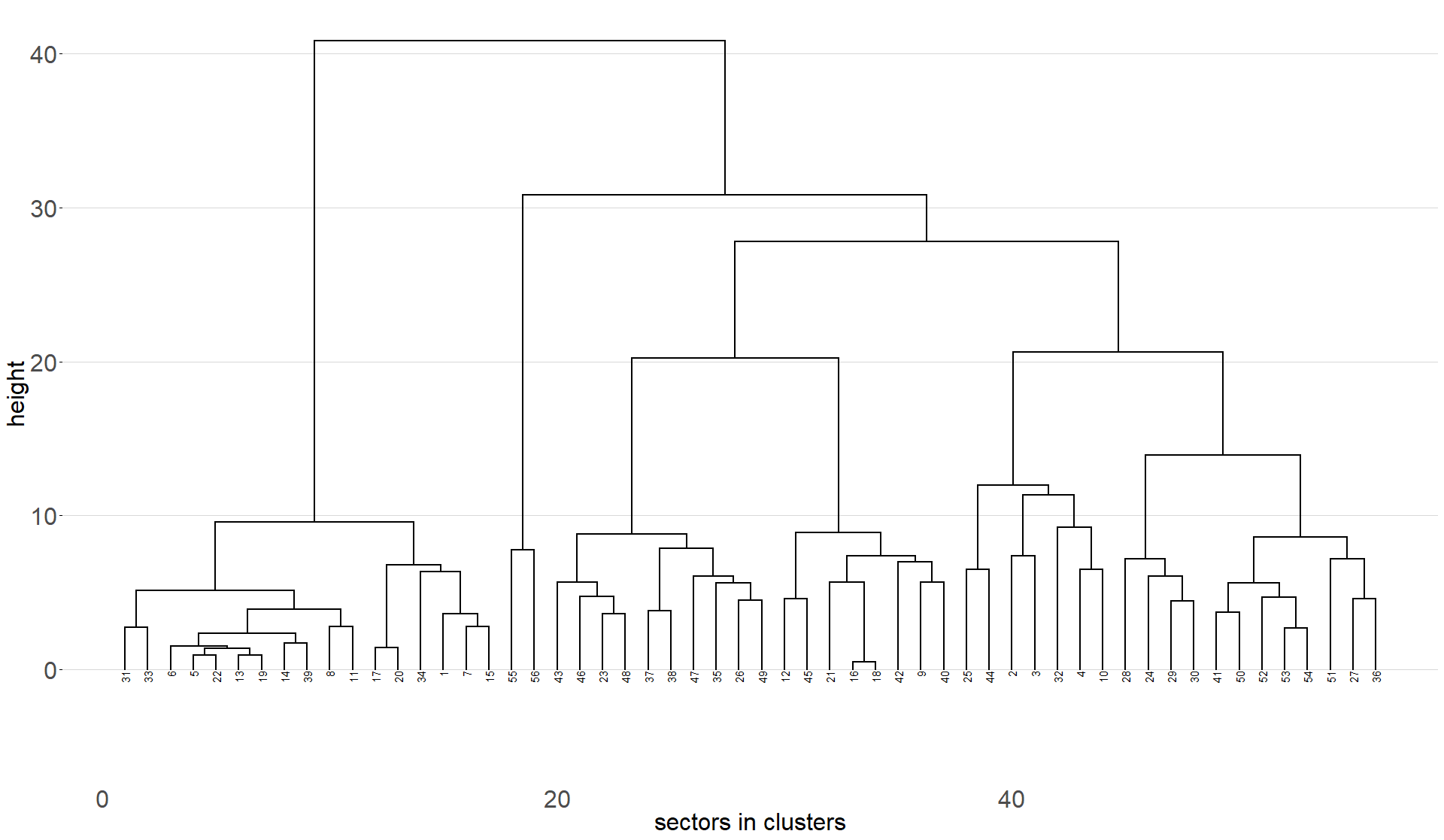}}\\
\subfloat[for weighted version]{\includegraphics[height=3.5in,width = 5.5in]{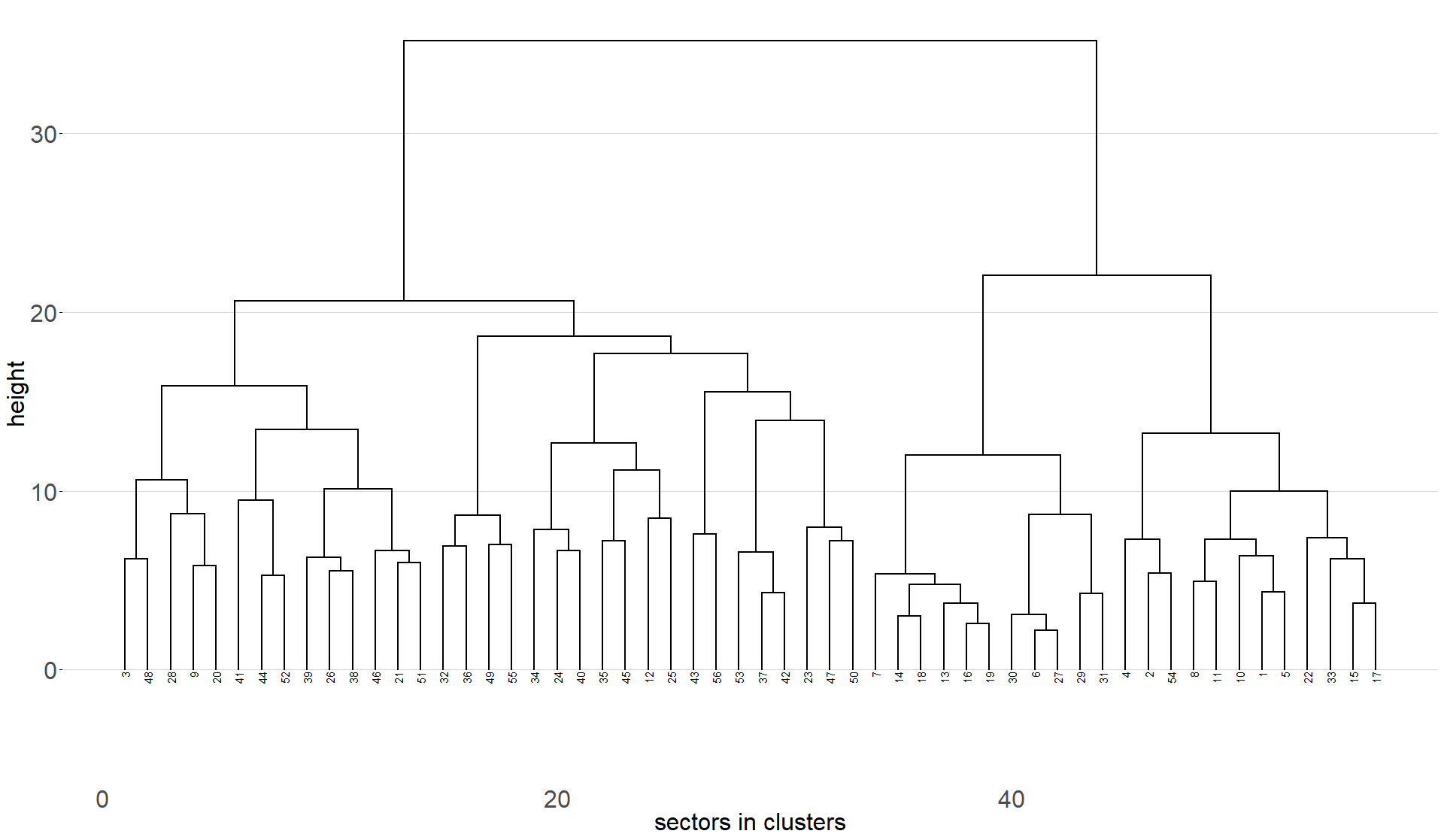}}\\
	\vspace{0.5cm}
  \caption{Dendrogram plot obtained from the layer-layer correlation matrices for the binary version (panel a) and the weighted version (panel (b)).}
  \label{dendogram_layer_correlations}
\end{figure}

\begin{figure} [H]
\centering
\captionsetup[subfloat]{farskip=0pt,captionskip=0pt}
\subfloat[for binary version]{\includegraphics[height=3.5in,width = 5.5in]{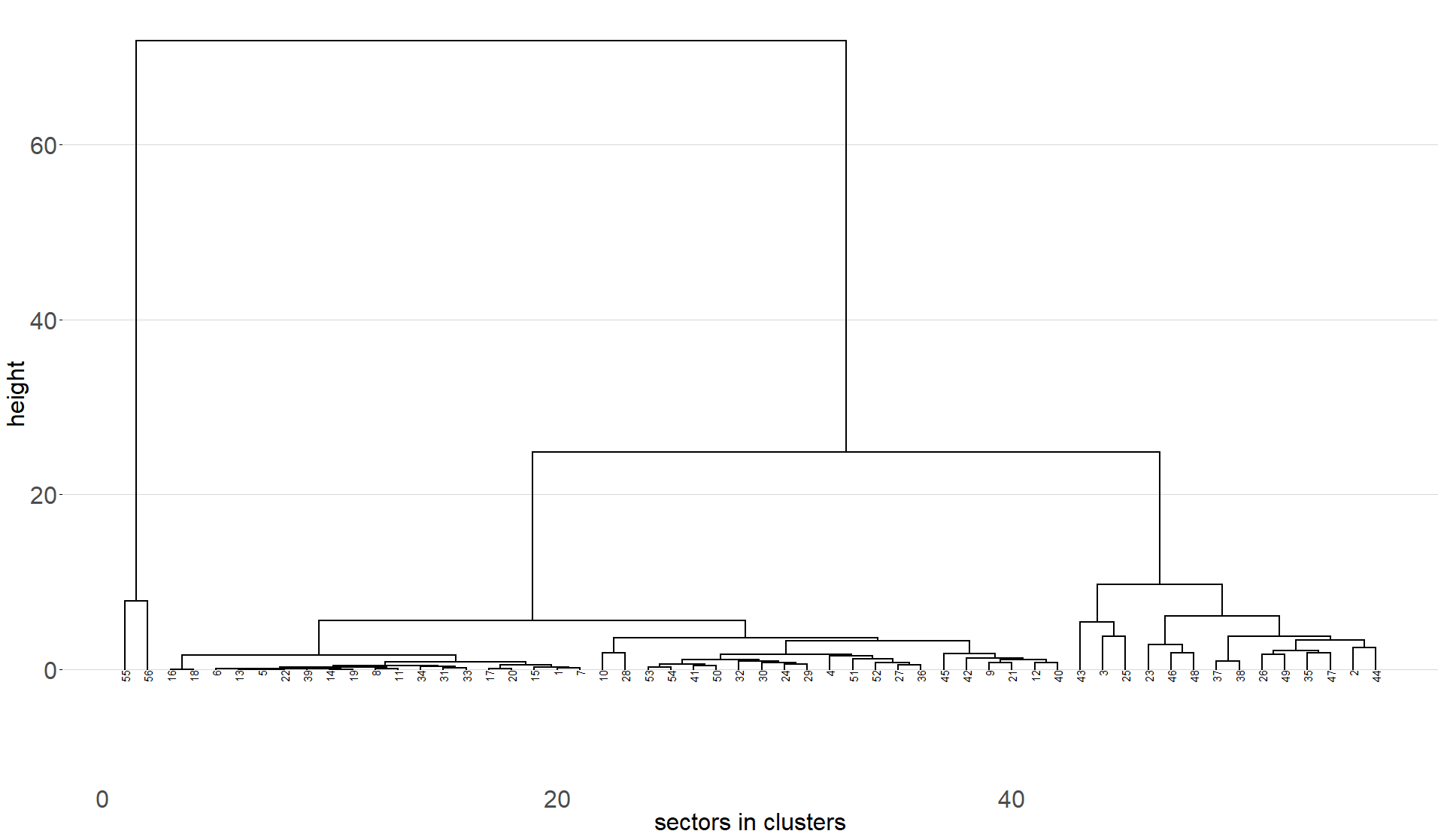}}\\
\subfloat[for weighted version]{\includegraphics[height=3.5in,width = 5.5in]{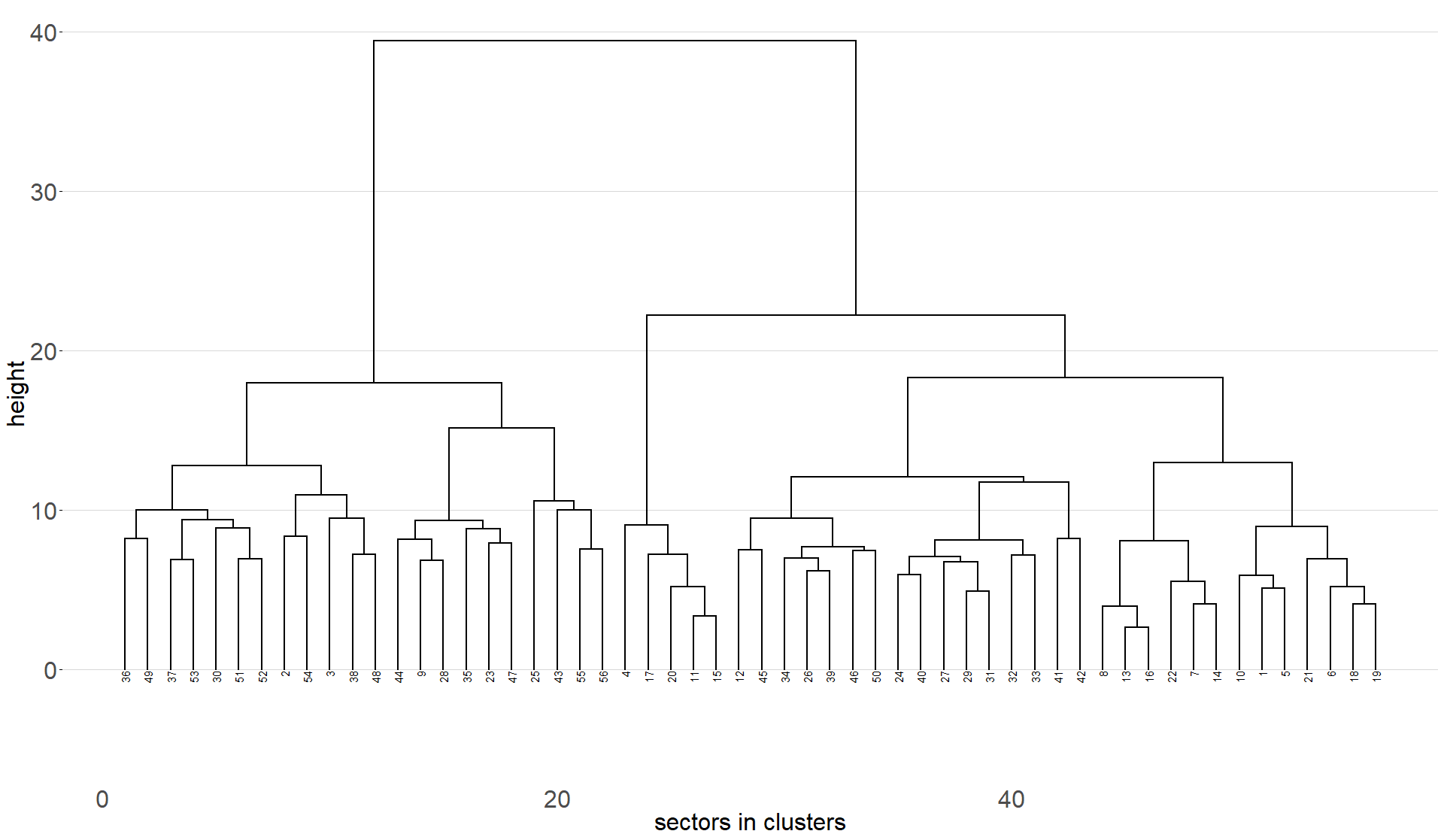}}\\
	\vspace{0.5cm}
  \caption{Dendrogram plot obtained from the layer-layer overlap matrices for the binary version (panel a) and the weighted version (panel (b)). }
  \label{dendogram_layer_overlaps}
\end{figure}

Comparing  across layers, we want to examine  whether layers exhibit similar internal structures.  In particular, if  trade relations among 44 countries  are strong in a particular layer \added{(sector)}, do they also exist and tend to form intensive relations in the other layers \added{(sectors)}?  To answer this question, we measure the overall overlap  coefficient and Pearson correlation coefficient for every pair of layers \citep [][]{Gemmetto_Garlaschelli_2014, Gemmetto_et_al_2016, Luu_Lux_2019}. In general, on the one hand, trade relations among $44$ countries in binary version are more similar across many couples of layers (see Figure \ref{layers_overlaps_correlations_bin}). This is not a very surprising result, since most of countries have a well diversified list of partners when they trade among themselves in each industry. On the other hand, results obtained from the weighted analysis show a more hierarchical structure in which several clusters of  layers exhibit a relatively higher level of similarity,  while  many other couples of layers are much less aligned  (see  Figures \ref{layers_overlaps_correlations_bin} and \ref{layers_overlaps_correlations_weight} as well as the dendrograms\footnote{The analysis of the layer correlation matrices ($\{R^{[\alpha,  \beta]}_{bin}\}$, $\{R^{[\alpha,  \beta]}_{w}\}$) and that of the layer overlap matrices ($\{O^{[\alpha,  \beta]}_{bin}\}$, $\{O^{[\alpha,  \beta]}_{w}\}$)  reveal that some layers are more correlated or overlapped. To provide a more detailed refinement and visualization of clustering \added{behaviors} among them,  we report in Figures \ref{dendogram_layer_correlations} and  \ref{dendogram_layer_overlaps}  the dendrograms obtained from these four matrices.  In these figures,  the euclidean distance method and average agglomeration method are used to plot the dendrograms.} shown in Figures \ref{dendogram_layer_correlations} and  \ref{dendogram_layer_overlaps}). From this perspective, the way countries trade among themselves is, therefore, not the same in all layers.

\clearpage

\added{In summary, our analysis of the basic network properties reveals that  the global trade  network in input-output possesses a complex architecture. The network exhibits a complex and mixed nature of the  trade  diversification as well as specialization of different countries in different  sectors.\\
First, binary and the weighted versions have distinct topological features. The results obtained from the binary analysis reveal that a number of industries establish trade links with many partners from different countries. Nevertheless, 
once the intensity of trade relations is taken into account, we observe a more hierarchical structure of interactions between sectors and countries.  The analysis of the weighted version also reveals a heterogeneity in the way they select their input providers and their output customers.  These observations are somewhat similar to those found from different empirical analyses of the aggregated World Trade network \citep [see][] {Garlaschelli_Loffredo_2005, Serrano_et_al_2007, Fagiolo_et_al_2008, Benedictis_Tajoli_2009}.} 

\added{In addition, the presence of an already well diversified trading list for many industries and countries in the densely connected binary version suggests that what mainly govern the temporal dynamics of the global trade network in input-output are not the extensive margin effects (i.e. obtaining new partners through new links). Instead, the intensive margin effects (i.e., when the intensities of existing trade linkages in the weighted version are adjusted and/or strengthened over time) may play a more prominent role in explaining the changes in the network structure\citep[][]{Felbermayr_Kohler_2006, Helpman_et_al_2008}.}
Furthermore, once various types of trade linkages are considered in a multilayer architecture, we find that the interactions between layers vary across pairs of  layers in the weighted version. On top of that, although many layers have strongly dissimilar internal structures,  few others  are somewhat more overlapped or correlated. \added{Therefore, the trade interrelations among countries, while being not homogeneous in general, may span over a certain number of industries.} \added{Interestingly, to a certain extent, this property  is also similar to what has been found in the multiplex trade network in which each layer is associated with a particular commodity \citep [see][]{Barigozzi_et_al_2010, Gemmetto_Garlaschelli_2014, Gemmetto_et_al_2016}. This again confirms a complex and mixed nature of the  diversification as well as the specialization of nations in different trading layers (industries or commodities) that cannot be detected if we only consider the aggregated trade network.}

\clearpage

\newpage
\section{Additional community structure results}
\label{Appendix D}
 We report some additional tables and figures related to the application of the proposed methodology of community detection.

\begin{table}[!h]
\footnotesize
	\begin{center}
		\begin{tabular}{lccccc}
			\hline\hline
			Countries & Number of Sectors& Number of Sectors & Number of Sectors & Community &  Gini \\ 
			 & in Community 1 & in Community 2 & Isolated & with highest & heterogeneity\\
			 
			 &  &  &  & number of sectors &  Index\\ 

			\hline
 
AUS	&33	&0	&18	&1	&64.9\% \\
AUT	&2	&1	&20	&13	&72.5\% \\
BEL	&3	&0	&18	&22	&83.4\% \\
BGR	&41	&0	&15	&1	&46.4\% \\
BRA	&0	&0	&13	&5	&46.3\% \\
CAN	&0	&35	&16	&2	&60.5\% \\
CHE	&2	&0	&20	&14	&72.6\% \\
CHN	&41	&1	&10	&1	&46.1\% \\
CYP	&0	&2	&23	&16	&81.1\% \\
CZE	&0	&30	&17	&2	&69.9\% \\
DEU	&37	&6	&11	&1	&55.1\% \\
DNK	&5	&0	&14	&10	&62.1\% \\
ESP	&1	&0	&10	&3	&40.9\% \\
EST	&0	&0	&20	&18	&77.5\% \\
FIN	&2	&0	&16	&6	&53.8\% \\
FRA	&1	&36	&10	&2	&58.0\% \\
GBR	&8	&33	&12	&2	&62.9\% \\
GRC	&38	&0	&18	&1	&54.0\% \\
HRV	&0	&0	&16	&7	&53.8\% \\
HUN	&0	&1	&27	&24	&87.7\% \\
IDN	&0	&0	&15	&9	&58.3\% \\
IND	&36	&0	&20	&1	&58.7\% \\
IRL	&1	&12	&29	&2	&94.0\% \\
ITA	&2	&0	&11	&4	&40.9\% \\
JPN	&39	&0	&9	&1	&50.2\% \\
KOR	&46	&0	&8	&1	&32.4\% \\
LTU	&0	&0	&23	&15	&74.7\% \\
LUX	&0	&2	&44	&39	&98.7\% \\
LVA	&0	&0	&18	&17	&76.1\% \\
MEX	&0	&33	&17	&2	&64.6\% \\
MLT	&0	&0	&30	&25	&95.0\% \\
NLD	&11	&2	&21	&23	&80.6\% \\
NOR	&28	&0	&19	&1	&74.1\% \\
POL	&0	&45	&11	&2	&35.4\% \\
PRT	&30	&0	&14	&1	&70.2\% \\
ROU	&0	&0	&16	&19	&78.3\% \\
RUS	&0	&28	&26	&2	&74.9\% \\
SVK	&0	&4	&19	&20	&80.2\% \\
SVN	&0	&0	&17	&12	&66.8\% \\
SWE	&3	&0	&10	&11	&63.9\% \\
TUR	&0	&0	&18	&8	&54.0\% \\
TWN	&2	&0	&14	&21	&79.3\% \\
USA	&0	&41	&12	&2	&46.2\% \\
ROW	&31	&6	&5	&1	&66.9\% \\ \\

\hline
\end{tabular}\end{center}
\caption  {In this Table, we report for each country the number of sectors that belong to the two largest communities (1 and 2, respectively), the number of sector that does not belong to any community (i.e. isolated), the indication of the community with the highest number of sectors of that country and the Gini heterogeneity index to measure the dispersion of sectors along the communities in the same country.}
\label{table_cou}
\end{table}

\begin{table}[!tbp]
\vspace{-1cm}
\scriptsize
	\begin{center}
		\begin{tabular}{lccccc}
			\hline\hline
		Sectors & Number of Countries& Number of Countries & Number of Countries & Community &  Gini \\ 
			& in Community 1 & in Community 2 & Isolated & with highest & heterogeneity\\
			
			&  &  &  & number of countries &  Index\\ 
			
			\hline

A01	&10	&5	&0	&1	&92.0\% \\
A02	&8	&5	&10	&1	&94.3\% \\
A03	&4	&1	&32	&1	&98.7\% \\
B	&21	&6	&7	&1	&74.8\% \\
C10-C12	&10	&5	&1	&1	&92.1\% \\
C13-C15	&4	&0	&35	&1	&98.9\% \\
C16	&9	&6	&4	&1	&92.7\% \\
C17	&6	&3	&19	&1	&96.8\% \\
C18	&8	&4	&6	&1	&94.5\% \\
C19	&17	&6	&8	&1	&82.4\% \\
C20	&11	&5	&16	&1	&91.8\% \\
C21	&9	&4	&18	&1	&94.3\% \\
C22	&6	&7	&21	&2	&95.1\% \\
C23	&11	&7	&4	&1	&90.1\% \\
C24	&8	&7	&13	&1	&93.0\% \\
C25	&9	&8	&6	&1	&91.4\% \\
C26	&6	&1	&30	&1	&97.5\% \\
C27	&5	&2	&32	&1	&98.2\% \\
C28	&5	&7	&19	&2	&95.5\% \\
C29	&6	&11	&14	&2	&90.8\% \\
C30	&4	&7	&27	&2	&96.3\% \\
C31-C32	&6	&4	&19	&1	&96.5\% \\
C33	&2	&2	&27	&1	&98.9\% \\
D35	&13	&6	&6	&1	&88.4\% \\
E36	&4	&1	&24	&1	&98.3\% \\
E37-E39	&2	&3	&22	&2	&98.3\% \\
F	&11	&8	&0	&1	&89.2\% \\
G45	&4	&6	&18	&2	&96.4\% \\
G46	&12	&8	&1	&1	&88.1\% \\
G47	&11	&9	&0	&1	&88.3\% \\
H49	&11	&8	&2	&1	&89.3\% \\
H50	&8	&3	&23	&1	&95.6\% \\
H51	&7	&7	&11	&1	&94.0\% \\
H52	&11	&8	&2	&1	&89.3\% \\
H53	&6	&6	&21	&1	&95.7\% \\
I	&11	&6	&2	&1	&90.6\% \\
J58	&6	&6	&7	&1	&95.0\% \\
J59-J60	&7	&5	&11	&1	&95.1\% \\
J61	&10	&7	&9	&1	&91.4\% \\
J62-J63	&9	&8	&7	&1	&91.5\% \\
K64	&11	&10	&0	&1	&87.3\% \\
K65	&9	&10	&7	&2	&89.7\% \\
K66	&6	&9	&12	&2	&93.1\% \\
L68	&11	&8	&1	&1	&89.2\% \\
M69-M70	&10	&7	&10	&1	&91.4\% \\
M71	&8	&7	&15	&1	&93.4\% \\
M72	&2	&4	&30	&2	&98.6\% \\
M73	&6	&8	&13	&2	&94.0\% \\
M74-M75	&6	&2	&23	&1	&97.3\% \\
N	&8	&9	&2	&2	&91.1\% \\
O84	&10	&8	&2	&1	&90.3\% \\
P85	&9	&7	&8	&1	&92.3\% \\
Q	&8	&4	&5	&1	&94.5\% \\
R-S	&10	&7	&4	&1	&91.1\% \\
T	&1	&0	&40	&1,7,10,16	&99.8\% \\
U	&0	&0	&44	&	&100.0\% \\

\hline
\end{tabular}\end{center}
\caption  {For each sector, the Table reports the number of countries that belong to the two largest communities (1 and 2, respectively), the number of countries that does not belong to any community (i.e. isolated), the indication of the community with the highest number of countries of that sector and the Gini heterogeneity index to measure the dispersion of countries along the communities in the same sector.}
\label{table_sec}
\end{table}

\begin{figure}
\includegraphics[height=7in,width=7in]{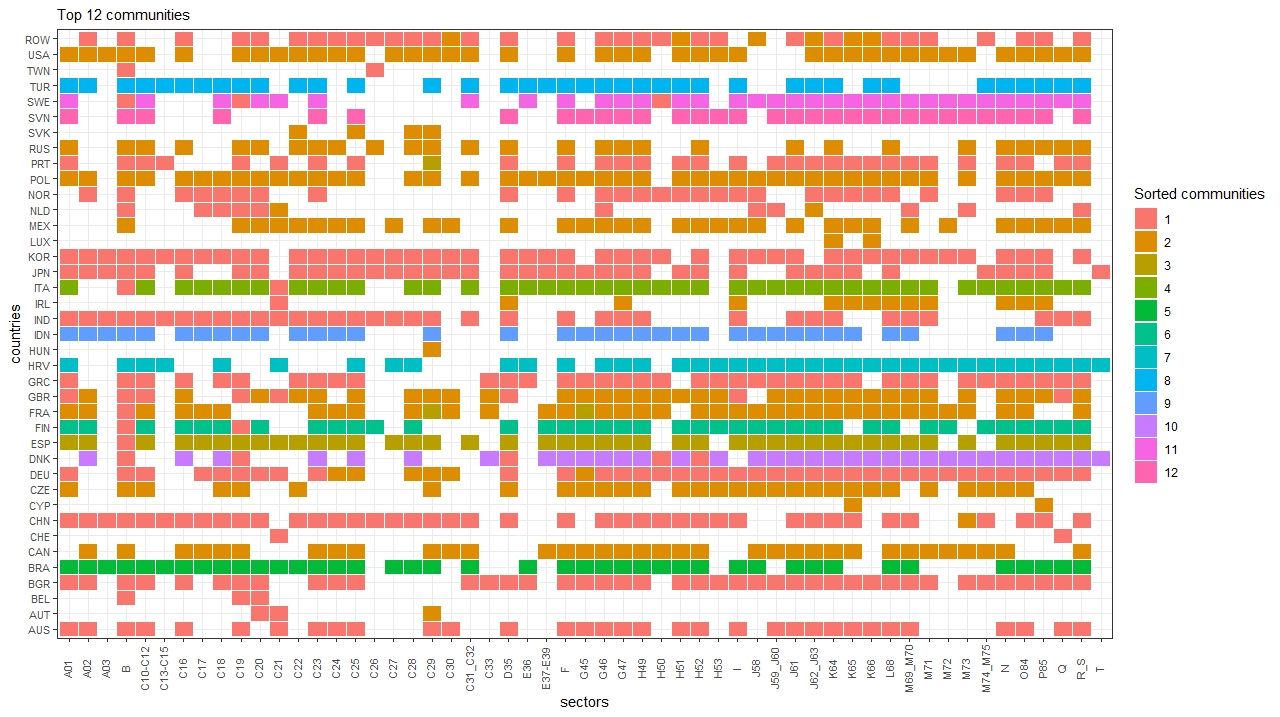}
	\caption{Membership (in terms of countries and sectors) of larger communities with at least 30 nodes. The communities have been sorted according to the number of nodes. Empty cells represent nodes that belong to smaller communities (lower than 30 nodes) or isolated nodes. The lists of sector names and  country names  are shown in Tables \ref{table_WIOD_industries} and  \ref{table_WIOD_countries}.}
\label{comm}
\end{figure}

\begin{figure} [H]
\centering
\captionsetup[subfloat]{farskip=0pt,captionskip=0pt}
\subfloat[Top industries in Community 1, 2000]{\includegraphics[height=3.2 in,width = 3.2in]{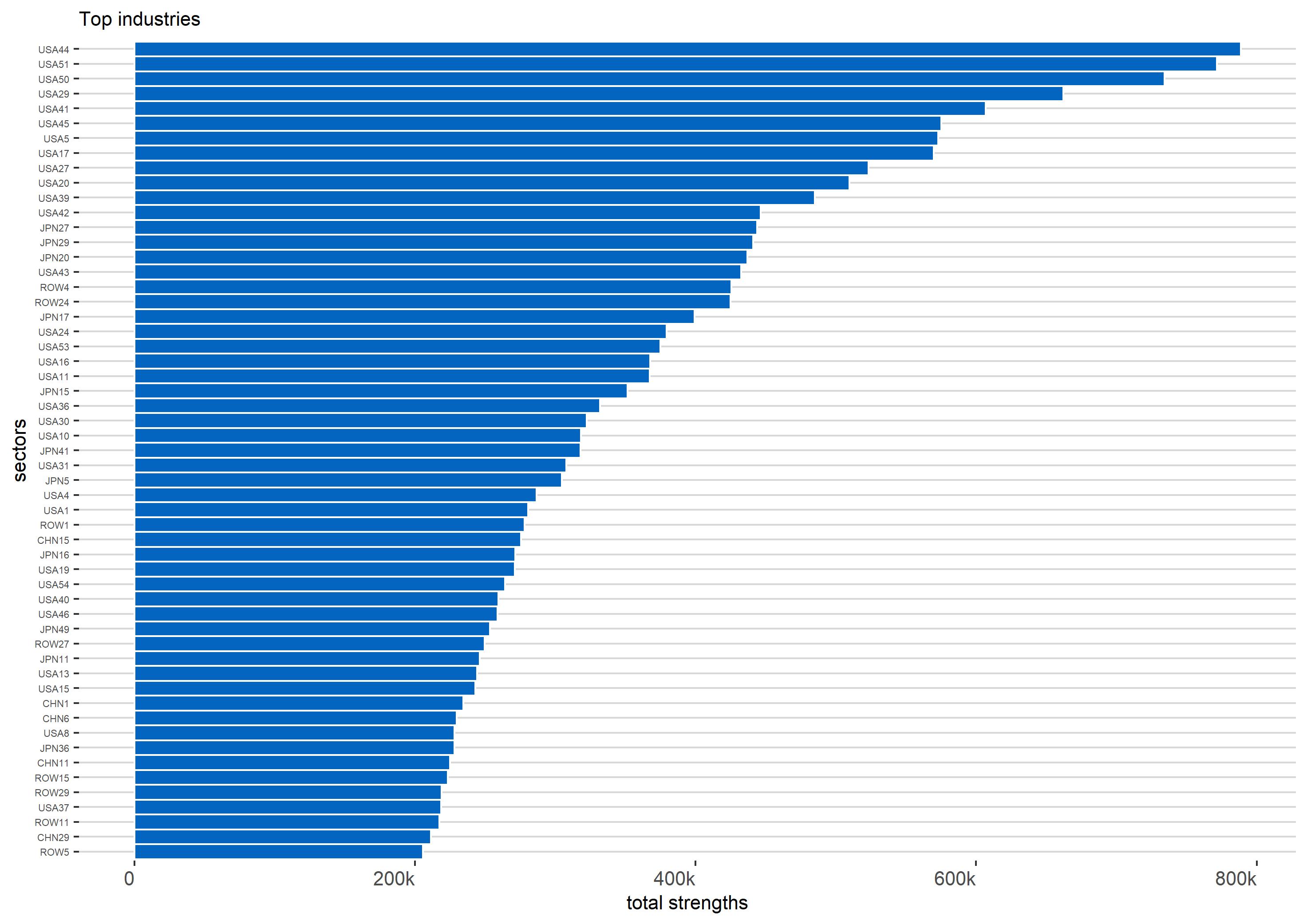}}
\subfloat[Top industries in Community 2, 2000]{\includegraphics[height=3.2  in,width = 3.2in]{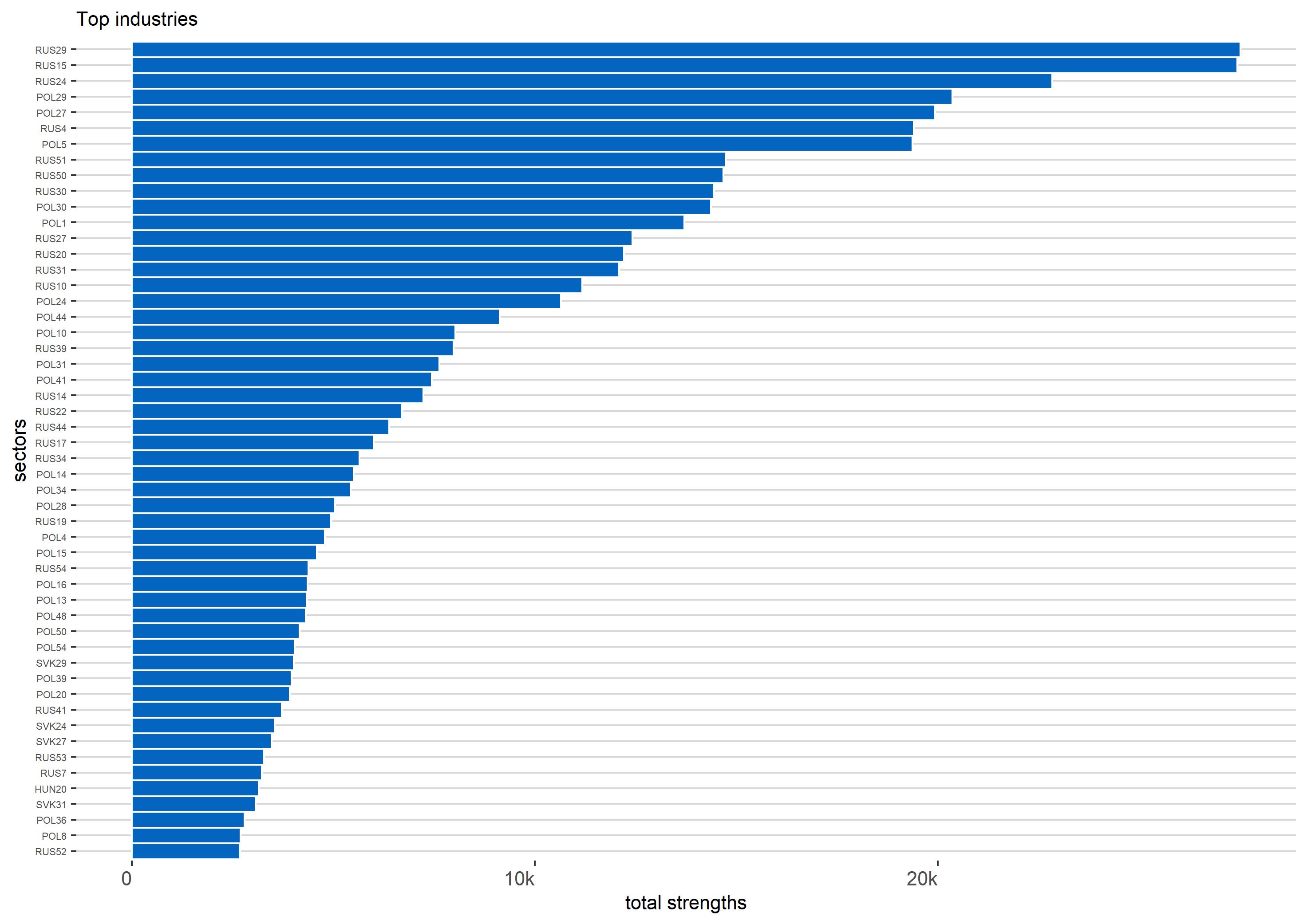}}
  \caption{Ranking of top industries in two important international communities in 2000, based on the (total) node strengths.}
    \label{top_sectors_two_largest_com_2000}
\end{figure}

\begin{figure} [H]
\centering
\captionsetup[subfloat]{farskip=0pt,captionskip=0pt}
\subfloat[Top industries in Community 1, 2002]{\includegraphics[height=3.2 in,width = 3.2in]{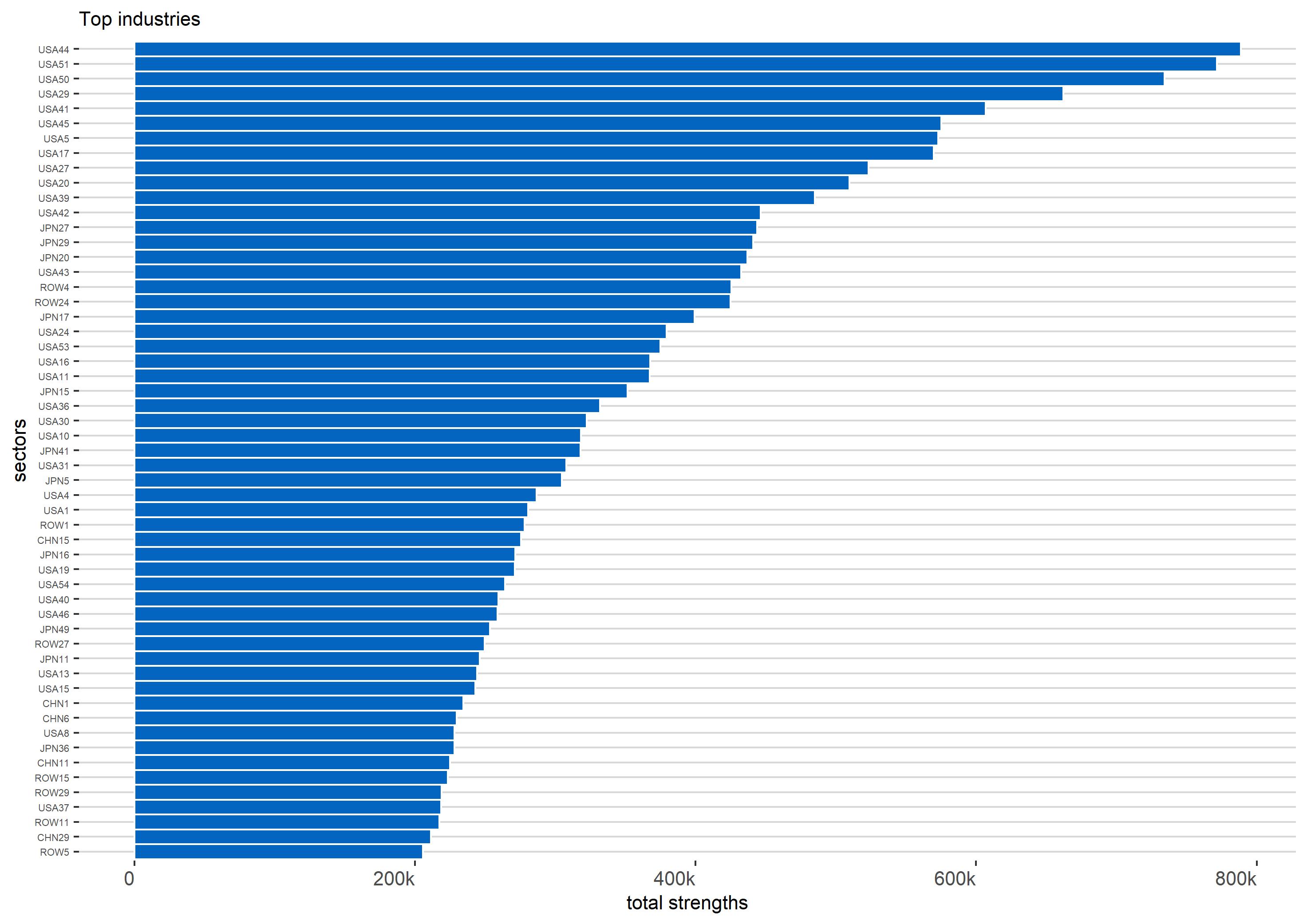}}
\subfloat[Top industries in Community 2, 2002]{\includegraphics[height=3.2  in,width = 3.2in]{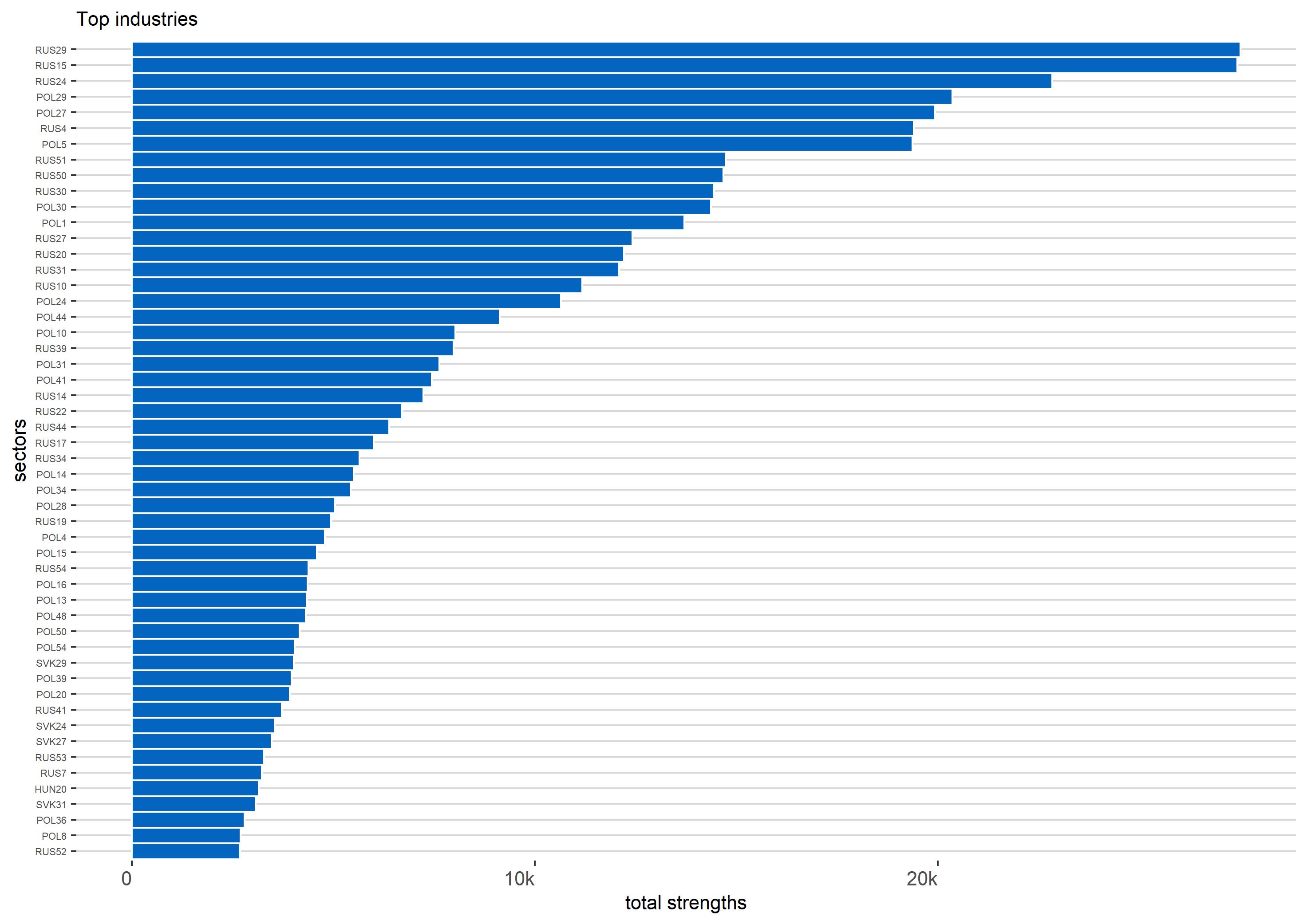}}
  \caption{\added{Ranking of top industries in two important international communities in 2002, based on the (total) node strengths.}}
    \label{top_sectors_two_largest_com_2002}
\end{figure}

\begin{figure} [H]
\centering
\captionsetup[subfloat]{farskip=0pt,captionskip=0pt}
\subfloat[Top industries in Community 1, 2004]{\includegraphics[height=3.5 in,width = 3.2in]{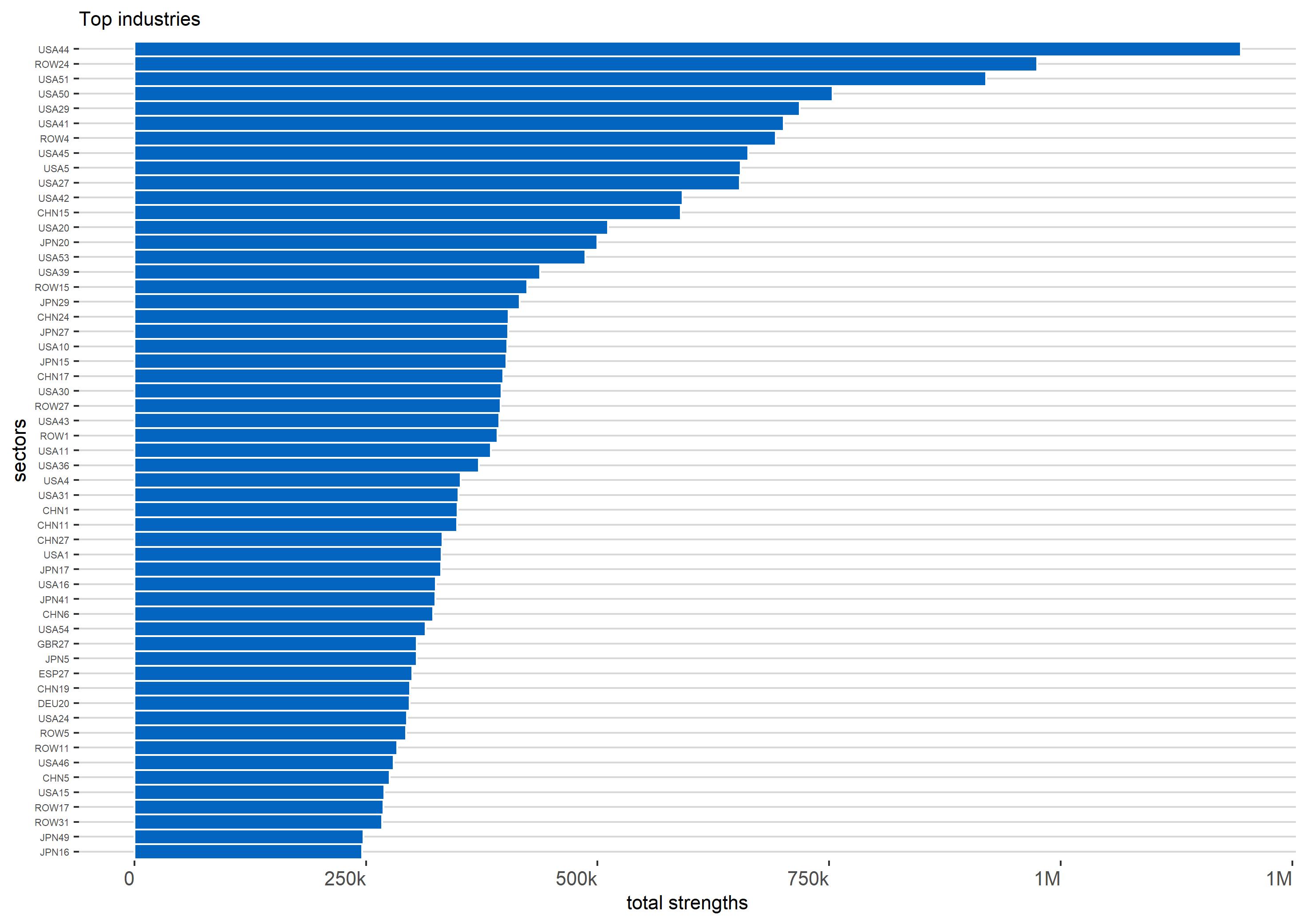}}
\subfloat[Top industries in Community 2, 2004]{\includegraphics[height=3.5 in,width = 3.2in]{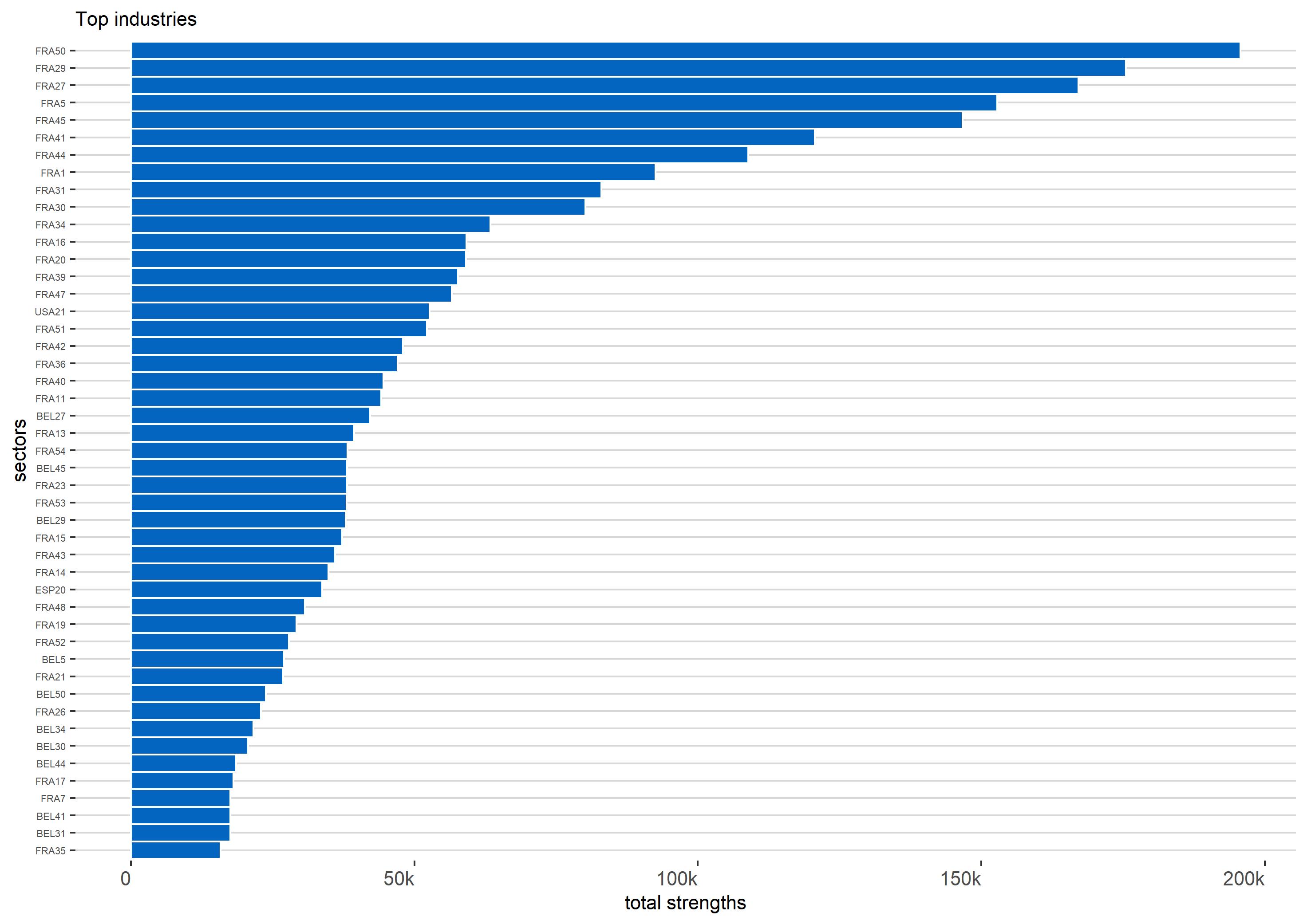}}
  \caption{Ranking of top industries in two important international communities in 2004, based on the (total) node strengths.}
    \label{top_sectors_two_largest_com_2004}
\end{figure}

\begin{figure} [H]
\centering
\captionsetup[subfloat]{farskip=0pt,captionskip=0pt}
\subfloat[Top industries in Community 1, 2009]{\includegraphics[height=3.5 in,width = 3.2in]{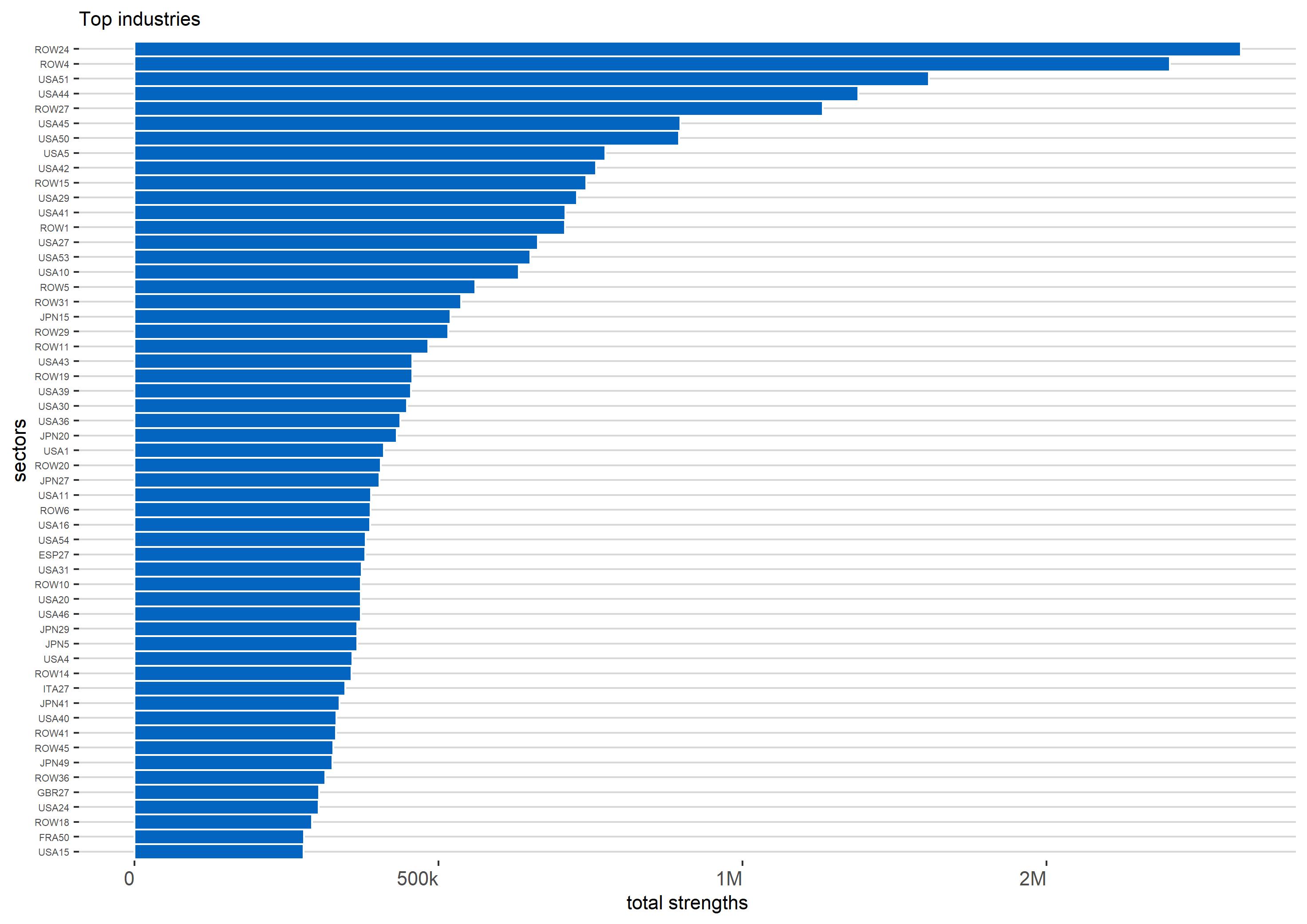}}
\subfloat[Top industries in Community 2, 2009]{\includegraphics[height=3.5 in,width = 3.2in]{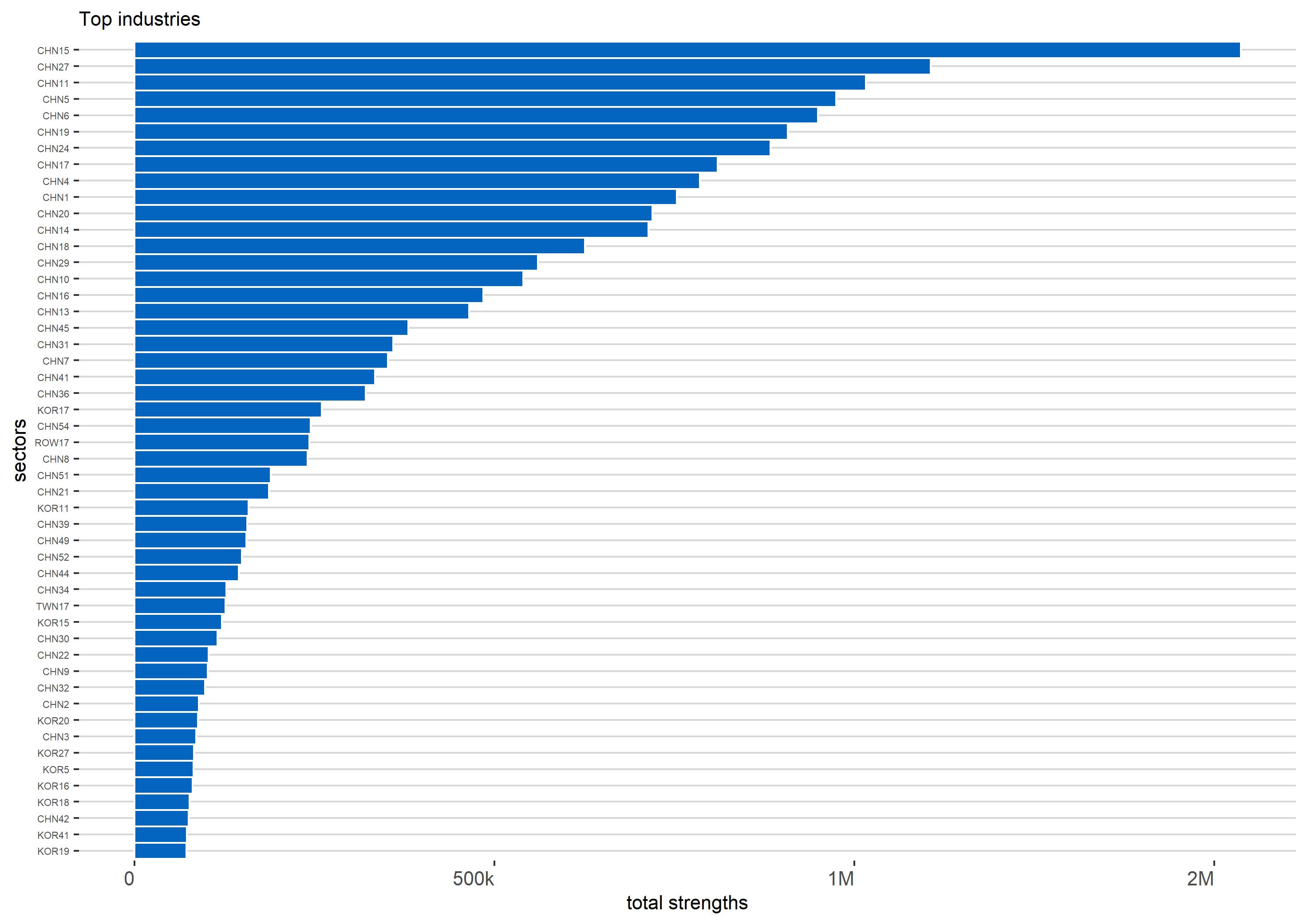}}
  \caption{Ranking of top industries in two important international communities in 2009, based on the (total) node strengths.}
    \label{top_sectors_two_largest_com_2009}
\end{figure}

\end{document}